\documentclass[apj]{emulateapj}
\usepackage{apjfonts}
\usepackage{graphicx}
\usepackage{psfig}
\begin{document}

\title{A New Class of Luminous Transients and A First Census \\
of Their Massive Stellar Progenitors}

\author{
Todd A.~Thompson\altaffilmark{1,2,3}, 
Jos{\'e}~L.~Prieto\altaffilmark{1,2},
K.~Z.~Stanek\altaffilmark{1,2}, 
Matthew~D.~Kistler\altaffilmark{2,4}, \\
John~F.~Beacom\altaffilmark{1,2,4}, \&
Christopher~S.~Kochanek\altaffilmark{1,2}
}

\altaffiltext{1}{Department of Astronomy, The Ohio State University, 140
W.\ 18th Ave., Columbus, OH 43210; thompson, prieto, kstanek, 
ckochanek@astronomy.ohio-state.edu.}

\altaffiltext{2}{Center for Cosmology \& Astro-Particle Physics, 
The Ohio State University, 191 W.\ Woodruff Ave., Columbus, OH 43210.}

\altaffiltext{3}{Alfred P.~Sloan Fellow}

\altaffiltext{4}{Department of Physics, The Ohio State University, 
191 W.\ Woodruff Ave., Columbus, OH 43210; kistler, 
beacom@mps.ohio-state.edu.}

\shorttitle{A New Class of Transients}
\shortauthors{Thompson et al.~2008}

\begin{abstract}

The progenitors of SN 2008S and the 2008 luminous transient in NGC 300
were deeply dust-enshrouded massive stars, with extremely red
mid-infrared colors and relatively low bolometric luminosities 
($\approx5\times10^4$\,L$_\odot$). The transients
were optically faint compared to normal core-collapse supernovae, with
peak absolute visual magnitudes of $-13\gtrsim M_V\gtrsim -15$, and their 
spectra exhibit narrow Balmer and [Ca II] emission lines.  
These events are unique among transient-progenitor pairs and hence constitute a new
class. Additional members of this class may include the M85 transient, SN 1999bw, 2002bu, 
and others.  Whether they are true supernovae or bright massive-star eruptions, we
argue that their rate is of order $\sim20$\% of the core-collapse supernova
rate in star-forming galaxies.  This fact is remarkable in light of the
observation that a {\it very} small fraction of all massive stars in any 
one galaxy, at any moment, have the infrared colors of the progenitors of SN 2008S
and the NGC 300 transient.  We show this by extracting mid-infrared
and optical luminosity, color, and variability properties of 
massive stars in M33 using archival imaging.  We find that the fraction of 
massive stars with colors consistent with the progenitors of SN 2008S
and the NGC 300 transient is $\lesssim10^{-4}$.  In fact, only 
$\lesssim10$ similar objects exist in M33 (and perhaps $\lesssim1$) 
--- all of which lie at the luminous red extremum of the AGB sequence.  
That these transients are simultaneously relatively {\it common} with respect 
to supernovae, while their progenitors are remarkably {\it rare} 
compared to the massive star population, implies that the dust-enshrouded phase is
a short-lived phenomenon in the lives of many massive stars.  This 
shrouded epoch can occur only in the last $\lesssim10^4$\,yr before explosion, 
be it death or merely eruption.  We discuss the implications of this finding 
for the evolution and census of ``low-mass'' massive stars (i.e.,
$\sim$$8-12$\,M$_\odot$), and we connect it with theoretical discussions 
of electron-capture supernovae near this mass range.  Other potential mechanisms,
including the explosive birth of massive white dwarfs, and massive
star outbursts are also discussed. A systematic 
census with (warm) {\it Spitzer} of galaxies in the local
universe ($D\lesssim10$ Mpc) for analogous progenitors
would significantly improve our knowledge of this channel to 
massive stellar explosions, and potentially to others with obscured progenitors.

\end{abstract}

\keywords{supernovae: general, individual (SN 2008S, 1999bw, 2002bu) 
--- surveys: stars, galaxies --- 
stars: evolution}


\section{Introduction}
\label{section:introduction}

Identifying the progenitors of core-collapse supernovae, the outbursts
of Luminous Blue Variables (LBVs), and other massive-star transients 
is essential for understanding the physics, demographics, variability,
evolution, and end-states of massive stars.  The problem of identifying
the progenitors of bright transients from massive stars is difficult, and
traditionally limited to serendipitous archival imaging of nearby
galaxies in the optical and near-infrared (e.g., Van Dyk et al.~2003;
Smartt et al.~2004; Li et al.~2007; see the extensive summary in
Smartt et al.~2009).  Progenitor searches are complemented by statistical 
studies of supernova environments within their host galaxies, which provide 
indirect evidence for associations between certain types of supernovae
and broad classes of progenitors (e.g., 
James \& Anderson 2006; Kelly et al.~2008;
Prieto et al.~2008; Anderson \& James 2008).  

A much more direct method for understanding the relation
between types of massive stars and their transients is to
catalog all of the massive stars in the local universe ($D\lesssim10$\,Mpc)
before explosion.
While surveys for bright optical transients in the local universe are 
well-developed (e.g., Li et al.~2001), a fairly complete census of the 
massive stars in nearby galaxies has only recently been proposed and undertaken 
(Massey et al.~2006; Kochanek et al.~2008).  Despite the technical 
challenges required by the depth, area, and cadence of the observations,
these surveys are critical for our understanding of the one-to-one 
correspondence between massive stars and their end-states, whether
they are successful or failed explosions (e.g., Kochanek et al.~2008).  
The long-term promise of these  surveys is to 
produce a catalog within which the characteristics of progenitors of 
future supernovae are listed pre-explosion, as in the 
case of SN 1987A (West et al.~1987; Menzies et al.~1987).   
They will provide an essential mechanism for understanding 
the direct causal mapping between individual progenitor types and 
their transients (see, e.g., Gal-Yam et al.~2007). 

Here, we describe a new link in this causal mapping. 
The discovery by Prieto et al.~(2008a) and Prieto (2008) of 
the dust-obscured progenitors of the luminous outbursts in 
NGC 6946 (SN 2008S; Arbour \& Boles 2008) and in NGC 300 (Monard 2008)
with {\it Spitzer}, opens up qualitatively new possibilities in the 
study of the connection between massive stars and their explosions. 
We show that these discoveries allow us to make a strong 
--- and perhaps unprecedented --- connection 
between a dust-enshrouded sub-population of massive stars and 
a new relatively common class of bright transients.  \\

The argument presented in this paper can be summarized in four points: \\

\noindent 1.~{\it The transients SN 2008S and NGC
300\footnote{Throughout this paper we denote ``the transient in NGC
300'' as ``NGC 300'' (e.g., ``the progenitor of NGC 300'')
unless we specifically refer to the host galaxy.}  
constitute a class.}  Both have peak absolute $V$-band
magnitude $M_V\approx-14\pm1$ ($\approx2-3$ magnitudes fainter than normal
core-collapse supernovae; e.g., Richardson et al.~2002), 
strong evidence for internal extinction in
their spectra, narrow emission lines (similar to low-luminosity
Type IIn supernovae; e.g., Filippenko 1997), and progenitors that are optically 
obscured and deeply dust-embedded (dust-reprocessed emission giving blackbody 
temperatures $\lesssim500$\,K), and with bolometric luminosities of
$\sim5\times10^4$\,$L_{\odot}$. In addition, they show little
infrared variability on a few-year baseline.  The details of this 
unique class of progenitor-transient pairs and its members, as well as a 
comparison with other classes of 
optical transients are presented in \S\ref{section:class}.  An in-depth
search for analogs to the progenitors of SN 2008S and NGC 300
in M33 is presented in \S\ref{section:m33}.\\

\noindent 2.~{\it Transients of this type are relatively common with 
respect to core-collapse supernovae.}  A total of
$\approx$\,22 core-collapse supernovae or supernova-like transients 
have been discovered within $\approx10$\,Mpc since 1999.  
Sixteen were confirmed supernovae, two were LBV eruptions, one  
was a Type IIn supernova (SN 2002bu
in NGC 4242, $D\approx8$\,Mpc; Puckett 
\& Gauthier 2002), whose relatively low peak magnitude 
($M_V\approx-15$; Hornoch 2002)  suggests 
a close similarity with the remaining two bright transients
(see \S\ref{section:other}), 
which are of primary interest in this paper:
SN 2008S ($D\approx5.6$\,Mpc) and NGC 300 ($D\approx1.9$\,Mpc), 
whose physical nature is uncertain.   They may be either true 
(but optically sub-luminous) supernovae, or a new class of massive 
star eruptions. Taken at face value, these numbers 
imply that the rate of SN 2008S-like transients is of order 
$\sim10$\% of the supernova rate.\footnote{Throughout this paper
we use ``supernova rate'' to mean the core-collapse supernova rate
unless we specifically mention the contribution from Type Ia
supernovae.}  Because of incompleteness,
the true rate is likely higher. We discuss the 
frequency of these events in detail in \S\ref{section:rates}.\\

\noindent 3.~{\it The progenitors of this class are extremely rare 
among massive stars at any moment, in any star-forming galaxy.}  
Although the bolometric
luminosities of the progenitors of SN 2008S and NGC 300 are unremarkable
for massive stars ($\sim5\times10^4$\,L$_\odot$), their colors put them in
a class consisting of less than $10^{-4}$ of all massive stars
($[3.6]-[4.5]\mu$m color $\gtrsim2.0$ and $\approx2.7$,
respectively). In a mid-infrared (MIR) color-magnitude diagram (see 
Figs.~\ref{fig:cmd} \& \ref{fig:cmd2}), these progenitors lie at the extremum 
of the AGB sequence in both luminosity and MIR color.  We 
refer to them as ``extreme AGB'' (EAGB) stars throughout this work.
Because of their relatively low bolometric luminosities, they 
are not $\eta$ Carinae, cool hypergiant, or classical LBV analogs
(see, e.g., Smith 2008).  
They are thus distinct from the ``supernova impostors,'' produced by 
bright outbursts of optically-luminous LBVs 
(e.g., SN 1997bs; Van Dyk et al.~2000, 2003). 
In \S\ref{section:m33} we present results from a comprehensive
survey of M33 for massive stars with properties  similar 
(in bolometric luminosity, obscuration, and 
variability) to the progenitors of SN 2008S and NGC 300.
We find remarkably few.  We compare with MIR surveys of the
LMC, SMC, and NGC 300.\\

\noindent 4.~{\it Conclusion: A large fraction of all massive stars 
undergo a dust-enshrouded phase within $\lesssim10^4$\,yr of explosion.}  
This is the most natural explanation for the facts of points (2) and (3) 
above. If these transients have a rate comparable to the supernova rate
($\sim20$\%; point 2), 
then the timescale for the obscured phase is determined by the ratio of 
the number of dust-obscured massive stars relative to the entire 
population ($\lesssim10^{-4}$; see \S\ref{section:m33}, 
\S\ref{section:discussion}) times the average lifetime of a massive 
star ($\sim10^7$\,yr).  
Importantly, from the rarity of SN 2008S-like progenitors 
alone ($\sim10^{-4}$ of massive stars; point 3), one would naively expect 
a comparable fraction of supernovae to have progenitors of this
type, if the dust-obscured phase occurs at a random time 
in the life of a massive star.  However, the relative frequency of these 
explosions (point 2) shows that this phase must come in the last 
$\lesssim10^4$\,yr, just before explosion.  Thus, there must be a 
causal relation between the occurrence of the
highly dust-enshrouded phase and eruption.  These points,
together with a discussion of the theory of the evolution of massive
stars, the potential connection with electron-capture supernovae,
white dwarf birth, and other hypotheses for the physical mechanism of 2008S-like 
explosions, as well as a call for a more comprehensive 
{\it Spitzer} survey for analogous sources within $D\approx10$\,Mpc, are presented in
\S\ref{section:discussion}.

\section{The Class}
\label{section:class}

We start by listing the objects we view as likely to represent
this new class of SN 2008S-like transients and progenitors.  The two objects
that define the class  --- SN 2008S (\S\ref{section:sn2008s}) and 
NGC 300 (\S\ref{section:ngc300}) --- are unique among transient-progenitor
pairs. The progenitors are relatively low-luminosity, have 
low variability, and are deeply dust-embedded on $\sim100$\,AU scales.
The transients are low-luminosity, with spectra exhibiting both narrow
Balmer lines (similar to low-luminosity IIn's and impostors),
and [Ca II] in emission, and have rapidly decaying lightcurves
compared to IIP supernovae.
The transient in M85 (\S\ref{section:m85}) and SN 1999bw 
(\S\ref{section:sn1999bw}) may also be members of the same class, 
but we cannot confirm the existence of a dust-obscured progenitor
similar to SN 2008S/NGC 300.   In \S\ref{section:other}, we contrast 
this class with other peculiar outbursts, such as the supernova 
impostors and low-luminosity Type IIP supernovae, and we note 
a number of other transients that are not excluded as members of
the SN 2008S-like class.

\subsection{SN 2008S}
\label{section:sn2008s}

SN~2008S in NGC 6946 ($D\approx5.6$\,Mpc; Sahu et al.~2006) was
discovered February 1.79 UT (Arbour \& Boles 2008).  Because of the presence
of narrow Balmer lines (${\rm FWHM} \approx 1000$~km~s$^{-1}$) it was
initially classified as a young Type IIn supernova. Stanishev et
al.~(2008) reported strong Na~D absorption with equivalent width of
5~\AA, indicating a high degree of internal extinction.  On the basis of
its relatively low luminosity ($M_{V}\simeq -14$) and peculiar spectrum
(including the presence of strong and narrow [Ca~II] 730~nm doublet in
emission), Steele et al.~(2008) proposed that SN~2008S was a {\em
supernova impostor} such as SN~1997bs (Van Dyk et al.~2000). 
Botticella et al.~(2009) present the late-time lightcurve of SN 2008S,
which shows evidence for a power-law time dependence with a slope 
indicative of being powered by the radioactive decay $^{56}$Co.
Following the suggestion by Prieto et al.~(2008, hereafter P08a)
and the discussion presented in  \S\ref{section:discussion}, 
Botticella et al.~argue that SN 2008S may have been an electron-capture 
supernova.

P08a used a deep pre-explosion archival 
Large Binocular Telescope (LBT) image of NGC 6946 to put stringent upper
limits on the optical emission from the site of the supernova (3$\sigma$
limits of $M_U>-4.8$, $M_B>-4.3$, and $M_V>-3.8$). The failure to detect
a progenitor in the optical led P08a to examine archival {\it Spitzer}
IRAC imaging of NGC 6946 from the SINGS Legacy Survey (Kennicutt et
al.~2003) for the progenitor. A point source at the location of SN 2008S 
was detected at 4.5, 5.8, and 8$\mu$m, but
undetected at 3.6$\mu$m and 24$\mu$m, leading to a lower
limit on the $[3.6]-[4.5]$ color of $\gtrsim2$.  The best-fitting
blackbody temperature to the SED was $\approx440$\,K.\footnote{A
blackbody provides a rather poor fit to the SED, perhaps indicating an
interesting grain size distribution in the obscuring material (P08a).} The
integrated luminosity was $\approx 3.5\times10^4$\,L$_\odot$, consistent
with a relatively low-mass massive-star progenitor with a zero-age main
sequence (ZAMS) mass of $M\approx10$\,M$_\odot$. The progenitor SED 
is shown in the left panel of Figure \ref{fig:sed}.  Simple arguments suggest that 
the obscuration was circumstellar, with an optical depth at visual
wavelengths larger than a few on a physical scale of order 150\,AU
(P08a). The explosion itself was also serendipitously observed by 
{\it Spitzer} 5 days after discovery (Wesson et al.~2008), with 
an infrared luminosity $\sim$50 times larger than the progenitor, 
suggesting a substantial amount of dust-reprocessing (see Fig.~2 of P08c).
The lack of variability of the progenitor on a $\sim3$\,yr
timescale argues that the obscuring medium was either a continuous wind
with a steady photosphere or a (implausibly?)  slowly
($\lesssim10$\,km s$^{-1}$) expanding shell.

\subsection{NGC 300}
\label{section:ngc300}

A luminous optical transient ($M_{V} \simeq -13$) in the nearby galaxy
NGC 300 ($D\approx1.9$\,Mpc; Gieren et al.~2005) was discovered by
Monard (2008), and reported by Berger \& Soderberg (2008).  The latter
used archival imaging with ACS/WFC onboard the {\it Hubble Space
Telescope} ({\it HST}) to put very tight limits on the optical emission
(see Fig.~\ref{fig:sed}).  These limits led them to suggest that the 
NGC 300 transient was analogous to the outburst V838 Monocerotis 
(Bond et al.~2003).

Similar to the case of SN 2008S, Prieto (2008, hereafter P08b)
discovered a deeply-embedded source in archival {\it Spitzer} imaging
(PI: R.~Kennicutt) at the location of the transient.  The source was
detected in all IRAC bands, as well as in MIPS 24$\mu$m, and had a
$[3.6]-[4.5]$ color of $\approx 2.7$. The SED implies a
$\approx 330$\,K blackbody (as with SN 2008S, a blackbody is a fairly 
poor fit to the SED) with a bolometric luminosity of
$\approx5.6\times10^4$\,L$_\odot$.  This finding confirms the massive stellar
origin of the NGC 300 transient, and is consistent with 
a relatively low-mass massive star (see Fig.~\ref{fig:sed}).  
Importantly, depending on the details of stellar models 
for ZAMS masses in the range of $\sim10$\,M$_\odot$, 
at fixed final bolometric luminosity, the inferred initial 
progenitor mass may be multiply-valued and a luminosity 
of $\approx5.6\times10^4$\,L$_\odot$ can imply a $\sim5$, $\sim8$
or $\sim11$\,M$_\odot$ progenitor (see Smartt et al.~2009, 
their Fig.~2; \S\ref{section:theory}).\footnote{
P08b assumed standard singly-valued masses, and given the measured luminosity
of the progenitor of NGC 300 quoted ($\sim10^5$\,L$_\odot$) 
they inferred an initial progenitor mass of 
$\sim15-20$\,M$_\odot$.\label{foot_p08b}}

The luminosity and blackbody temperature of the progenitor of NGC 300
suggest an obscuring medium with physical scale of order 300\,AU. 
The deep limits on the optical emission from the
progenitor with {\it HST} suggest an optical depth at $V$ considerably
larger than unity ($\sim8-10$).\footnote{Assuming a spherical homogeneous 
medium and a Galactic dust-to-gas ratio, this optical depth at $V$ requires 
a total mass of obscuring material of roughly $\sim 2\times10^{-3}$\,M$_\odot$ 
on a scale of 300\,AU.}

The fact that the transient in NGC 300 and SN 2008S were similar, both
in their luminosities (both relatively faint with respect to typical
supernovae with $M_V\approx-14$) and spectra (with narrow Balmer lines
and strong [Ca~II] 730~nm doublet in emission; Bond et al.~2008), and that their
progenitors were similar (highly dust-obscured, relatively modest
bolometric luminosities) led P08b to propose that NGC 300 and SN
2008S share a common origin: the explosion --- whether supernova or
outburst --- of a massive star enshrouded in its own dust.
As we show below, the fact that their progenitors are so rare
among massive stars implies that just two events (although, 
see \S\ref{section:m85} \& \ref{section:sn1999bw}) are sufficient
to define a class. 

\subsection{M85}
\label{section:m85}

The optical transient in the Virgo galaxy M85 (NGC 4382) was discovered
in early 2006 by the Lick Observatory Supernova Search Team (KAIT; Li et
al.~2001) and discussed in Kulkarni et al.~(2007) and Pastorello et
al.~(2007) and may also be a member of the class defined by SN 2008S and
NGC 300.  The transient had peak $R$-band absolute magnitude of
$\approx-12$ with a $\sim80-100$ day plateau, similar to low-luminosity
Type IIP supernovae. Optical limits constrain the progenitor to be
less than $\approx 7$\,M$_\odot$ or highly obscured (Ofek et al.~2008;
see also Pastorello et al.~2007).\footnote{Ofek et al.~(2008) quote a limiting absolute
magnitude at F850LP($z$) with HST of $>-6.2$.} 
The low optical luminosity, plateau, and redward spectral evolution of
the transient led 
Kulkarni et al.~(2007) to propose that it was analogous to the outburst V838 Mon.
In contrast, Pastorello et al.~(2007) argued that it was a low-luminosity
Type IIP supernova (see \S\ref{section:other}).

Prieto et al.~(2008c, hereafter P08c) again searched archival {\it
Spitzer} imaging and discovered an infrared source at the site of the
optical transient, but, by chance, taken 8.8 days before the discovery
in the optical by KAIT.  The source was detected in all IRAC bands and
was associated with the transient itself, and not the progenitor, since
archival {\it Spitzer} imaging from 2004 (PID 3649; PI P.~C\^ot\'e) 
does not reveal a point source at the location of the M85 transient.  
Using these images, we derive 3$\sigma$ upper limits of $4\times10^5$\,L$_\odot$ and 
$2\times10^{5}$\,L$_{\odot}$ at 3.6\,$\mu$m and $4.5$\,$\mu$m,
respectively, for the progenitor.
The transient was also detected in the mid-IR seven months after the
initial discovery by Rau et al.~(2007), the fluxes having decreased by a
factor of $\approx5$ over that time.  The bright infrared transient
discovered by P08c is adequately fit by a blackbody with temperature of
$\approx800$\,K with luminosity $\approx2\times10^6$\,L$_\odot$.  The
optical and NIR photometry of Kulkarni et al.~(2007) indicate a second
component to the SED with a blackbody temperature of $\approx3900$\,K
and with a luminosity of $\approx5\times10^6$\,L$_\odot$.

The cooler re-radiated dust emission arises from a region of order
$300-400$\,AU in physical scale.  Assuming that the optical emission did
not vary in the $\approx$8.8 days between the IR discovery and the optical
discovery, the ratio of the power in these two blackbody components
implies that the optical depth at $V$ ($\tau_V$) is less than unity.  
The physical scale of the obscuring medium indicates that it is likely 
circumstellar, and the result of a mass-loaded wind.
In addition, the luminosity of the transient 
suggests that any pre-existing dust within $\sim100-200$\,AU would have been 
destroyed during the explosion.  Given the fact that the optical depth 
to the source scales as $r^{-1}$ in a freely-expanding wind, it is
not implausible that $\tau_V$ to the progenitor was a factor of 
$\sim10-20$ larger before explosion.  These estimates suggest a
pre-explosion obscuring medium similar in its gross properties 
to the SN 2008S and NGC 300 progenitors.

The M85 transient also showed narrow Balmer lines
in emission, as well as some Fe II lines, similar to SN 2008S and NGC 300. 
Because of the strong evidence for obscuration of the progenitor, as evidenced 
by the bright IR transient, and the similarity of the spectra, P08c proposed 
that these outbursts share a common origin and that their obscured progenitors 
may give rise a new class of 2008S-like transients.

We emphasize that because the character of the progenitor is not known
(except for the optical and IR limits),
the connection to SN 2008S- and NGC 300-like events is plausible
rather than certain.  Nevertheless, if the M85 transient was associated with an embedded 
massive star, the IR limits we derive are consistent
with the luminosities derived for the 2008S and NGC 300 progenitors.

\subsection{SN 1999bw}
\label{section:sn1999bw}

The Lick Observatory Supernova Search reported in April 1999 the
discovery of a possible supernova in the galaxy NGC 3198 (Li 1999). The
optical spectrum of the transient, dominated by narrow Balmer lines in
emission (Garnavich et al.~1999; Filippenko et al.~1999), and its low
$V$-band absolute magnitude at maximum of $\simeq -13$ ($D\approx 13.7$\,Mpc; 
Freedman et al.~2001) led Li et al.~(2002) to propose that
this transient was an LBV-like outburst. Like SN 2008S and NGC 300, 
its spectrum showed [Ca II] in emission.   Additionally, an infrared source coincident
with the optical position of the transient was detected in archival {\it
Spitzer} imaging obtained with IRAC by the SINGS Legacy Survey five
years after the discovery of SN 1999bw (Sugerman et al.~2004). The
source was detected in all IRAC bands and the SED was well-fit by a
450\,K blackbody with an integrated luminosity of $\approx 1.4\times
10^{5}$\,L$_\odot$, which translates into a blackbody scale of $\sim
300$\,AU.\footnote{The luminosity and blackbody scale have been adjusted 
to the distance employed here.}
We have checked archival IRAC images obtained in December 2005
(PID 20320; PI B.~Sugerman), 1.5\,yr after the detection in the SINGS
images, and we confirm that the MIR source is indeed the transient,
since the fluxes have declined by a factor of more than 3 during this
time.

The combination of a low optical luminosity at maximum, an optical
spectrum dominated by narrow Balmer lines in emission, the presence of [Ca II]
emission, and a luminous
infrared emission detected with {\it Spitzer}, make SN 1999bw similar
to SN 2008S, NGC 300,  and the transient in M85.  However, as in the case
of M85, we emphasize that because there is no information on the
progenitor we cannot be sure that SN 1999bw was of the same class 
as SN 2008S and NGC 300.

\subsection{The Connection to Other Transients}
\label{section:other}

As we discuss in detail in \S\ref{section:m33}, perhaps the primary
distinguishing characteristic of this class of transients is their
deeply embedded progenitors. Since we are unable to confirm the presence
of such progenitors for the M85 transient or SN 1999bw, we are unable to make a
direct analogy with SN 2008S and NGC 300, and instead rely on the fact
that the transients themselves provides strong evidence for obscuration on
few-hundred AU scales. 

In our effort to understand which cataloged transients might belong to 
the SN 2008S/NGC 300-like class we have examined archival imaging
of many recent supernovae, as well as archetypal peculiar supernovae, 
including supernova impostors,
LBV outbursts, and low-luminosity Type IIP supernovae.
Here, we provide a brief discussion in an effort to orient 
the reader.

As we mentioned in \S\ref{section:introduction} (point 2),
the low-luminosity Type IIn SN 2002bu is an interesting transient 
that may be a member of the class defined by SN 2008S and NGC 300.  
We checked archival {\it Spitzer} data of the host galaxy NGC 4242
(PID 69; PI G.~Fazio) taken 2 years after discovery and we find a 
bright infrared point source detected in all IRAC bands within
$0\farcs6$ of the supernova position.   Also, two epochs
of MIPS data (PID 40204; PI R.~Kennicutt), obtained in 2008, 
6 years after explosion, and separated by just 6 days, 
reveal a 24\,$\mu$m source at the position of the supernova.
This is qualitatively similar to 
the case of M85 and SN 1999bw, but because only a single 
post-explosion IRAC epoch exists, and because the 
two MIPS epochs are separated by such a short time,
we are unable to  definitively confirm that the MIR 
source is associated with SN 2002bu. 

SN 1997bs is an intriguing example of an object that does not fit into
this class, although its peak absolute magnitude, lightcurve, color, and
some spectral features are comparable to SN 2008S and NGC 300. In this
case, a luminous un-obscured progenitor has been identified in the
optical ($M_V\approx-7$) and the transient itself has been argued to be
the outburst of an LBV (Van Dyk et al.~2003). Of interest is the fact that 
no object has been subsequently identified in the optical at the site of 
the transient (Van Dyk 2005). We have checked archival {\it Spitzer} data 
of the host galaxy (NGC~3627) obtained in 2004 ($\sim 7$~years after 
discovery) by SINGS and we do not detect a bright MIR source at the site 
of SN 1997bs, in contrast with SN 2008S, M85, and SN 1999bw. 
The event SN 2003gm is also interesting in this context since it had
photometric and spectroscopic evolution similar to SN~1997bs, 
and also showed an optically luminous progenitor ($M_{V} \approx -7.5$; 
Maund et al.~2006).  The fact that both SN 1997bs and 2003gm had bright 
unobscured progenitors is our primary reason for excluding them
from the class defined by SN 2008S and NGC 300.

Historical LBV eruptions in nearby galaxies that have been initially
classified as supernovae are also worth mentioning here. These include
SN 1954J (e.g., Smith et al.~2000) and SN 2002kg (e.g., Maund et
al.~2006; Van Dyk et al.~2006) in NGC 2403, SN 1961V (e.g., Humphreys
2005) in NGC 1058, and SN 2000ch in NGC 3432 (Wagner et al.~2004). As in
the cases of SN 1997bs and SN 2003gm, a very important common difference
between these objects and SN 2008S or NGC 300 is that they all had
optically luminous progenitors with absolute magnitudes $\lesssim -6$,
consistent with originating from very massive stars. These
transients also have other properties that are not consistent with SN
2008S-like explosions: (1) their peak absolute magnitudes range between
$-18 \gtrsim M_{V} \gtrsim -9$, (2) the transient timescales vary widely
from a few days to years, and (3) they are not luminous MIR sources in
archival {\it Spitzer} data.

Other optical transients classified as sub-luminous Type IIP supernovae
that might potentially fall into the class defined by SN 2008S and NGC
300 include SN 1994N, SN 1997D, SN 1999eu, 1999br, 2001dc, 2003Z, and
2005cs (e.g., Pastorello et al.~2004, 2006). We note, however, that
there are fundamental differences in the spectra of low-luminosity
Type IIP SNe compared with SN 2008S-like transients. In particular, 
sub-luminous Type IIP SNe show Balmer lines with strong P-Cygni 
absorption profiles and velocities of a few thousand km s$^{-1}$, 
as is observed in more luminous Type IIP SNe.  This stands in sharp contrast 
with the Balmer lines in SN 2008S-like transients, which are fairly narrow 
(${\rm FWHM} \sim 1000$~km~s$^{-1}$) and which do not show strong P-Cygni 
absorption features.  In this way, SN 2008S-like transients 
most closely resemble the spectra of low-luminosity Type IIn SNe 
and LBV outbursts.

In addition to the very interesting case of 2002bu, there were five
other low-luminosity transients classified as ``impostors'' or ``unknown''
that could have been LBV outbursts, but for which no progenitor has been identified,  
and which might be 2008S-like: NGC 4656, SN 2001ac, 2006bv, 2006fp, and 2007sv 
(see \S\ref{section:rates}).
However, as we have emphasized and as the sources discussed above imply,
the properties of the progenitor cannot be deduced from
the character of the optical outburst alone (e.g., contrast SN 1997bs and SN
2003gm with SN 2008S).  Thus, in order to understand the causal mapping
between progenitor and explosion, a census of the progenitors must first
be completed.  What is clearly needed is a comprehensive survey for bright 
MIR sources in all nearby galaxies ($\lesssim10$\,Mpc) with (warm) 
{\it Spitzer}, analogous to the survey proposed by Kochanek et al.~(2008)
in the optical.
In the next section we discuss our search for deeply-embedded progenitors
in M33.  We discuss a more complete census in \S\ref{section:discussion}.

\section{Rates}
\label{section:rates}

The absolute rate of transients analogous to NGC 300 and SN 2008S
is uncertain.  Current samples of supernovae over the last decade
in the local volume within 10, 20, and 30 Mpc allow us to make 
only a rough estimate of the true rate.  A systematic transient search in the local
volume is crucial to solidify these numbers.  Nevertheless, we 
estimate that 2008S-like transients occur with a frequency 
equivalent to $\sim20$\% of the Type II supernova rate.
We discuss the observed rates in \S\ref{section:count} below, 
and then we enumerate several arguments suggesting that the 
sample of SN 2008S-like transients may be highly incomplete
in the local universe (\S\ref{section:incomplete}).

\subsection{Observed Counts}
\label{section:count}
\subsubsection{$D\lesssim10$\,Mpc}
\label{section:10}

In addition to SN 2008S and NGC 300, 
$\approx$\,20 other core-collapse supernovae or supernova-like transients 
have been discovered within $\approx10$\,Mpc since 1999.\footnote{Throughout 
the discussion here we exclude Type Ia supernovae.}  
Sixteen were confirmed supernovae; they are SN 1999em,
1999ev, 1999gi, 1999gq, 2002hh, 2002ap, 2003gd, 2004am, 2004dj,
2004et, 2005af, 2005at, 2005cs, 2007gr, 2008bk, and 2008ax.  Two were
bona fide LBV eruptions (SN 2000ch, Wagner et al.~2004; 2002kg, Schwartz et
al.~2003; Weis \& Bomans 2005; Maund et al.~2006; Van Dyk et al.~2006).  
One was a Type IIn supernova potentially of the 
SN 2008S class (SN 2002bu in NGC 4242, $D\approx8$\,Mpc; Puckett 
\& Gauthier 2002; Hornoch 2002).  Finally, the transient in NGC 4656
also had some of the spectral characteristics of low-luminosity IIn supernovae
(e.g., narrow H$\alpha$ in emission), but only 
reached an absolute magnitude of $\approx-11.5$,
 (Rich et al.~2005; Elias-Rosa et al.~2005).

Taken at face value, with no correction for incompleteness,
these numbers imply that $2/22\approx9\%$ or 
$3/22\approx14\%$ (including SN 2002bu) of all optically bright
transients are SN 2008S-like.  Removing the two bona fide 
LBV outbursts (2000ch and 2002kg) for comparison with the 
supernova sample proper, if SN 2008S and NGC 300 are supernovae, 
they represent $\approx10$\% and $\approx$15\% 
(again, with SN 2002bu) of the sample.

\subsubsection{$D\lesssim20$\,Mpc}
\label{section:20}

A similar exercise can be carried out within the larger volume of 
$\sim20$\,Mpc.  With a peak absolute magnitude of $\sim-14$, SN 2008S-like
transients would have an apparent magnitude of 17.5 at $D=20$\,Mpc, without
including a correction for extinction intrinsic to the transient.  In fact,
SN 2008S and NGC 300 had $A_V\approx1.2$ (P08a) and $A_V\sim0.3-1.2$ (Bond et al.~2008),
respectively.

Using the Smartt et al.~(2009) compilation, we find 29 IIP,
4 IIb, 15 Ib/c, 2 IIn (1998S \& 2002bu),  and 2 IIL supernovae
in the last decade.  There are 6 classified as ``LBV eruptions/impostors'' (1999bw, 2000ch, 2002kg,
2003gm, 2007sv, and NGC 4656\footnote{Note that NGC 4656 is not included in Smartt et al.~(2009).}),
but only 2000ch and 2002kg have strong evidence for an LBV progenitor.  Whether the
yellow supergiant progenitor of 2003gm survived the explosion, and hence whether 
2003gm was in fact a supernova, has not yet been definitively established (Maund et al.~2006).
Similarly, although SN 2007sv, which reached an absolute magnitude of $\approx-14$, bears
some similarity to 1997bs, the nature of its progenitor has not been established
(Duszanowicz et al.~2007; Harutyunyan et al.~2007).
Finally, there are 3 events whose nature is classified in Smartt et al.~(2009)
as unknown: M85, NGC 300, and SN 2008S.   

With no correction for incompleteness, and taking only NGC 300 and
SN 2008S, this compilation implies an overall rate of $2/61\approx3.3$\% within 20\,Mpc.
Including M85 and 2002bu in the sample of 2008S analogs
doubles the rate.  Including 1999bw, 2007sv, and NGC 4656
brings the overall observed rate of SN 2008S-like transients to $\sim10$\%
within $D\lesssim20$\,Mpc.

\subsubsection{$D\lesssim30$\,Mpc}
\label{section:30}

With a limiting magnitude of $\approx18-19$, amateur and 
professional supernova surveys could in principle find SN 2008S-like 
transients with $M_V\approx-14$ to a luminosity distance of 
$\sim25-40$\,Mpc (again assuming no extinction).  Objects of interest 
discovered over the last $\sim10$ years with faint absolute magnitudes 
in this distance range, and with the
spectral characteristics of IIn supernovae analogous to 
SN 2008S include 2001ac, 2006bv, and 2006fp.

A total of 92 core-collapse supernovae and 7 LBV eruptions 
appear in the recent compilation of Smartt et al.~(2009).
Three events --- M85, NGC 300, and SN 2008S --- are classified 
as ``unknown.''  Of the 7 ``LBV eruptions,'' only 2000ch, 2002kg,
and 2003gm have LBV-like progenitors.  The remainder ---
1999bw, 2001ac, 2006fp, and 2007sv --- have little or no progenitor
information, spectra that resemble IIn's, and {\it may} be
SN 2008S analogs. Given the number of confirmed
NGC 300 and SN 2008S-like analogs (just 2) and those suspected 
of belonging to this class (M85, 1999bw, 2002bu), as well as
those tradiationally labelled ``LBV outbursts,'' but with no 
strong confirmation (4656, 2001ac, 2006bv, 2006fp, 2007sv) the overall rate ranges from 
$2/102\approx2$\% to $10/102\approx10$\% when measured within 30\,Mpc.

\subsection{Arguments for Incompleteness \& Some Implications}
\label{section:incomplete}

To summarize the previous subsections, 
conservatively taking NGC 300 and SN 2008S as the only examples of their type ever
observed (that is, excluding all other low-luminosity transients), the observed rate
is $\sim9$\%, $\sim3$\%, and $\sim2$\% within 10, 20, and 30 Mpc volumes, respectively, 
with respect to all bright optical transients when averaged over the last 10 years.

It is difficult to estimate the degree of uncertainty in these
numbers since the surveys that find local
supernovae are a combination of professional and amateur, with complicated
and unquantitified selection functions for transient identification.
Most surveys responsible for transient discoveries in the local 
universe do not have detailed descriptions of completeness in the 
literature.  On the contrary, the large majority of the transients 
in the local volume are discovered by amateurs.\footnote{Of the 13, 17 and 13 
local ($D \lesssim 30$\,Mpc) supernovae (including Type Ia's) discovered in 
2006, 2007, and 2008, only 1, 4 and 4 were discovered by professional searches.}   
Nevertheless, it 
is possible to make an estimate of completeness that gives a sense 
of how large the correction to the rate of SN 2008S-like transients 
may be.  

As an example, B.~Monard typically quotes $\pm0.2$ magnitude (unfiltered)
photometric errors on discovery observations of SNe detected at 
$\approx16-17$ magnitude (e.g., Monard 2006).  This photometric error 
translates to a signal-to-noise ratio of ${\rm S/N}\approx5$.  In order to 
estimate a lower limit on the incompleteness, we can compare
this value for ${\rm S/N}$ with the mean detection efficiency for
Type Ia supernovae in the SDSS-II Supernova Survey, which employs a well-tested
photometric pipeline that uses difference imaging to subtract off
the host galaxy (see Dilday et al.~2008).  Their Figure 7 shows that 
for ${\rm S/N}\approx5$, the detection  efficiency is $\approx0.5$
in Sloan $gri$.  The detection efficiency drops to $\sim0.1-0.2$ for 
${\rm S/N}\approx2$, corresponding to a photometric error of $\pm0.4$ 
magnitudes. 

\begin{figure*}
\centerline{\includegraphics[width=15cm]{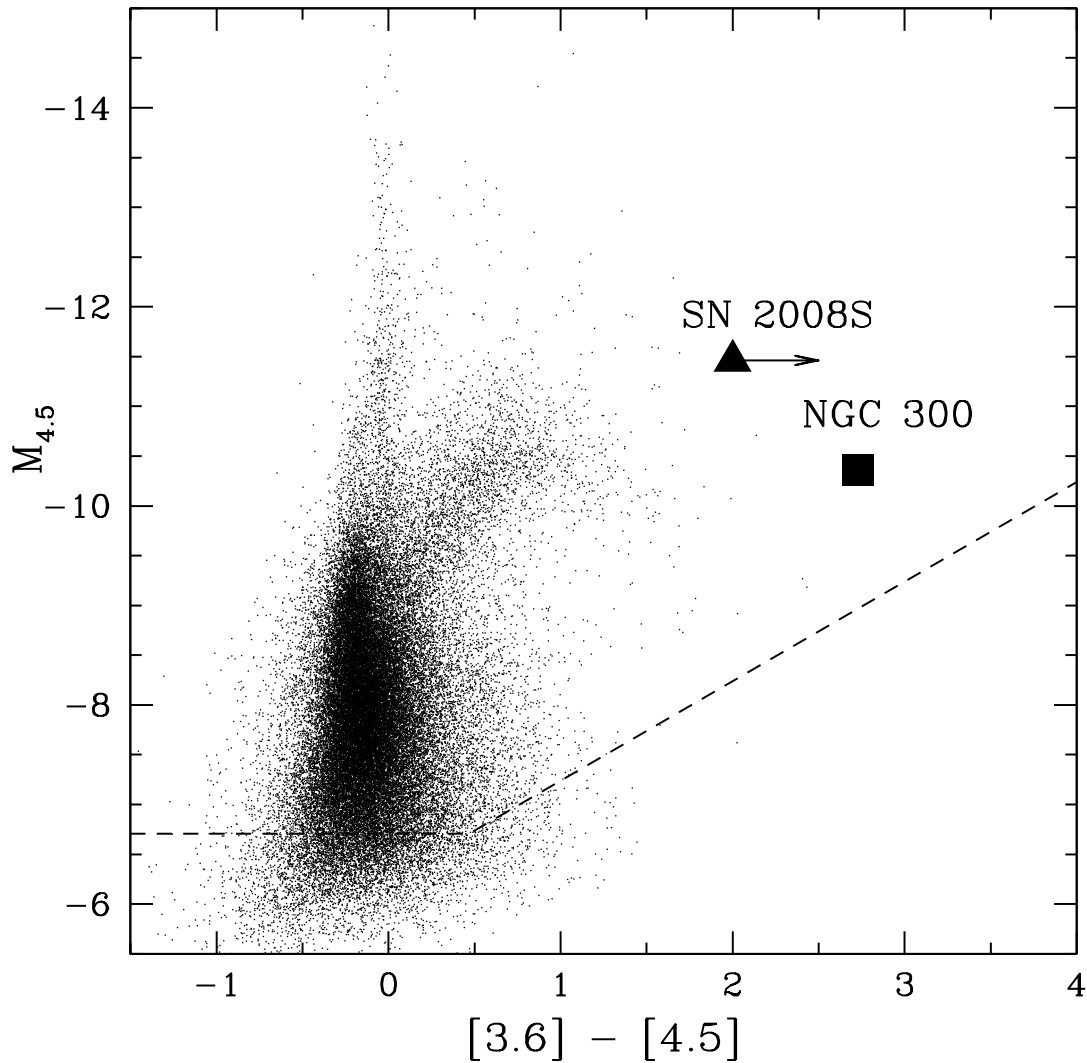}}
\caption{Mid-infrared color-magnitude diagram for M33. 
Absolute magnitude at 4.5$\mu$m
is plotted versus $[3.6]-[4.5]$ color for all detected sources 
(3$\sigma$ limits denoted by dashed lines; see \S\ref{section:catalog};
Table \ref{table:m33}). 
For comparison, the positions of the progenitors of NGC 300 (filled square)
and SN 2008S (filled triangle; lower limit) are also plotted (P08ab).  
Stars analogous to these progenitors are exceedingly rare. Compare
with Figure \ref{fig:cmd2}.
The main sequence, AGB, and EAGB stars are clearly visible.  
Note the ``spur'' in the data extending to fainter
$M_{4.5}$ and redder color at $[3.6]-[4.5]\approx1$ and
$M_{4.5}\approx-10.5$, originating at the red extremum of the AGB
population.  To our knowledge, this is the first time that such a
feature has been identified in a MIR CMD.}
\label{fig:cmd}
\end{figure*}

Momentarily ignoring the difference in the shape of Type Ia lightcurves with respect to 
those of 2008S and NGC 300, the results of Dilday et al.~(2008) imply that 
surveys achieving limiting magnitudes of $\approx18-19$ are of order $\sim10\%$ 
complete for SN 2008S-like transients with $M_V\approx-14$ 
at $25-40$\,Mpc (${\rm S/N}\sim1-2$).  Thus, a factor of $\sim10$ correction
should be applied to the 30\,Mpc sample in \S\ref{section:30} from the
Smartt et al.~(2009) catalog for the true rate
of low-luminosity IIn supernovae like SN 2008S.  Within 20\,Mpc the
correction for incompletness is likely a factor of $\sim5$, and 
within 10\,Mpc the incompleteness is probably closer to a factor of 
$\sim2$.  Similar corrections should be applied to the observed
rate of true LBV eruptions.

There is another argument for incompleteness at the factor of 
$\sim2$ level within 10\,Mpc for SN 2008S-like transients.
As summarized by Horiuchi et al.~(2008) (their Section II.B),
the observed rate of supernovae of all types within 30\,Mpc
yields a ratio of Type Ia supernovae to core-collapse
supernovae that is significantly in excess of the cosmic ratio measured
at high redshift ($0.5\lesssim z\lesssim1$; Dahlen et al.~2008). 
Indeed, the ratio within $30$\,Mpc is large enough that we would 
expect to have seen several Type Ia supernovae within 10\,Mpc
in the last 10 years.  Yet, none have been found.  This fact
implies that the ratio of Ia to core-collapse supernovae 
has been over-estimated within 30\,Mpc because normal 
core-collapse supernovae are intrinsically fainter than Type Ia's.
Thus, the sample of 
{\it normal} core-collapse supernovae is incomplete within
30\,Mpc at the factor of $\sim1.5-2$ level, even though these objects 
typically have peak absolute visual magnitudes of $M_V\approx-16$ 
to $-18$.  Naively, analogs to SN 2008S with $M_V\approx-14$ would be 
$\sim6$ times {\it more} incomplete than normal supernovae at 
$D\approx30$\,Mpc.  Of course, the actual incompleteness correction 
depends on the overall extinction correction for the transient population
and on the cadence of the observations since the lightcurve declines much 
faster for 2008S-like transients than for supernova of type IIP.  
This implied nearly order-of-magnitude incompleteness correction 
at $D\approx30$\,Mpc for 2008S-like events strongly indicates that the 
sample is incomplete at order unity within $D\approx10$\,Mpc.

Additionally, a plot of the discovery rate of all supernovae (Ia's included)
within 30\,Mpc over the last 10 years shows an {\it increasing}
trend, super-Poisson variance, and strong dependence on the 
results and observing strategy of a single survey (LOSS; Li et al.~2000).  
There is also an asymmetry between the rate of discovery 
in the northern and the southern sky in excess of a simple 
extrapolation of star formation from the catalog of 
Karachentsev et al.~(2004).  Finally, there is an unquantified
bias against small star-forming galaxies in the local universe.
These points further solidify the case that the normal 
core-collapse supernova rate is incomplete, which implies
that we are missing 2008S analogs in abundance in the 
local universe.

Taking yet another angle on the question of overall rate, 
we may consider the {\it a posteriori} statistics of the
events SN 2008S and NGC 300 themselves.  Taking the overall supernova
rate as $\sim1-2$ yr$^{-1}$ within $10$\,Mpc implies a probability
of $\sim4-8\times10^{-4}$ of seeing two events in a single year
if the overall rate of 2008S-like transients is 2\% of the supernova rate.
We consider this uncomfortably small.  Similarly, if we were to 
take the true SN 2008S-like transient rate to be $\sim100$\% of 
local supernova rate, we would be forced to explain the fact that 
only $\sim2$ such events have been seen within 10\,Mpc in the last 10 
years.  Given the discussion of incompleteness above, an overall
rate of $\sim20$\% with respect to supernovae gives a reasonable
chance of seeing two in one year and of seeing only a handful on a
10 year baseline.

Taken together, these arguments imply that the sample
of transients in the local universe {\it when averaged over the
last 10 years} is highly incomplete.  We
suggest that the incompleteness correction is a factor of 
$\sim2$, $\sim5$, and $\sim10$ for $M_V\approx-14$ transients 
at 10 ($\approx 16.0$ mag), 20 ($\approx 17.5$ mag), and 30\,Mpc ($\approx 18.4$ mag),
even before accounting for the potentially higher average extinction
of these transients relative to normal supernovae.
These estimates are consistent with the Richardson et al.~(2002),
who conclude that low-luminosity supernovae with $M_B\gtrsim-15$ may constitute more 
than 20\% of the overall supernova population (for related arguments, see Schaefer 1996; 
Hatano, Fisher, \& Branch 1997; Pastorello et al.~2004).
There are several immediate implications:
\begin{enumerate}
\item The true rate of SN 2008S-like transients is $\sim20$\% of the core-collapse supernova rate.
However, we emphasize that lower and higher values at the factor of $\sim2$ level 
are not excluded until a more thorough census has been made.
\item The true rate of massive star eruptions (LBV-like and otherwise) is similarly incomplete. 
The observed rate of ``LBV eruptions'' within 10 and 20\,Mpc in \S\ref{section:10} and \S\ref{section:20}
implies that they are $\sim1-3$ times more common than SN 2008S-like transients.  Thus, the true
rate of massive LBV eruptions is likely $\sim20-60$\% of the core-collapse supernova 
rate.\footnote{We thank the anonymous referee for pointing this out.} 
\item The observed rate of core-collapse supernovae is incomplete at the factor of $\sim2$ level
for $D\lesssim30$\,Mpc.
\end{enumerate}

\begin{figure*}
\centerline{\includegraphics[width=15cm]{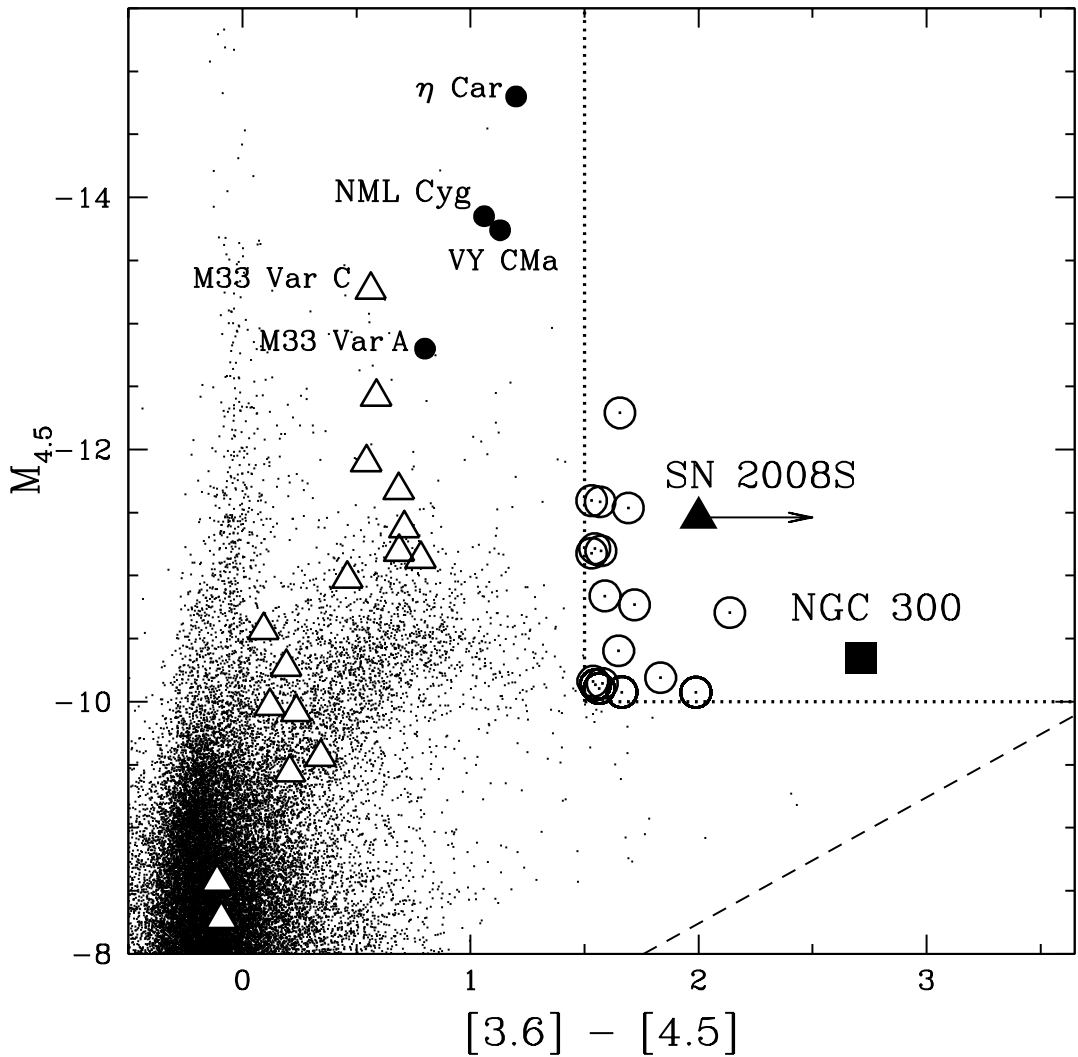}}
\caption{
Same as Figure \ref{fig:cmd}, but focused on the red and
bright region of interest. Here, the $[3.6]-[4.5]>1.5$ and $M_{4.5}<-10$ 
selection for extremely 
red and bright objects is shown explicitly by the dotted lines, as are the 
18 EAGB sources  in M33 that meet these criteria 
(large open circles; see \S\ref{section:sources}). 
We denote them ``S1-S18,'' ordered by color.  The reddest source,
S1, is shown in Figure \ref{fig:s1}.
Most of these sources are highly variable (see
Figs.~\ref{fig:rms}, \ref{fig:agb1}, \& \ref{fig:agb2}). 
SEDs are shown in Figure \ref{fig:sed} (see Table \ref{table:eagb}).
The EAGB population is clearly distinct from the 
optically-luminous LBV candidates from the catalog of M07, 
shown by the open triangles  (see \S\ref{section:m33}; Table \ref{table:lbv}).  
Most of the LBV candidates are considerably more bolometrically luminous and
much less variable at 4.5$\mu$m than the EAGB sample
(see Figs.~\ref{fig:sed}, \ref{fig:rms}, \ref{fig:lbv1}, \& \ref{fig:lbv2}).   
A number of cool hyper-giants such as VY CMa, NML Cyg, and M33 Var A, 
as well as $\eta$-Carina, are shown for comparison.  The brightest 
LBV candidate is M33 Var C.  The LBV candidates appear to be bimodal in 
MIR color (right panel of Fig.~\ref{fig:sed}). }
\label{fig:cmd2}
\end{figure*}

\section{A First Census}
\label{section:m33}

Because of the implied frequency of events similar to SN 2008S and NGC
300 (\S\ref{section:introduction}, point 2) and the interesting character
of their progenitors, we searched for analogous sources in archival 
{\it Spitzer} imaging of nearby galaxies. Our goal was to identify the 
underlying sub-population of massive stars from which these progenitors 
emerge, to characterize their properties and frequency, and to catalog 
them for future study.

The key characteristics of the progenitors of SN 2008S and
NGC 300 are that they are optically-obscured and deeply embedded, with
very red MIR colors, that their bolometric luminosities are indicative
of relatively low-mass massive stars ($L\approx4\times10^4$ and
$\approx5.6\times10^4$\,L$_\odot$, respectively), and that the several
epochs on NGC 6946 revealed that the progenitor of SN 2008S was not
highly variable in the $\approx 3$ years before explosion (P08a).  We 
discuss the variability of the progenitor of NGC 300 in \S\ref{section:var}
below based on just two pre-explosion epochs.  Although we are only 
able to derive a lower limit to its RMS variation at 4.5\,$\mu$m, 
like the progenitor of SN 2008S, we find that it too is consistent 
with  a low level of variability.

For a first census, we searched for the deepest archival {\it Spitzer}
observations of a nearby relatively massive bright star-forming galaxy,
with already extant optical catalogs. The Triangulum galaxy M33 is a
perfect test case.  It has an
absolute $B$-band magnitude of $M_B\approx-19.2$, a distance
of $\approx0.96$\,Mpc (distance modulus $\mu=24.92$; Bonanos et al.~2006), 
and it has extensive optical (e.g., Hartman et al.~2006; Massey et al.~2006, 
hereafter M06), H$\alpha$ (Massey et al.~2007, hereafter M07), and MIR 
and FIR imaging (McQuinn et al.~2007, hereafter Mc07). 
A similarly rich dataset exists for several other
local galaxies, including the Magellanic clouds (e.g., Blum et al.~2006;
Bolatto et al.~2007) and M31 (e.g., Mould et al.~2008). An analysis similar
to that described below will be the subject of future work 
(but, see \S\ref{section:other_galaxies}).

In this Section, we present the MIR color-magnitude and color-color
diagrams for all cataloged point sources in  M33 obtained 
from multi-epoch archival {\it Spitzer}/IRAC
(Fazio et al.~2004) observations (PI R.~Gehrz; PID 5). This dataset allows
us to search for and identify stars analogous to the progenitors of SN 2008S 
and NGC 300. We also present a variability study for both the reddest sources 
we find in M33 (all extreme-AGB ``EAGB'' stars) and for the optically-selected LBV 
candidates from M07, which are detected in the MIR imaging.  We find that the 
population of the reddest sources --- those most likely to be true analogs 
of the progenitors of SN 2008S and NGC 300 --- is completely distinct from 
the population of LBV candidates in the primary metrics of color, magnitude, 
and variability.  Indeed, we find very few sources with the properties of 
the SN 2008S and NGC 300 progenitors.

Section \ref{section:catalog} describes our procedure for extracting
point sources and the resulting catalog.  In \S\ref{section:sources}
we present the color-magnitude diagram for M33 and we discuss 
the reddest sources vis \'a vis
the optically-selected LBV candidates from M07.  In \S\ref{section:sed}
and \S\ref{section:var} we discuss their SEDs and variability, respectively.
Finally, in \S\ref{section:other_galaxies} we discuss a preliminary
search for similar sources in NGC 300, the LMC, and the SMC.

\subsection{Catalog}
\label{section:catalog}

We coadded six epochs of mid-IR imaging of M33 obtained between Jan.~9,
2004 and Feb.~4, 2006 with IRAC (3.6-8.0$\mu$m; see Mc07
for details of the observing program). We produced the coadds in all
four IRAC channels from the flux calibrated mosaics provided by the {\it
Spitzer Science Center} (post-BCD data). Our final mosaics cover an area
of $\sim 33\arcmin \times 33\arcmin$ ($1650\times 1650$~pixels, with
$1\farcs2$/pix) centered on M33, this is approximately within 
$R_{25}$ ($\simeq35\arcmin$; de Vaucouleurs et al.~1991). 
We performed source detection and PSF fitting photometry on the
coadds using the DAOPHOT/ALLSTAR package (Stetson 1992). The
PSF magnitudes obtained with ALLSTAR were transformed to
Vega-calibrated magnitudes using simple zero point shifts derived from
aperture photometry (using a 12$\arcsec$ radius), performed in the
original images, of $10-20$ bright and isolated stars in each band. We
estimate errors in the photometric transformations to the Vega system of
$0.04$~mag in 3.6~$\mu$m, $0.05$~mag in 4.5~$\mu$m, $0.07$~mag in
5.8~$\mu$m, and 0.07~mag in 8.0~$\mu$m channel. For the 
3.6 and 4.5\,$\mu$m bands, the detection limits (3$\sigma$) in the coadds are
$[3.6]\simeq18.9$~mag and $[4.5]=18.2$~mag, respectively. The sample
becomes incomplete at $[3.6]\simeq [4.5] > 17.1$, approximately
0.5 magnitudes deeper than Mc07. A total of $\approx80,000$
sources are detected in either 3.6 or 4.5\,$\mu$m. We cross-matched 
the two catalogs using a 0.5~pixel ($0\farcs6$) matching radius
to obtain a final catalog with  $\approx53,200$ individual sources 
detected at both 3.6 and 4.5\,$\mu$m (see Table \ref{table:m33}).  

There are several reasons for producing new point source 
and variability (see \S\ref{section:var}) catalogs,
given the already existing catalog from Mc07: (1) because we 
are specifically looking for objects with extreme colors, we wanted
to be able to relax the criterion for point source detection in 
all IRAC bands (in the Figures that follow, all sources of
interest are detected with 3$\sigma$ confidence); (2) we wanted to be able
to derive our own upper limits in each band for the same reason; 
(3) we wanted full IRAC SEDs, whereas the catalog of Mc07 does not 
provide data at 5.8\,$\mu$m; (4) we wanted to combine the images 
used by Mc07 with a sixth archival epoch (PI R.~Gehrz; PID 5); 
(5) for the reddest stars, we wanted to derive full six-epoch 
lightcurves for a more  complete measure of variability.  The 
resulting MIR color-magnitude diagram looks different from that 
in Mc07.  Of the 18 sources we discuss extensively below, only 
2 were detected at both 3.6 and 4.5 $\mu$m in the catalog of 
Mc07.

\begin{figure}[t]
\centerline{\includegraphics[width=8cm]{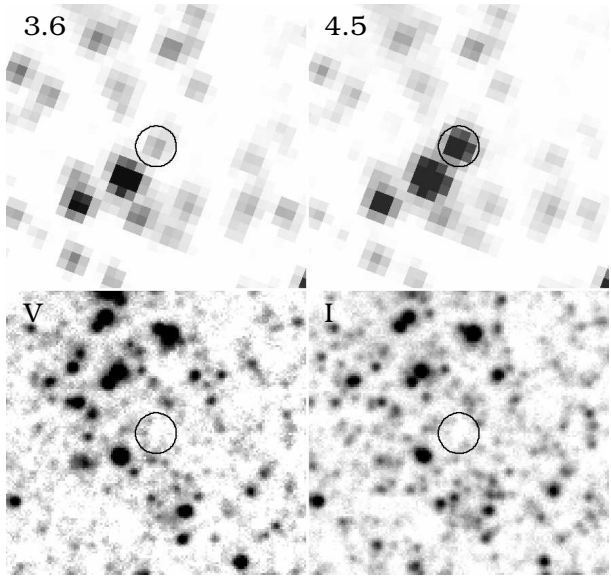}}
\caption{Image showing the point source S1 at 3.6\,$\mu$m, 4.5\,$\mu$m, 
as well as $V$ and $I$ band (M06).  This is the reddest of the 18 sources 
in $[3.6]-[4.5]$ color selected in Figure \ref{fig:cmd}.  It is 
optically obscured (see Fig.~\ref{fig:sed}).  S1 is also highly
variable in color and 4.5\,$\mu$m magnitude (see Figs.~\ref{fig:comp}
\& \ref{fig:rms}).}
\label{fig:s1}
\end{figure}

\begin{figure*}[t]
\centerline{\includegraphics[width=15cm]{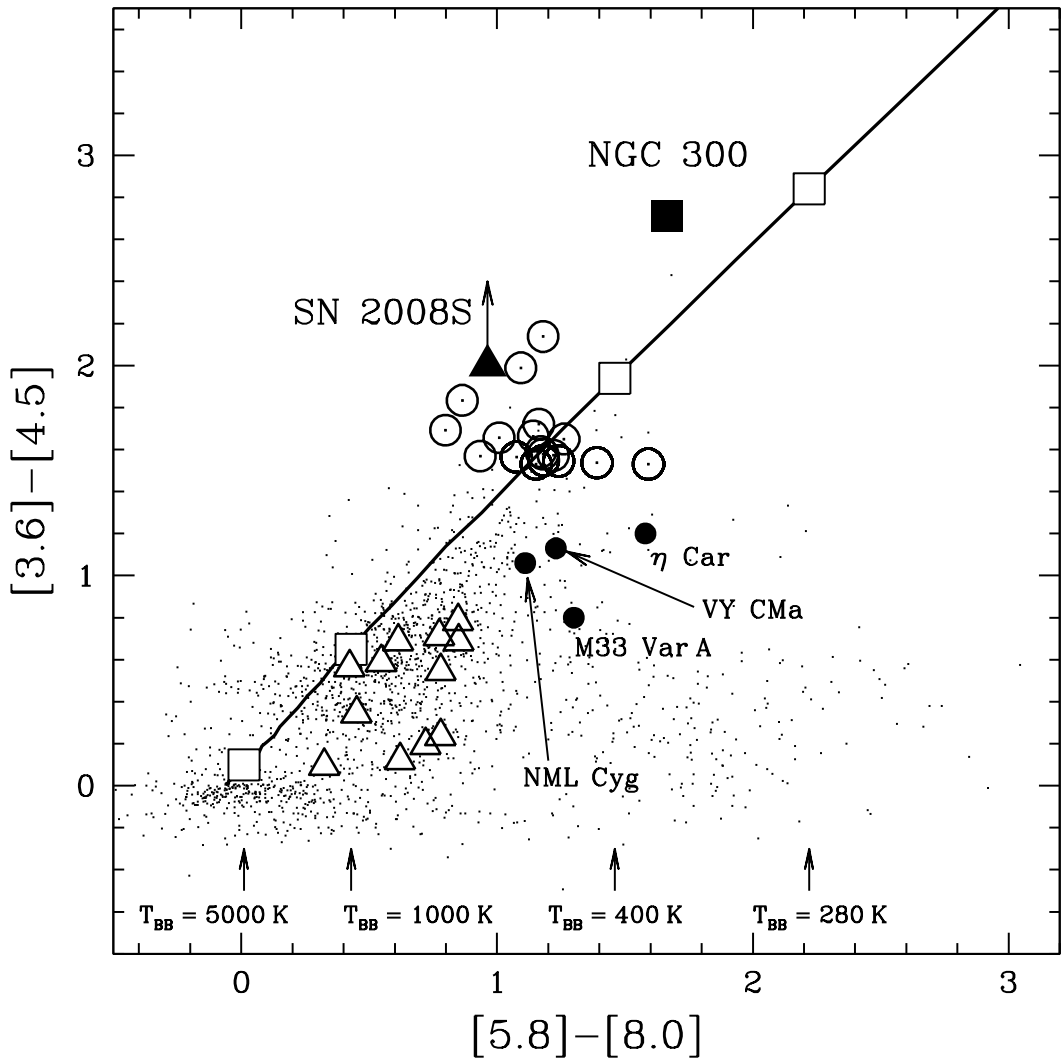}}
\caption{Color-color diagram showing $[5.8]-[8.0]$ versus $[3.6]-[4.5]$
colors for all ($\approx1800$) the sources detected in all four IRAC
bands. Symbols are the same as in Figure \ref{fig:cmd}.  
The solid line and open squares show the expectation for a
blackbody of temperature $T_{\rm BB}=5000$, 1000, 400, and 280\,K. The small
filled points with $[5.8]-[8.0]\approx1.6-1.8$ and $[3.6]-[4.5]\gtrsim2$
are not sufficiently bright at 4.5$\mu$m ($M_{4.5}<-10$) to be included
in the sample defined by the dotted lines in 
Figure~\ref{fig:cmd2}.}
\label{fig:cc}
\end{figure*}

\begin{figure*}[t]
\centerline{\includegraphics[width=15cm]{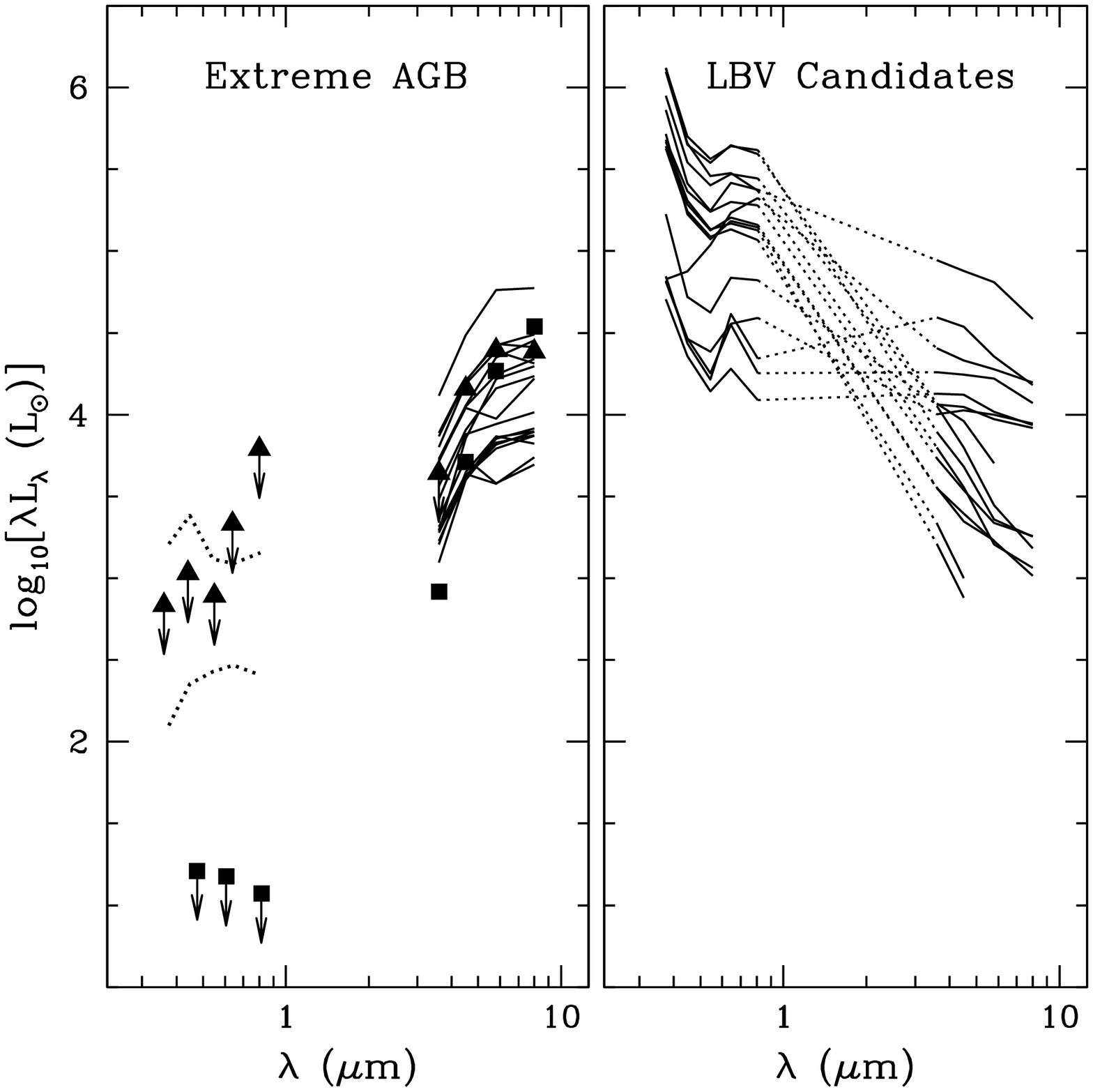}}
\caption{{\it Left Panel:} Spectral energy distributions of the 18
EAGB stars in M33 with $[3.6]-[4.5]>1.5$
and $M_{4.5} < -10$ (see Fig.~\ref{fig:cmd2}; Table \ref{table:eagb}). 
The lower dotted line shows the best lower
limits obtained in the optical, whereas the upper dotted
line shows the worst lower limits at $U$ and $I$ and the 
two $BVR$ detections (see discussion \S\ref{section:sed}; Table \ref{table:eagb}).
The SEDs of the SN 2008S (filled triangles) and NGC 300 (filled squares)
progenitors are also shown. For all the sources in M33 we assume a
total extinction of $E(B-V)=0.1$~mag.
{\it Right Panel:} Spectral energy distributions of the 16 LBVs detected
in MIR from the H$\alpha$ selected catalog of M07 (see Fig.~\ref{fig:cmd2}; 
Table \ref{table:lbv}). The relative increase in $\lambda L_{\lambda}$ in 
the $R$-band (0.6~$\mu$m) is due to the
presence of strong H$\alpha$ emission.}
\label{fig:sed}
\end{figure*}

\subsection{The Color-Magnitude Diagram}
\label{section:sources}

The primary result of this effort on M33 was the production of the
MIR color-magnitude diagram (CMD), shown in Figure~\ref{fig:cmd},
which shows the $[3.6]-[4.5]$ color for all the sources detected at both 3.6
and 4.5\,$\mu$m as a function of absolute magnitude at 4.5\,$\mu$m,
$M_{4.5}$. The dashed line marks the $3\sigma$ completeness limit in
this plane.  The {\it Spitzer} colors and magnitudes of the progenitors
of SN 2008S and NGC 300 are shown for comparison (filled triangle and
filled square, respectively; P08a, P08b). Note that there are remarkably 
few objects inhabiting the bright and (very) red region of the CMD.  
Among the $\sim5\times10^4$ massive stars in M33 (see \S\ref{section:discussion}),
$\approx 18$, $186$, and 567 have both $M_{4.5}<-10$ and $[3.6]-[4.5]$
color larger than 1.5, 1.0, and 0.7, respectively, which correspond to
blackbody temperatures of $\approx 500$, 700, and 1000\,K, respectively.
A total of 2264 point sources are detected with $[3.6]-[4.5]\leq 0.7$
and $M_{4.5}<-10$. 

Figure \ref{fig:cmd2} shows an expanded view of the brightest MIR sources.
In addition to the progenitors of SN 2008S and NGC 300, we include
several well-studied LBVs (e.g., $\eta$-Carina), and a number of cool
hyper-giants (VY CMa, NML Cyg, and Var~A in M33) for comparison. The MIR
magnitudes and colors for $\eta$-Carina, VY CMa, and NML Cyg were
synthesized from ISO spectra (Sloan et al.~2003). The magnitudes of M33
Var A were obtained from Humphreys et al.~(2006). Also included are the
16 sources matched between the catalog of MIR sources presented here and
the LBV sample of M07, obtained from narrow-band H$\alpha$
imaging, using a $0\farcs6$ matching radius. The larger
circles within the dotted lines show the 
EAGB stars, which we discuss in detail below.

The primary point of Figures~\ref{fig:cmd} and \ref{fig:cmd2} is to show 
that there are {\it very few} massive stars with the colors and MIR luminosities 
of the progenitors of SN 2008S and NGC 300. Although it is difficult to
identify a quantitative criterion for inclusion in the class defined by
the progenitors of SN 2008S and NGC 300, it is clear from Figure
\ref{fig:cmd} that the number of analogs in color and magnitude
is very small with respect to the total
number of massive stars in M33.  For example, if to be included as an
analog to the SN 2008S and NGC 300 progenitors we require that
$M_{4.5}$ be brighter than or equal to the progenitor of NGC 300 and we
require that the color be redder than the lower limit on the progenitor of 
SN 2008S, we find a {\it single} source.  If we require $M_{4.5}<-10$ and color
redder than SN 2008S, we find just two sources. Casting the net more widely,
for the purpose of having a sample larger than one or two objects and in an
effort to be conservative, we use $M_{4.5}<-10$ and $[3.6]-[4.5]>1.5$ to 
identify a sample of 18 EAGB stars (large open circles).  We discuss the 
spectra and variability properties of these sources in \S\ref{section:sed} 
and \S\ref{section:var}.

Our choice of the cuts $M_{4.5}<-10$ and $[3.6]-[4.5]>1.5$ to identify
objects of interest is somewhat arbitrary.  
Because our argument in this paper relies on the fact that analogs to 
the progenitors of SN 2008S and NGC 300 are intrinsically rare 
(\S\ref{section:introduction} \& \S\ref{section:discussion}), this 
issue deserves discussion.
The magnitude limit is straightforward: it is meant to select objects that have 
bolometric luminosities indicative of massive stars ($\gtrsim8$\,M$_\odot$).
In \S\ref{section:sed} below, we show that  $M_{4.5}<-10$ is conservative;
only half of the 18 sources selected have bolometric luminosities large 
enough to be massive stars.  Note that had we required $M_{4.5}$ to be 
brighter than or equal to the NGC 300 progenitor, we would exclude 8 of 
our 18 sources (see Fig.~\ref{fig:sed}).   Our goal was to not miss any 
deeply-embedded massive stars and the criterion 
$M_{4.5}<-10$ accomplishes that goal.

The cut on color is more complicated.  We were motivated by 
several factors.  First, we wanted to avoid the AGB 
sequence blueward of 1.5, where the density of points increases 
dramatically and where the sample would consist largely of Carbon 
stars (see Figs.~\ref{fig:cmd} \& \ref{fig:cmd2}). Second,
this cut essentially eliminates contamination from background
active galaxies (Stern et al.~2005).
Third, the $[3.6]-[4.5]>1.5$ cut gives us a reasonable number of 
objects to assess individually --- it is neither too many, nor
too few (again, had we taken $[3.6]-[4.5]>2.0$ the sample would 
consist of just one or two objects). 

These considerations leave us open to the potential criticism that 
optically-obscured  massive stars may exist in the region 
$M_{4.5}<-11.5$ (again, avoiding the AGB feature in the CMD) and with 
$0.5<[3.6]-[4.5]<1.5$.  To address this, we have examined 
the 45 sources that occupy this region of the CMD.  Four 
are identified with the LBV candidate catalog from M07,
which we discuss more below.  Sixteen of the remaining 41
sources have bright optical counterparts from the catalog 
of M06.  Of the 25 sources that do not appear in M06,
16 are optically-detected, but at flux levels where the 
M06 catalog is highly incomplete.  That leaves 9 
optically-obscured sources that have MIR luminosities 
indicative of massive stars.  
Like the very brightest of the 18 sources within the
dotted lines in Figure \ref{fig:cmd2}, about half of these have very large 
MIR luminosities, considerably larger than the progenitors of SN 2008S
and NGC 300.  We conclude that in this region there is just a handful 
of sources that might be true analogs to these progenitors ---
and we note that these have colors $\sim0.5-1.0$ magnitudes bluer than the
lower limit on the progenitor of SN 2008S.  Indeed, 
self-obscuration to the extent of the NGC 300 and SN 2008S 
progenitors is exceedingly rare for the most luminous stars, 
as evidenced by the lack of objects in the upper right corner
of Figures \ref{fig:cmd} and \ref{fig:cmd2}.

Finally, we note that we have searched for 4.5\,$\mu$m
sources without 3.6\,$\mu$m detections
that would lie within the dotted lines in  Figure \ref{fig:cmd2} 
and we find just one source.  Close inspection of the images
reveals a marginal $3.6$\,$\mu$m detection and 
$[3.6]-[4.5]\approx1.5$.

Despite this long discussion of color and magnitude selection,
the primary point of Figures \ref{fig:cmd} and \ref{fig:cmd2}
still stands: there are remarkably few massive stars in M33
that have the color and luminosity of the progenitors of 
SN 2008S and NGC 300.  As we discuss in \S\ref{section:sed},
all are consistent with relatively low-mass massive stars.

The second point to note from Figures \ref{fig:cmd} and \ref{fig:cmd2} 
is that the EAGB stars we have selected are not optically-luminous LBVs, 
$\eta$-Carina analogs, or cool hyper-giants. Indeed, all of the 
LBV candidates (open triangles) have
$[3.6]-[4.5]\lesssim0.8$, and about half have colors $\lesssim0.3$.  The
cool hyper-giants and $\eta$-Carina-like objects are also bluer than the
EAGB population, and considerably brighter than the progenitors
of SN 2008S and NGC 300. Indeed, the latter are most naturally
associated in this diagram with the luminous red extremum of the AGB 
population, hence our use of ``extreme-AGB'' (EAGB) stars.

The color-color diagram for all the sources detected in the four IRAC
bands is shown in Figure \ref{fig:cc}. The symbols are the same as in
Figure~\ref{fig:cmd2}. The small points with extremely
red $[5.8]-[8.0]$ colors are relatively dim, with $M_{4.5}>-10$ 
and are likely young stellar objects (YSOs; e.g., Bolatto et al.~2007).
The strong deviation of the SN 2008S and NGC 300 progenitors from the 
blackbody curve (solid line) reinforces the fact that the SEDs of 
these sources are not well-fit by a simple blackbody (P08a).
Despite this, the SN 2008S and NGC 300 progenitors, EAGBs, LBV candidates, 
and cool-hypergiants do not stand out as separate populations in 
$[5.8]-[8.0]$ color.  This is important because it means that 
$[5.8]-[8.0]$ color alone cannot be used as a metric for inclusion or 
exclusion from the class of SN 2008S/NGC300-like  progenitors.

\begin{figure*}[t]
\begin{center}
\centerline{\includegraphics[width=8.5cm]{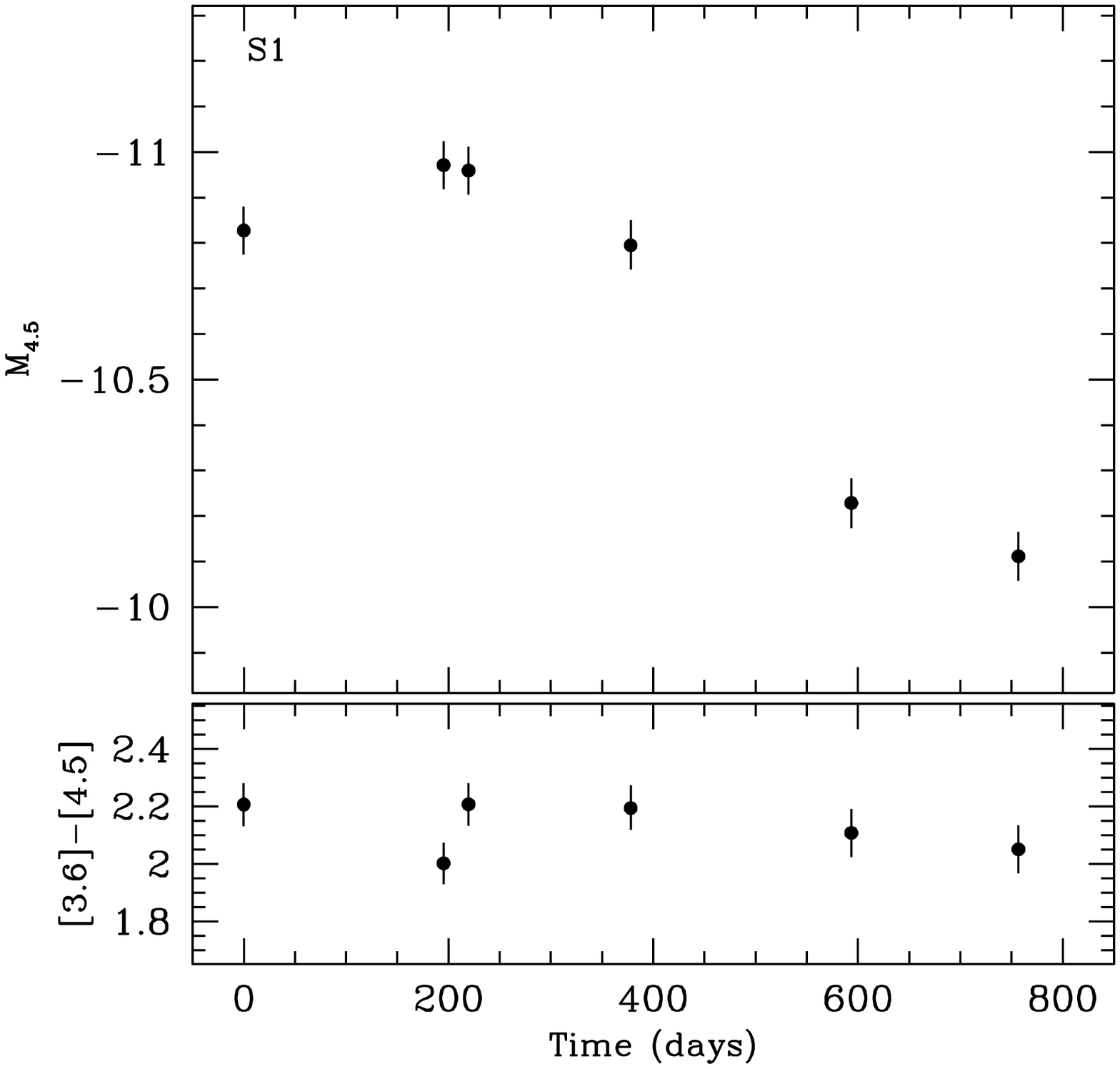}
\includegraphics[width=8.5cm]{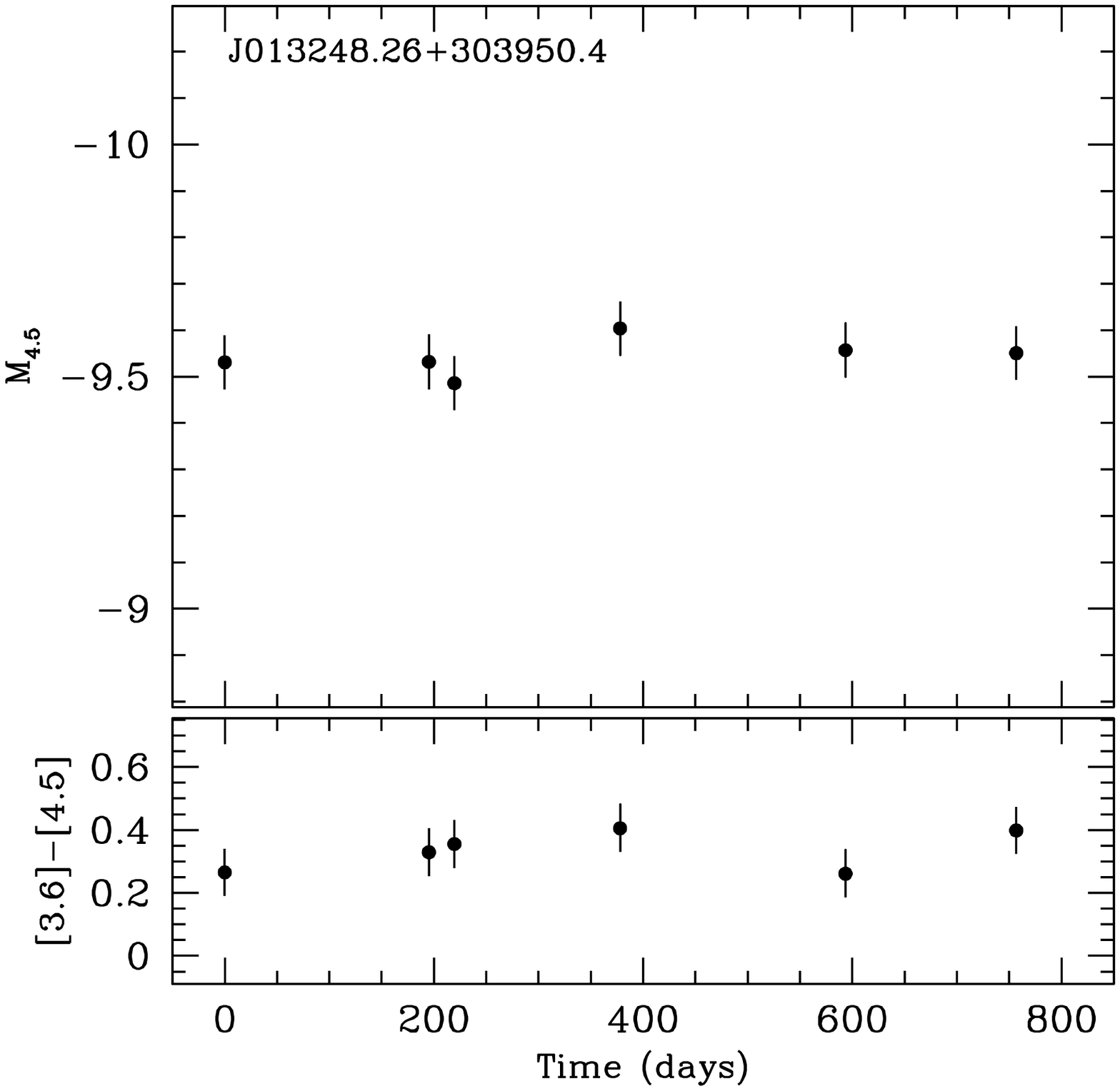}}
\caption{{\it Left panel:} Lightcurve for S1 (see Fig.~\ref{fig:s1}), the reddest of the 18 EAGB
stars selected in Figure \ref{fig:cmd2}. Note the high degree of variability,
which for this source, is {\it inconsistent} with the progenitor of SN 2008S
and (potentially) NGC 300 (see Fig.~\ref{fig:rms}).
Lightcurves for all of the EAGB stars are shown in Figures \ref{fig:agb1}
and \ref{fig:agb2}.  Although not strong in this example, most of the 
sources exhibit correlated color-magnitude variations.
{\it Right panel:} 4.5\,$\mu$m lightcurve for one of the LBV candidates
from M07 (see Figs.~\ref{fig:lbv1} \& \ref{fig:lbv2} for the complete set).  
The large majority of the 16 LBV candidates are not highly variable at 
4.5\,$\mu$m, although there are two exceptions (Fig.~\ref{fig:rms}).
}
\label{fig:comp}
\end{center}
\end{figure*}

\subsection{Spectral Energy Distributions}
\label{section:sed}

The limits on the optical emission from the progenitors of NGC 300
and SN 2008S are tight, and effectively rule out an optically
unobscured massive star (Berger \& Soderberg 2008; P08a; Fig.~\ref{fig:sed}).
For the purposes of finding analogs to these sources, it is critical
to derive the optical luminosities and/or upper limits for the 
18 EAGBs identified in \S\ref{section:sources}.  Here, we 
present these results and we compare the derived optical-to-MIR SEDs 
of the EAGBs to the optically-bright (narrow band H$\alpha$-selected)
LBV candidates from M07.

As part of the {\it Survey of Local Group Galaxies Currently Forming
Stars} (SLGG), M06 presented a catalog of $\sim 150,000$ point sources
in M33 with well-calibrated $UBVRI$ photometry obtained from
observations at the KPNO 4m telescope with the Mosaic imager. We use the
published photometric catalog and images from M06 to complement the 
{\it Spitzer} MIR photometry described in \S\ref{section:sources}.

We first cross-matched the
positions of the EAGB stars with the M06 photometric catalog.
Importantly, we do not find any optical counterpart to the EAGB
sources using a matching radius of $0\farcs5$. 
Since the completeness of the M06 catalog starts to decline rapidly at
$V\simeq 22$\,mag, we analyze the images independently to look for faint
optical counterparts to the EAGB stars. We use SEXtractor (Bertin \&
Arnouts 1996) with a low detection threshold ($2\sigma$ above the local
background) to detect and measure aperture photometry (using a small
aperture of 3~pixels radius) of all the sources detected in the
KPNO/Mosaic $UBVRI$ images of M33 from M06. We calibrate the photometry
relative to the magnitudes in the catalog of M06. Using a radius of
$0\farcs5$ to cross-match the MIR positions of the EAGB stars with our
multi-band catalogs of faint optical sources, we only detect two EAGB
stars in the $BVR$ bands (again, 2$\sigma$). The remaining 17 sources do not 
have optical counterparts. We estimate $3\sigma$ upper limits on the $UBVRI$
magnitudes using the local background RMS at the positions of the EAGB
stars. The median $3\sigma$ upper limits are: $U=24.0$, $B=24.0$,
$V=23.5$, $R=23.0$, $I=22.5$.  The data for each of the 18 EAGB stars
is listed in Table \ref{table:eagb}.

In order to convert MIR and optical magnitudes to
fluxes, we used zeropoints in Reach et al.~(2005) and Cohen et
al.~(2003) for the EAGBs and LBV candidates, respectively. 
The luminosities of all sources were
calculated assuming a constant reddening of $E(B-V)=0.15$~mag and a
distance of $\mu=24.92$ (Bonanos et al.~2006).  The reddening 
correction is motivated by the uncorrected $B-V$ color-magnitude
diagram of M06, which shows that the bluest sources only reach
$B-V\approx-0.2$, instead of $\approx-0.33$, as would be expected
from an un-reddened massive star.  In addition, 
Bonanos et al.~(2006) also quote an average reddening
correction of 0.1 mag to massive stars for M33.  A larger
adopted reddening correction increases our upper-limits 
for the optical fluxes of the EAGB stars.

The primary result of this procedure is
Figure~\ref{fig:sed}, which shows the SEDs 
of the EAGB stars (left panel) and LBV candidates (right panel).
The dotted lines in the left panel show the range of upper
limits (and one $BVR$ detection, as described above) for the 
18 EAGB stars at $UBVRI$ (see Table \ref{table:eagb}).  
The filled triangles show the 
optical and 3.6\,$\mu$m upper limits for the progenitor of 
SN 2008S. The solid squares show the MIR detections 
and optical upper limits for the NGC 300 progenitor.
These sources should be contrasted with the optically-luminous
LBV candidates from M07 (right panel).  Note that 3 of the LBVs 
do not have 5.6 and 8.0\,$\mu$m detections.  The fact that one of 
the non-detections would appear to have a 5.6 and 8.0\,$\mu$m
flux larger than some of the other detections is a consequence of
the locally higher MIR diffuse flux near that particular object.
 
There are a number of points to take away from the two panels of 
Figure \ref{fig:sed}.  First, 9 of the 18 EAGB stars we identified
in Figure \ref{fig:cmd} do not have bolometric luminosities indicative
of massive stars; they have $L_{\rm bol}\lesssim2\times10^4$\,L$_\odot$.
Thus, these are not likely to be true analogs to the SN 2008S and NGC 300 progenitors.
Second, all of the sources are highly optically-obscured, with 
$\lambda L_\lambda[V]/\lambda L_\lambda[4.5\,\mu{\rm m}]\sim10^{-2}$.
Third, these sources are qualitatively different from the more bolometrically
luminous LBV candidates (right panel).  The LBVs are interesting in their own right, 
dividing approximately into two classes: (1) relatively optically-dim with a 
strong MIR excess and (2) optically-bright with little MIR excess, if at 
all.  This division is also evidenced by their positions in the 
CMD (Fig.~\ref{fig:cmd}), which indicates a bimodality in 
MIR color.  For a possible analog to LBVs with a MIR excess, 
see Smith (2007).

\subsection{Variability}
\label{section:var}

Because four epochs of archival {\it Spitzer} data were available for
NGC 6946 in the three years prior to the discovery of SN 2008S, P08a
investigated potential variability of the progenitor. They found that
there was remarkably little and showed that this fact could be used to
constrain the motion of the obscuring medium, under the assumption of a
geometrically-thin, but optically-thick shell (see \S\ref{section:class}; P08a).

Motivated by this result, and by the fact that six epochs of archival
data over two years exist for M33, we investigated the MIR variability of
all the sources detected at 3.6 and 4.5\,$\mu$m.
To generate lightcurves, we used the difference imaging analysis package
ISIS, based on the techniques of Alard \& Lupton (1998), Alard (2000).\footnote{
See http://www2.iap.fr/users/alard/package.html.}  For a discussion,
see Hartman et al.~(2004).  

There is a striking difference between the 4.5\,$\mu$m light curves of the
18 EAGB stars and the LBV candidates from M07. Most of the EAGB stars are 
highly variable, both in magnitude ($0.1 \lesssim {\rm RMS (mag)} \lesssim 0.8$) 
and in $[3.6]-[4.5]$ color.  In contrast, the
majority of the LBVs are not variable (${\rm RMS} \lesssim 0.1$\,mag);
only two sources show clear variability (both are blue in $[3.6]-[4.5]$). 
To illustrate these differences, we show in Figure~\ref{fig:comp} the light 
curve at 4.5~$\mu$m and color variations of the reddest source in our EAGB sample
S1 (left panel) and an LBV candidate (right panel). For completeness, we present 
all of the 4.5~$\mu$m light curves and color variations of the 18 EAGB stars 
in Appendix~\ref{appendix:agb} (Figures~\ref{fig:agb1} \& \ref{fig:agb2}),
and of the LBVs in Appendix~\ref{appendix:lbv} (Figures~\ref{fig:lbv1} \&
\ref{fig:lbv2}).

\begin{figure*}[t]
\centerline{\includegraphics[width=15cm]{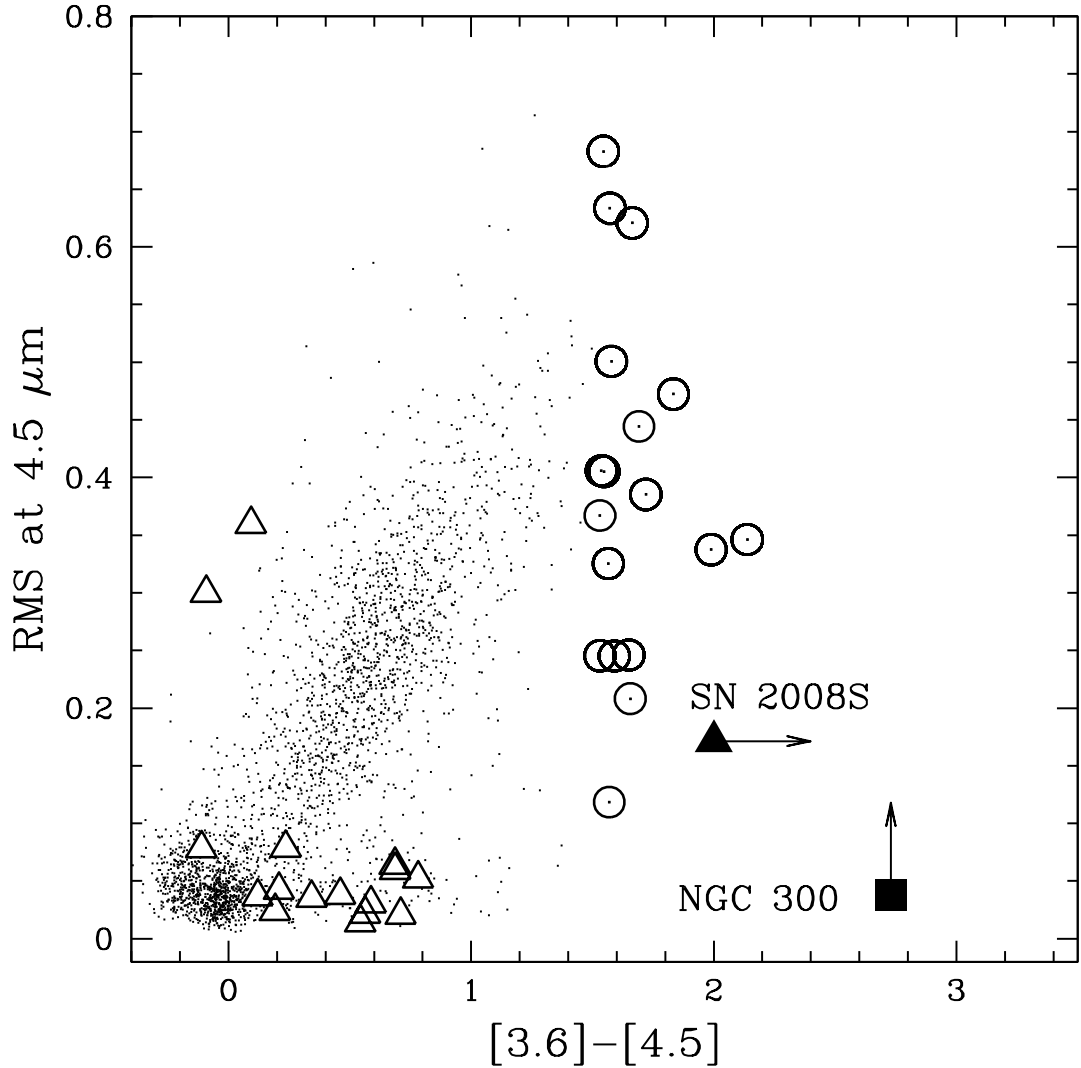}}
\caption{RMS variation in 4.5$\mu$m magnitude as a function of
$[3.6]-[4.5]$ color for all of the sources detected in 3.6$\mu$m and
$4.5$$\mu$m with $M_{4.5}<-10$ (points) and the 16 LBV candidates (open
triangles). As in Figure \ref{fig:cmd2}, the 18
EAGB stars are denoted with larger open circles.  As
noted in Figures \ref{fig:agb1} \& \ref{fig:agb2}, as well as
\ref{fig:lbv1} \& \ref{fig:lbv2}, the extreme-AGB stars are highly
variable, whereas all but two of the LBV candidates are not.
Variability of the progenitors of SN 2008S and NGC 300 is also shown.
For the former, the data are taken from P08a (their Fig.~2).  For
NGC 300 only two epochs are available and hence the value of the
RMS ($\approx0.05$) is a lower limit on the variability of the
progenitor.}
\label{fig:rms}
\end{figure*}

Figure~\ref{fig:rms} summarizes these findings.  It shows the measured 
RMS at 4.5~$\mu$m as a function of $[3.6]-[4.5]$ color for all the bright 
sources with $M_{4.5} < -10$. The symbols are the same as in 
Figures \ref{fig:cmd2} and \ref{fig:cc}.
There is a clear correlation evident between the RMS (or amplitude) and 
color for the AGB stars (see also Mc07).  

For comparison, the RMS variation of the progenitor of SN 2008S 
derived from its 3-year light curve is also shown. In
the case of the NGC 300 progenitor we can only put a lower limit on its
RMS variation, because only two epochs of archival {\it Spitzer} imaging
exist.  Nevertheless, it is striking that both the SN 2008S and the NGC 300
progenitors are consistent with very little variation in the few years
preceding their explosions.  In particular, SN 2008S is inconsistent 
with the clear trend among the AGB stars to become more variable as they
become redder.  Only a handful of the EAGB stars vary so little,
which suggests that a {\it lack of variability among an otherwise 
variable EAGB star population} may be used as a selection 
criterion for analogs to the SN 2008S and NGC 300 progenitors.
As an example, requiring the RMS to be $\lesssim0.3$\,magnitudes, 
we find $5$ sources.  They are S12, S7, S9, S17, and S8, in order of
increasing RMS (see Table \ref{table:eagb}).  
Sources S8 and S9 are among the lower luminosity sources in 
the left panel of Figure \ref{fig:sed}, and are therefore not likely 
true SN 2008S analogs.  In contrast, S7 is the brightest of our EAGB 
stars.  Finally, the least variable source (S12) has 
$M_V\approx-11.6$, which is quite close to the SN 2008S progenitor,
even though it is $\sim0.5$ magnitudes bluer.

Although we have identified a few rare and interesting sources, 
the sparsity of data in Figures \ref{fig:cmd2} and
\ref{fig:rms} with $[3.6]-[4.5]>1.5$ makes it difficult to construct a 
strict quantitative joint criterion in the space of luminosity, color, and 
variability for inclusion in the class of SN 2008S-like progenitors.
Based on the discussion above, as in \S\ref{section:sources} and \S\ref{section:sed},
we expect a total of $\sim1-10$ in M33.
A more complete multi-epoch survey of EAGB stars in the local universe 
may fill in the region in Figure \ref{fig:rms} between the AGB locus
and the SN 2008S and NGC 300 progenitors.  Most importantly, it might 
make clear a quantitative criterion for ``SN 2008S-like'' 
in the RMS-color plane.

\subsection{Other Galaxies}
\label{section:other_galaxies}

Blum et al.~(2006) and Bolatto et al.~(2007) present {\it Spitzer}
point source catalogs for the LMC and SMC, respectively.  We have
searched these catalogs for sources that satisfy the selection
criteria $M_{4.5}<-10$ and $[3.6]-[4.5]>1.5$ used to identify
the 18 EAGB stars discussed throughout this section.  In the 
catalog of Blum et al.~(2006), we find 9 sources.  Three  
are coincident with 2MASS sources and 5 appear in the IRAS catalog. 
Although more careful follow-up is clearly required, a subset of 
these 9 sources may be EAGB stars.  In the catalog 
of Bolatto et al.~(2007) for the SMC, we find a single 
source that satisfies $M_{4.5}<-10$ and $[3.6]-[4.5]>1.5$.
Finally, we have also completed a cursory search for EAGB stars 
in archival imaging of NGC 300 (PI R.~Kennicutt; ID 40204), 
and we find just 4 potential sources. 

In sum, even the relatively conservative criteria  
$M_{4.5}<-10$ and $[3.6]-[4.5]>1.5$ pick out remarkably few
stars in any galaxy --- M33 is not peculiar in this regard.
Given the fact that we expect only a 
fraction of the 18 sources in M33 to be bona fide analogs 
to the progenitors of SN 2008S and the transient in NGC 300
(based on luminosity, color, and variability; see \S\ref{section:sed}, 
\S\ref{section:var}),  a more careful look at the sources of interest 
in the LMC ($\approx9$), SMC ($\approx1$),  and NGC 300 ($\approx4$) 
is likely to further decrease the total number of sources of interest
in these systems. As emphasized in 
\S\ref{section:introduction} and in \S\ref{section:discussion} below,
the scarcity of SN 2008S-like progenitors with respect to the total
massive star population is remarkable in light of the 
fact that SN 2008S-like transients are likely to be relatively common 
with respect to the overall supernova rate.

\section{Discussion}
\label{section:discussion}

We have shown that in the primary metrics of color, luminosity,
and variability, stars analogous to the progenitors 
of SN 2008S and NGC 300 are exceedingly rare in star-forming galaxies.  
They have luminosities characteristic of low-mass massive stars,
are deeply dust-obscured with extremely red MIR colors, and show
little MIR variability (see Figs.~\ref{fig:cmd}, \ref{fig:cmd2}, 
\ref{fig:sed}, and \ref{fig:rms}).  In luminosity and color they are distinct
from the population of optically-luminous LBV candidates selected 
from M06.  Although many of the reddest objects selected as EAGB
stars in Figure \ref{fig:cmd2} are highly variable, the few ($\sim1-5$)
least-variable sources most closely resemble the SN 2008S and NGC 300 
progenitors.  In this way (but in only this way), they are similar
to the LBV candidates.

In this section we discuss the implications of our finding that 
stars with characteristics analogous to the progenitors of SN 2008S
and NGC 300 are rare.  
In \S\ref{section:numbers} we estimate the overall fraction of massive
stars that are deeply dust-embedded and the lifetime of stars 
in that state.  In \S\ref{section:theory}, we connect
with the evolution of massive stars, including the possibility
that SN 2008S-like transients are the result of electron-capture
supernovae or massive white dwarf birth.  Section \ref{section:survey} discusses how many 
EAGB stars can be found within the local universe ($D\lesssim10$\,Mpc) 
using {\it Spitzer}.

\subsection{Numbers \& Rates}
\label{section:numbers}

The total number of analogs to the progenitors of SN 2008S and NGC 300
in M33 is uncertain.  This uncertainty comes primarily from the fact that 
we are unable to identify an absolute quantitative criterion for inclusion
into this progenitor class based on our three primary metrics: 
color, luminosity, and variability.  To be conservative,
we have identified 18 sources in the region  $M_{4.5}<-10$ and $[3.6]-[4.5]>1.5$
of the CMD that satisfy the minimal criteria of being bright and
extremely red (larger open circles in Fig.~\ref{fig:cmd2}).  
However, a very strict cut in color and magnitude (i.e., all sources redder 
than the lower limit to SN 2008S and brighter than NGC 300) yields just two 
sources.  Among the 18 selected sources, we have shown that roughly half do 
not have bolometric luminosities indicative of massive stars.  That is, they
do not have luminosities as large as one would expect for stars who
are traditionally thought to end their lives as supernovae 
($L_{\rm bol}\gtrsim4\times10^4$\,L$_\odot$).
It is important to note, however, that at fixed final luminosity of a 
massive star its initial ZAMS stellar mass 
may be multi-valued, implying progenitors with either 
$\sim5-7$, $\sim8-9$, 
and $\sim11-14$\,M$_\odot$ for $L\approx5-10\times10^4$\,L$_\odot$.  
(see Fig.~2 of Smartt et al.~2009; \S\ref{section:theory}; see
footnote \ref{foot_p08b}).
Of course, the explosions SN 2008S and NGC 300 may not be 
true supernovae, but rather a new class of bright eruptions from 
obscured massive stars (see \S\ref{section:theory}). Nevertheless, comparing 
these 18 sources to the progenitors of SN 2008S and NGC 300 in Figure 
\ref{fig:sed} we would argue that only roughly half belong to this
progenitor class based on SED alone.  Finally, Figure \ref{fig:rms}
shows that only $\approx1-5$ of the 18 sources vary as {\it little} 
as the progenitors of SN 2008S and (potentially) NGC 300.  Importantly, 
16 of the 18 sources satisfy the criterion of being highly 
optically-obscured (see Table \ref{table:eagb}).

In summary, very few massive stars have the color, luminosity, and 
variability of the SN 2008S and NGC 300 progenitors.  Our best guess 
is that the number of true analogs may be as few as zero and as large
as $\sim10-20$.  We denote this number in M33 --- the number of {\it true}
analogs --- as $N_{\rm EAGB}$.
A larger sample of stars, culled from a larger
multi-epoch study of local star-forming galaxies (see \S\ref{section:survey}) 
is clearly needed to fill in the parameter space in the extreme red 
and bright side of the CMD.  This is the most robust way to understand
$N_{\rm EAGB}$ and its uncertainty.

In order to evaluate the {\it fraction} of stars in M33 that might be analogs
to the progenitors of SN 2008S and NGC 300, and so constrain the rate of
production of such objects, we must first estimate the total number of
massive stars in M33 ($N_\star$; i.e., with $M_{\rm ZAMS}\gtrsim8$\,M$_\odot$).
This number can be estimated in several ways: (1)
extinction-corrected H$\alpha$ luminosity (e.g., Hoopes et al.~2001; Hoopes \& Walterbos 2000;
Greenawalt 1998), (2) dust-reddening corrected UV continuum luminosity
(for {\it GALEX} observations, see Thilker et al.~2005), 
(3) total number of main-sequence optical point
sources detected with $M_V\lesssim-2$ (appropriate for stars with ZAMS 
masses above $\approx9-10$\,M$_\odot$; Lejeune \& Schaerer 2001), 
or (4) total number of red supergiants (RSGs)
($M_V\lesssim-3.5$; M06) times the ratio of the lifetime of a massive star 
to the time spent as an RSG ($t_\star/t_{\rm RSG}\approx10$; e.g., 
Schaller et al.~1992).  
Using the latter method, and selecting RSGs with $V-R>0.5$ and $M_V<-3.5$ 
from the catalog of M06, we find $\approx5400$ sources, implying 
$N_\star\approx5.4\times10^4$.  Taking  a more conservative color cut of 
$V-R>0.7$ and $M_V<-3.5$, we find that  $N_\star\approx3.5\times10^4$.  
Similar estimates in the range of $N_\star\approx 3-6\times10^4$
are obtained using method (3) with the M06 catalog, although this 
estimate suffers significantly from incompleteness.
We take $N_\star=5\times10^4$ as a fiducial number and include it in 
our scalings below. Note that estimates of the
total star formation rate in M33 range from $\sim0.3$ to $\sim0.7$
M$_\odot$ yr$^{-1}$, consistent with the UV, H$\alpha$, and FIR luminosities 
(e.g., Gardan et al.~2007 and references therein),
implying a supernova rate for the galaxy of $\sim0.005$\,yr$^{-1}$
(e.g., Gordon et al.~1998).

Taking $N_{\rm EAGB}\sim5$ and $N_\star\approx5\times10^4$,
we find that a fraction 
\begin{equation}
f_{\rm EAGB}=\frac{N_{\rm EAGB}}{N_\star}\sim1\times10^{-4}
\left[\frac{N_{\rm EAGB}}{5}\right]
\left[\frac{5\times10^4}{N_\star}\right]
\label{feagb}
\end{equation}
of the massive stars in M33 may be analogs to the progenitors of 
SN 2008S and NGC 300. 

As noted in \S\ref{section:introduction} (point 2) and \S\ref{section:rates}, 
only a fraction of
all massive stars go through this highly dust-enshrouded phase, and
produce transients like SN 2008S and NGC 300.   Since, by assumption,
roughly all of the massive stars in any galaxy become normal core-collapse
supernovae (but, see Kochanek et al.~2008), the rate of SN 2008S-like explosions 
can be characterized by their fractional rate with respect to the 
overall supernova rate.  This fraction is determined by dividing
the observed rate of SN 2008S-like transients by the total number of 
supernovae within some volume, times an incompleteness 
correction that accounts for the fact that SN 2008S-like transients are intrinsically 
less optically luminous.  Based on the numbers presented in \S\ref{section:rates}, 
we estimate that $f_{\rm SN}\approx0.2$, 
although higher and lower values are not excluded.  For example, it is possible 
that SN 2008S-like transients are intrinsically rare and the fact that 
NGC 300 and SN 2008S occurred in the same year was simply chance.
Although we cannot exclude this possibility, we note that such
an explanation appears improbable in the face of what is known about 
the rarity of their progenitors ($f_{\rm EAGB}$; eq.~\ref{feagb}).
Conversely, it is possible that the incompleteness correction exceeds 
the factor of $\sim2$ advocated in \S\ref{section:rates} within 10\,Mpc
and that such transients are indeed common with respect to supernovae.  
However, it then becomes 
increasingly difficult to explain why no more SN 2008S-like 
transients were observed within 10\,Mpc in the last 10 years. 
There is no way to circumvent these uncertainties without a 
more complete census of progenitors and outbursts.  

As stated in \S\ref{section:introduction}, the simplest explanation
for the fact that SN 2008S- and NGC 300-like transients are 
simultaneously common with respect to supernovae ($f_{\rm SN}\sim0.2$) and that 
their progenitors are very rare by number at any moment, in any 
star-forming galaxy ($f_{\rm EAGB}\sim10^{-4}$) with respect to massive stars, 
is that a significant fraction of all massive stars ($\sim0.2$) go through a 
brief evolutionary epoch in which they are highly dust-obscured, 
just before explosion.
Taking the average lifetime of massive stars with ZAMS masses 
in the range of $9-10$\,M$_\odot$ to be $t_\star\approx3\times10^7$\,yr 
(e.g., Schaller et al.~1992), we find that
the duration of this dust-obscured phase is 
\begin{equation}
t_{\rm EAGB}
\sim
1\times10^4
\left[\frac{t_\star}{10^{7.5}\,\,\rm yr}\right]
\left[\frac{N_{\rm EAGB}}{5}\right]
\left[\frac{5\times10^4}{N_\star}\right]
\left[\frac{0.20}{f_{\rm SN}}\right]\,\,{\rm yr}.
\label{teagb}
\end{equation}
We consider the uncertainty in $f_{\rm SN}$ to be at the factor of 
two level and the uncertainty in $N_\star$ to be at the level of 
a factor of 1.5.  However, as we have stressed,  
$N_{\rm EAGB}$ may be as much as a factor of 5 or more lower ($t_{\rm EAGB}\lesssim10^3$\,yr), 
or a factor of $\sim2-4$ higher ($t_{\rm EAGB}\sim6\times10^4$\,yr). 
To improve these numbers significantly, 
a careful monitoring program for optical transients like SN 2008S within 
the local universe ($D\lesssim10$\,Mpc), 
coupled with a survey of all local galaxies for bright MIR point 
sources with (warm) {\it Spitzer} (see \S\ref{section:survey})
should be undertaken.  The combination of watching for more
transients of this type and associating them with individual
progenitors whose luminosities and variability have been cataloged
will significantly decrease the uncertainty in both $f_{\rm SN}$ and 
$N_{\rm EAGB}$, and significantly increase our understanding of
the causal mapping between progenitors and their outbursts. \\

\subsection{Connection to The Evolution of Massive Stars}
\label{section:theory}

The relation between final luminosity and initial stellar
mass may be triply-valued (Smartt et al.~2009) at $\sim5-7$, $\sim8-9$, 
and $\sim11-14$\,M$_\odot$ for $L\approx5-10\times10^4$\,L$_\odot$.  
This relation is of course uncertain, particularly in the mass range 
singled out by the bolometric luminosity of the 2008S and NGC 300 progenitors
near $\sim10$\,M$_\odot$.  It is likely further complicated by 
binarity, and by the mass-loss history, metallicity, and rotation of 
massive stars.  Because the absolute rate of these outbursts as well as 
whether or not they should be associated with the death of the
progenitor are still uncertain, we consider a number of potential
scenarios below.  We list a subset of the possibilties in order of 
increasing progenitor mass.

\subsubsection{Massive White Dwarf Birth: $M\approx6-8$\,M$_{\odot}$}

We have referred to the progenitors of SN 2008S and NGC 300 
throughout this work as extreme (``E'') AGB stars because they 
lie at the red extremum of the AGB sequence in the MIR 
color-magnitude diagram (Figs.~\ref{fig:cmd} \& \ref{fig:cmd2}).
Taken literally, these stars may indeed be the progenitors
of the most massive O-Ne-Mg white dwarfs, undergoing explosive 
core-envelope separation as they transition to proto-planetary 
nebulae (e.g., Riera et al.~1995; Garc{\'{\i}}a-Hern{\'a}ndez et al.~2007).  
Perhaps the 2008S and NGC 300 progenitors were then akin to the most 
massive highly-evolved carbon- or oxygen-rich AGB stars (Kwok 1993).

Based on analogy with local proto-planetary nebulae, we would
expect bi-polar explosion morphology and eventually the emergence
of a hot ionizing continuum source as the newly-born white dwarf
begins its cooling phase (perhaps similar to Hen 3-1475/IRAS 17423-1755;
Riera et al.~2003).  The initial luminosity of the central
source would be of order $\sim5\times10^4$\,L$_\odot$ for a 
white dwarf near the Chandrasekhar mass and it should cool on a  
timescale comparable to $\sim10^5$\,yr.  Thus, the 
bolometric luminosity of the transient should eventually decrease back 
to approximately pre-outburst levels.  The primary distinguishing characteristic
of this particular scenario is the (eventual) emergent hot continuum source
and emission lines, bi-polar morphology, 
and  the fact that the bolometric luminosity should not decrease to pre-outburst
levels in the next decades (e.g., Kwok 1993).

\subsubsection{Electon-Capture Supernova: $M\approx9$\,M$_\odot$}
\label{section:ecsn}

The timescale estimated in equation (\ref{teagb})
is of the right order of magnitude to be associated with the 
onset of carbon burning in relatively low-mass massive stars.
This is traditionally a very difficult phase to model (see the 
summary in Woosley et al.~2002; Siess 2006; Poelarends et al.~2008).

One of the most intriguing explanations for the physics of SN 2008S-like
transients is that they result from electron-capture SNe (ecSNe)
of O-Ne-Mg cores of relatively low-mass massive stars (Miyaji et al.~1980).  
While speculative, this explanation accounts for many of the 
observed characteristics of both the transients and their 
progenitors.  In particular, it accounts for the fact that 
the progenitors of NGC 300 and SN 2008S were relatively 
low luminosity and deeply embedded.  Here we follow the scenario 
detailed by Poelarends et al.~(2008) (see also Nomoto 1984, 1987; Ritossa et al.~1996, 
Seiss 2006, 2007; as well as Chugai 1997; Wheeler et al.~1998; 
Woosley et al.~2002; Eldridge et al.~2007; Wanajo et al.~2009).

We know from the properties of the observed progenitors and our analysis 
of the luminous stars in M33 that the progenitors are extreme AGB
stars.  In these systems, the combination of thermal pulses due to
He shell burning and dredge-up produces a massive, dusty wind
(for lower luminosity and less enshrouded analogs, see the 
work on carbon stars in the Magellanic clouds by 
Groenewegen et al.~2007, as well as van Loon et al.~2005, 2006).
In the Poelarends et al.~(2008) models, the mass loss peaks
at nearly $10^{-4}$\,M$_\odot$ yr$^{-1}$ for stars of mass
$M\simeq 9M_\odot$ and luminosity $L\simeq 10^5 L_\odot$,
and then drops precipitously for slightly more massive stars
which can support core Neon burning, and which eventually 
become normal iron-core core-collapse supernovae (ccSNe) 
(see Fig.~13 from Poelarends et al.~2008).
The thermal pulses driving the mass loss occur at very high
rates for the EAGB stars (on timescales of years), suggesting that 
mass loss may appear as a steady wind, as seems to be required for 
the small variability in the lightcurves of the SN 2008S and NGC 300 
progenitors (see Fig.~\ref{fig:rms}; discussion in P08a), 
rather than as impulsive ejections of optically-thick shells
expected for normal carbon stars.  
The low degree of variability seen in SN 2008S (particularly when
contrasted with the EAGB stars in Fig.~\ref{fig:rms})
might also be explained by the onset of {\it core} carbon 
burning as the final phases of stellar evolution commence.
Note that the work of Nomoto (1984) and (1987) implies that
the EAGB envelope and mass-loss would be carbon-enhanced.

For a narrow mass range near $9$\,M$_\odot$, the balance between
mass loss and the growth of the core allows the core to become
unstable to collapse without igniting core Neon burning, leading
to an ecSNe.  This occurs only for a small model-dependent mass 
range near $9$\,M$_\odot$ (see also, e.g., Nomoto 1984; Seiss 2006, 2007).   
Poelarends et al.~(2008) estimated a mass range
of $M_{\rm min}\simeq9$\,M$_\odot$ to $M_{\rm max}\simeq9.25$\,M$_\odot$.
For a standard Salpeter IMF, and assuming 
that all stars from $M_{\rm max}$ to 20\,M$_\odot$ form
ccSNe, then the ecSN fraction is just 
$\approx6$\% ($9.0\leq M\leq9.25$\,M$_\odot$).
This is below our (albeit, uncertain) fiducial estimate
for the rate of SN 2008S-like transients relative to the 
normal ccSN rate, $f_{\rm SN}\sim0.2$. However, other studies
have found somewhat broader mass ranges for ecSNe.  
For example, Seiss (2007) find that 
$M_{\rm max}-M_{\rm min}\approx1-1.5$\,M$_\odot$, 
which implies a fractional rate for ecSNe more 
in accord with our nominal estimate for $f_{\rm SN}$.

While the observed rate of SN 2008S-like transients is uncertain,
we have argued that they represent a modest fraction of the normal 
ccSN rate, consistent with a limited progenitor mass range.  
The relatively low luminosity of their progenitors implies that 
they are low-mass massive stars, potentially near the 
boundary between electron-capture and normal core-collapse supernovae.
In addition, Kitaura et al.~(2006) argue that ecSNe should be sub-luminous 
compared to normal ccSNe, because of their low Ni yields, potentially 
explaining the low luminosity of SN 2008S-like transients.  Finally, 
for the fiducial ecSN model of Poelarends et al.~(2008) (see also Nomoto 1984), 
the  AGB phase lasts for $\approx4\times10^4$\,yr, which although short with
respect to the lifetime of the star itself, is of order 
$t_{\rm EAGB}$ in equation (\ref{teagb}) based on the 
number of analogs to the SN 2008S and NGC 300 progenitors 
in M33.  Of course, these timescales need not be identical,
since the fiducial ecSN progenitors of Poelarends et al.~may 
evolve significantly in color as a result of mass-loss
during their super-(``extreme'') AGB phase, becoming 
increasingly like the SN 2008S and NGC 300 progenitors as they
approach the end of their lives.

Fortunately, this speculative explanation has at least one simple
and testable prediction: there should be no surviving progenitor 
once the transient fades.  This may be testable in the optical, since 
most of the dust enshrouding the progenitor was likely destroyed by the
explosion, but observations in the MIR will be required to be
certain that a new shroud has not formed.  A second test is to
find strong evidence in the late-time lightcurve for synthesized
$^{56}$Ni.  This may be difficult both because
ecSN may produce little $^{56}$Ni (Kitaura et al.~2006), and
because dust, whether that remaining from the EAGB phase or
dust formed in the ejecta (e.g., P08c for M85), may make it difficult 
to correctly measure the late-time decay rate.  {\it Spitzer}
may again be key in constraining the nature of these events
because of this obscuration. A third test is to ensure 
that {\it Spitzer} has surveyed all the nearby galaxies so that future 
examples of these transients can be causally connected to their 
deeply obscured progenitors.  Finally, as with ccSNe, very nearby 
ecSNe ($D\lesssim300$\,kpc) should produce neutrino signatures 
characteristic of neutron star formation, which detectors such 
as SuperKamiokande and its successors would observe (see, e.g., Thompson et al.~2003;
Kistler et al.~2008).

The recent paper by Botticella et al.~(2009) in part corroborates the interpretation
discussed in Prieto et al.~(2008a) and proposed here that SN 2008S may
be an electron-caputre supernova. They present the late-time quasi-bolometric 
lightcurve of SN 2008S, which shows evidence for a power-law time dependence 
with a slope indicative of being powered by the radioactive decay $^{56}$Co.
Although this need not uniquely signal ecSN as the physical mechanism
(see \S\ref{section:sn}), it provides some evidence for core-collapse,
and something perhaps akin to normal neutron star formation (Kitaura et al.~2006).
A lightcurve with similar cadence and photometric coverage has recently 
been published in Bond et al.~(2009) for NGC 300.

\subsubsection{Intrinsically Low-Luminosity Iron Core-Collapse Supernova: 
$M\sim10-12$\,M$_{\odot}$}
\label{section:sn}

Heger et al.~(1997) discuss a mechanism for generating a potentially 
obscuring ``superwind'' via pulsational mass-loss in red supergiants
between 10 and 20\,M$_\odot$ during the last $10^4$\,yr before 
explosion.  The prediction of enhanced AGB-like obscuration (they 
compare directly with luminous OH/IR stars),
the evolutionary timescale (compare their $10^4$\,yr with our eq.~\ref{teagb}), 
and the secular {\it increase} in the fundamental mode pulsation period as 
the star approaches death (see their Fig.~7) are all in good agreement
with the requirements on the 2008S and NGC 300 progenitors we discuss
in this paper. 

The physical mechanism of iron core-collapse supernovae is unknown
(Rampp \& Janka 2000; Liebend\"orfer et al.~2001; Thompson et al.~2003;
Buras et al.~2003; Burrows et al.~2006).
Recent observations hint that low-luminosity Type IIP supernovae
may be more common than previously thought (e.g., Chugai \& Utrobin 2000;
Pastorello et al.~2004, 2007), 
particularly when one accounts for the incompleteness corrections discussed in 
\S\ref{section:rates}.  Because the mechanism of supernovae
has yet to be conclusively identified, it is difficult to 
interpret the diversity in inferred $^{56}{\rm Ni}$ yield 
physically.  In fact, that diversity may be 
larger than previously thought, and we are only now 
appreciating the existence of a very low-luminosity 
tail to the Type IIP luminosity function.  If so, it is 
natural to imagine that these low-luminosity core-collapse
events might have analogs that occur in the very dusty circumstellar 
medium of their massive stellar progenitors, as in Heger et al.~(1997),
and thus may give rise to events like SN 2008S and NGC 300.

This scenario yields many of the predictions of the ecSN scenario
discussed in \S\ref{section:ecsn}.  Indeed, even with a complete
sampling of ``ec-'' and ``cc-'' supernovae, it may 
be difficult to disentangle the two populations since many of 
the predictions --- radioactive decay powered lightcurves,
potentially embedded progenitor, no ``postgenitor''  --- are the 
same in both. 

\subsubsection{Massive Star Outburst: $M\approx10-15$\,M$_{\odot}$}

On the basis of the relatively low luminosity of their progenitors, 
we view the ecSN and massive white dwarf birth 
scenarios discussed above as the most probable explanation SN 2008S and NGC 300. 
Nevertheless, there is of course the possibility that they are instead 
a new class of outbursts from relatively low-mass massive stars,
potentially analogous to the pulsational instabilities discussed in 
 Poelarends et al.~(2008) or Heger et al.~(1997).
The majority of the true ``LBV'' eruptions with documented progenitors
(e.g., 1997bs, 2002kg) came from optically bright massive stars significantly
more bolometrically luminous than the progenitors of SN 2008S and NGC 300.
  As we have shown in Figures
\ref{fig:cmd2} and \ref{fig:sed},
the EAGB population is separate from the sources traditionally classified
as LBVs: they are less bolometrically luminous and much more dust-obscured.  
These facts suggest that if these transients
were the outbursts of massive stars then they are distinct from 
from the classical supernova impostors.
If these events are not supernovae, but merely outbursts, then their
existence is likely connected to the physics of the transition between 
stars that become ecSNe and/or ccSNe, and those that do not.
The degree of dust-obscuration at outburst is a crucial 
clue to their evolution.   A simple prediction 
of this possibility is that the progenitors of SN 2008S and NGC 300 
should eventually be re-discovered in the optical and/or infrared 
after the outburst emission has faded.  For further arguments on the 
nature of SN 2008S and NGC 300 related to this discussion, see
Bond et al.~(2008), Smith et al.~(2009), and Berger et al.~(2008).

\subsection{A More Complete Census}
\label{section:survey}

Equation (\ref{feagb}) implies that a fraction $\sim1\times10^{-4}$ 
of the massive star population 
in any given galaxy appears to be in the evolutionary state that led to 
the explosions observed as SN 2008S and NGC 300.
The simplest explanation, adopted throughout this work, is that 
the deeply dust-enshrouded phase marks the last $t_{\rm EAGB}\lesssim10^4$\,yr
(eq.~\ref{teagb}) in the life of a fraction $f_{\rm SN}\approx0.2$ of the massive 
star population. 

Compilations of star formation and supernova rates in the local
universe (e.g., Ando et al.~2005) suggest that the latter is 
$\approx2$\,yr$^{-1}$ within 10\,Mpc 
(see \S\ref{section:introduction}, point 2), implying that there
are $\sim5\times10^6$ massive stars and $\lesssim10^3$ 
EAGB stars within this volume.  If the lifetime in the
pre-explosion, highly dust-obscured phase is $t_{\rm EAGB}$ (eq.~\ref{teagb}),
we would expect to see one SN 2008S-like transient every few 
years, in accord with our estimate for $f_{\rm SN}$. 

A multi-epoch survey of all the local star-forming galaxies 
within 10\,Mpc with {\it Spitzer} would allow for a comprehensive 
census of EAGB stars.  It would significantly increase our knowledge 
of the variability properties and SED evolution of these objects, and
it might allow us to define more strict criteria for inclusion in the
class of SN 2008S/NGC 300-like progenitors.  It would therefore 
decrease the considerable uncertainty in $N_{\rm EAGB}$ in equations
(\ref{feagb}) and (\ref{teagb}).  Coupled with the supernova surveys
in the local volume, such a study would improve our knowledge of the 
fraction $f_{\rm SN}$ of stars that eventually go through the deeply embedded
phase just before explosion.  Of course, the most intriguing 
possibility is that the number of true analogs to the progenitors
of SN 2008S and NGC 300 is in fact $N_{\rm EAGB}\sim0-1$ in M33 and that 
$t_{\rm EAGB}\lesssim {\rm few}-10^3$\,yr.\footnote{The lower limit here comes
from the fact that pre-explosion imaging of the SN 2008S and NGC 300
progenitors establishes a few year baseline.} 
If so, the final catalog of 
EAGB stars that would be produced by a {\it Spitzer} survey 
would have just $\sim50-100$ members.  These could be followed up repeatedly,
since, given these numbers one would expect to wait just $\sim10$
years before one of these individual sources exploded. 
This would give a direct observational
link in the causal mapping between a sub-population of
massive stellar progenitors and their explosions,
connecting them with a short timescale. Indeed, the ability 
to identify an individual star as marked for imminent 
death (or eruption) would be an astonishing consequence of this work.

\acknowledgments

We thank B.~Monard, H.~Y\"uksel, A.~Gal-Yam, N.~Smith, R.~Humphreys, J.~J.~Eldridge, 
K.~Sellgren, and A.~Gould for helpful conversations, and the anonymous referee, whose
comments materially improved the manuscript.
We extend our gratitude to the SINGS Legacy Survey for making their data 
publicly available. 
This work is based in part on archival data obtained with the SST, which is 
operated by the JPL, Caltech under a contract with NASA. This research has 
made use of NED, which is operated by the JPL and Caltech, under contract 
with NASA and the HEASARC Online Service, provided by NASA's GSFC. 
T.A.T.~gratefully acknowledges 
the Aspen Center for Physics where a portion of this work was completed
and an Alfred P.~Sloan Fellowship for financial support.  
J.L.P.~and K.Z.S.~are supported by NSF grant AST-0707982. M.D.K.~and
J.F.B.~are supported by NSF CAREER grant PHY-0547102.



\begin{table}
\begin{scriptsize}
\begin{center}
\caption{MIR Catalog for 53,194 Point Sources in M33
\label{table:m33}}
\begin{tabular}{lccccccccc}
\hline \hline
\\
\multicolumn{1}{c}{RA} &
\multicolumn{1}{c}{Dec} &
\multicolumn{1}{c}{$[3.6]$} &
\multicolumn{1}{c}{$\sigma_{3.6}$} &
\multicolumn{1}{c}{$[4.5]$} &
\multicolumn{1}{c}{$\sigma_{4.5}$} \\
\multicolumn{1}{c}{(deg)} &
\multicolumn{1}{c}{(deg)} &
\multicolumn{1}{c}{(mag)} &
\multicolumn{1}{c}{} &
\multicolumn{1}{c}{(mag)} &
\multicolumn{1}{c}{} \\
\\
\hline
\hline
\\
23.03815 & 30.82294  &  18.70 &  0.21  &  18.19 &  0.12  \\
23.04489 & 30.82397  &  17.67 &  0.07  &  18.24 &  0.10  \\
23.04532 & 30.81396  &  18.15 &  0.14  &  18.59 &  0.15  \\
23.04906 & 30.81863  &  15.69 &  0.05  &  16.19 &  0.07  \\
23.04955 & 30.82077  &  17.52 &  0.08  &  17.82 &  0.13  \\
23.05103 & 30.82549  &  16.79 &  0.06  &  17.13 &  0.07  \\
23.05131 & 30.81658  &  17.96 &  0.09  &  18.33 &  0.12  \\
23.05277 & 30.79529  &  16.50 &  0.07  &  17.13 &  0.07  \\
23.05287 & 30.80321  &  17.45 &  0.06  &  17.94 &  0.09  \\
23.05303 & 30.82159  &  16.96 &  0.05  &  17.38 &  0.07  \\
23.05314 & 30.78955  &  18.73 &  0.11  &  18.41 &  0.19  \\
23.05333 & 30.81954  &  18.14 &  0.07  &  18.13 &  0.11  \\
23.05470 & 30.79727  &  17.57 &  0.06  &  18.04 &  0.12  \\
23.05473 & 30.81597  &  17.84 &  0.07  &  18.29 &  0.10  \\
23.05481 & 30.80367  &  17.86 &  0.06  &  18.14 &  0.10  \\
\dots    &  \dots    & \dots & \dots   & \dots & \dots \\
\\
\hline
\hline
\end{tabular}
\end{center}
\end{scriptsize}
\end{table}

\begin{table}
\begin{scriptsize}
\begin{center}
\caption{Photometry for the 18 EAGBs in M33
\label{table:eagb}}
\begin{tabular}{lcccccccccccccccccccccc}
\hline \hline
\\
\multicolumn{1}{c}{Name} &
\multicolumn{1}{c}{RA} &
\multicolumn{1}{c}{Dec} &
\multicolumn{1}{c}{$U$\tablenotemark{a}} &
\multicolumn{1}{c}{$B$\tablenotemark{a}} &
\multicolumn{1}{c}{$V$\tablenotemark{a}} &
\multicolumn{1}{c}{$R$\tablenotemark{a}} &
\multicolumn{1}{c}{$I$\tablenotemark{a}} &
\multicolumn{1}{c}{$[3.6]$} &
\multicolumn{1}{c}{$[4.5]$} &
\multicolumn{1}{c}{$[5.8]$} &
\multicolumn{1}{c}{$[8.0]$} \\
\multicolumn{1}{c}{} &
\multicolumn{1}{c}{(deg)} &
\multicolumn{1}{c}{(deg)} &
\multicolumn{1}{c}{(mag)} &
\multicolumn{1}{c}{(mag)} &
\multicolumn{1}{c}{(mag)} &
\multicolumn{1}{c}{(mag)} &
\multicolumn{1}{c}{(mag)} &
\multicolumn{1}{c}{(mag)} &
\multicolumn{1}{c}{(mag)} &
\multicolumn{1}{c}{(mag)} &
\multicolumn{1}{c}{(mag)} \\
\\
\hline
\hline
\\
S1 &  23.45485 &  30.85704 & 23.47 & 23.76 & 23.37 &  22.90 & 22.47 & 16.35 & 14.21 & 12.55 & 11.37 \\
S2 &  23.44397 &  30.79731 & 23.96 & 24.15 & 23.55 &  23.15 & 22.47 & 16.84 & 14.85 & 13.52 & 12.42 \\
S3 &  23.56813 &  30.87755 & 23.91 & 23.97 & 23.48 &  23.10 & 22.50 & 16.56 & 14.73 & 13.42 & 12.55 \\
S4 &  23.43452 &  30.57106 & 23.87 & 23.74 & 23.21 &  22.71 & 22.02 & 15.87 & 14.15 & 13.23 & 12.07 \\
S5 &  23.40436 &  30.51738 & 22.99 & 23.10 & 22.97 &  22.71 & 22.10 & 15.08 & 13.38 & 12.12 & 11.32 \\
S6 &  23.49194 &  30.82791 & 24.10 & 23.84 & 23.42 &  22.89 & 22.09 & 16.51 & 14.84 & 13.46 & 12.32 \\
S7 &  23.55640 &  30.55211 & 24.52 & 24.50 & 23.86 &  23.31 & 22.75 & 14.29 & 12.63 & 11.18 & 10.17 \\
S8 &  23.46408 &  30.64138 & 22.79\tablenotemark{b} & 23.11\tablenotemark{b} & 22.32\tablenotemark{b} &  21.75\tablenotemark{b} 
& 20.96\tablenotemark{b} & 16.17 & 14.52 & 14.14 & 12.87 \\
S9 &  23.53817 &  30.73269 & 21.75\tablenotemark{b} & 21.93\tablenotemark{b} & 22.14\tablenotemark{b} &  22.10\tablenotemark{b} 
& 21.89\tablenotemark{b} & 15.67 & 14.08 & 12.69 & 11.51 \\
S10 &  23.29877 &  30.59901 & 24.31 & 23.97 & 23.67 &  23.30 & 22.75 & 16.35 & 14.78 & 13.55 & 12.37 \\
S11 &  23.29777 &  30.50744 & 23.82 & 23.64 & 23.53 &  23.05 & 22.83 & 15.30 & 13.72 & 12.47 & 11.26 \\
S12 &  23.55236 &  30.90564 & 24.20 & 23.96 & 23.50 &  23.21 & 22.62 & 14.90 & 13.33 & 12.01 & 11.08 \\
S13 &  23.37907 &  30.70096 & 24.05 & 24.14 & 23.60 &  23.03 & 22.41 & 16.38 & 14.82 & 13.44 & 12.36 \\
S14 &  23.26238 &  30.34469 & 23.96 & 23.95 & 23.48 &  23.03 & 22.47 & 16.33 & 14.79 & 13.61 & 12.43 \\
S15 &  23.39709 &  30.67737 & 24.13 & 24.02 & 23.29 &  22.80 & 22.23 & 15.25 & 13.71 & 12.22 & 10.97 \\
S16 &  23.43722 &  30.64242 & 23.75 & 23.39 & 22.79 &  21.99 & 21.39 & 16.30 & 14.76 & 14.15 & 12.76 \\
S17 &  23.47176 &  30.67430 & 23.81 & 23.67 & 23.03 &  22.51 & 21.60 & 15.27 & 13.74 & 13.15 & 11.56 \\
S18 &  23.34248 &  30.64602 & 24.21 & 23.95 & 23.65 &  23.22 & 22.63 & 14.86 & 13.32 & 12.03 & 10.88 \\
\\
\hline
\hline
\end{tabular}
\tablenotetext{a}{Except where otherwise noted, all $UBVRI$ data in this table are upper limits.}
\tablenotetext{b}{Source detections. Magnitudes from M06.}
\end{center}
\end{scriptsize}
\end{table}

\begin{table}
\begin{scriptsize}
\begin{center}
\caption{Photometry for the 16 LBV Candidates in M33 from M07
\label{table:lbv}}
\begin{tabular}{lcccccccccccccccccccccc}
\hline \hline
\\
\multicolumn{1}{c}{Identifier} &
\multicolumn{1}{c}{$U$} &
\multicolumn{1}{c}{$B$} &
\multicolumn{1}{c}{$V$} &
\multicolumn{1}{c}{$R$} &
\multicolumn{1}{c}{$I$} &
\multicolumn{1}{c}{$[3.6]$} &
\multicolumn{1}{c}{$[4.5]$} &
\multicolumn{1}{c}{$[5.8]$} &
\multicolumn{1}{c}{$[8.0]$} \\
\multicolumn{1}{c}{(from M07)} &
\multicolumn{1}{c}{(mag)} &
\multicolumn{1}{c}{(mag)} &
\multicolumn{1}{c}{(mag)} &
\multicolumn{1}{c}{(mag)} &
\multicolumn{1}{c}{(mag)} &
\multicolumn{1}{c}{(mag)} &
\multicolumn{1}{c}{(mag)} &
\multicolumn{1}{c}{(mag)} &
\multicolumn{1}{c}{(mag)} \\
\\
\hline
\hline
\\
J013248.26+303950.4 & 16.13 &17.32 &17.25 &17.01 &16.89 &15.70 &15.36 &15.02 &14.57 \\
J013324.62+302328.4 & 18.66 &19.49 &19.58 &19.27 &19.53 &14.26 &13.55 &13.04 &12.27 \\
J013333.22+303343.4 & 18.30 &19.30 &19.40 &18.44 &18.89 &13.09 &12.50 &12.20 &11.65 \\
J013335.14+303600.4 & 15.54 &16.53 &16.43 &16.30 &16.14 &14.45 &14.35 &14.48 &14.15 \\
J013341.28+302237.2 & 15.18 &16.24 &16.29 &16.28 &16.32 &15.08 &14.96 &15.07 &14.45 \\
J013350.12+304126.6 & 15.76 &16.85 &16.82 &16.43 &16.30 &12.22 &11.65 &11.06 &10.64 \\
J013406.63+304147.8 & 15.12 &16.26 &16.08 &15.86 &15.76 &14.41 &13.95 &13.83 &\nodata\tablenotemark{a} \\
J013410.93+303437.6 & 15.13 &16.13 &16.03 &15.87 &15.70 &14.84 &14.65 &14.69 &13.97 \\
J013416.10+303344.9 & 16.32 &17.17 &17.12 &16.96 &16.85 &15.69 &15.48 &15.01 &\nodata\tablenotemark{a} \\
J013422.91+304411.0 & 16.36 &17.28 &17.21 &17.14 &17.07 &16.24 &16.35 &\nodata\tablenotemark{a} &\nodata\tablenotemark{a} \\
J013424.78+303306.6 & 16.21 &16.97 &16.84 &16.72 &16.54 &15.24 &15.00 &14.74 &13.97 \\
J013426.11+303424.7 & 18.38 &19.23 &18.97 &18.59 &18.27 &14.57 &13.79 &13.09 &12.24 \\
J013429.64+303732.1 & 16.24 &17.12 &17.11 &17.05 &16.93 &16.55 &16.64 &\nodata\tablenotemark{a} &\nodata\tablenotemark{a} \\
J013442.14+303216.0 & 18.35 &18.20 &17.34 &16.89 &16.44 &13.56 &13.02 &12.39 &11.62 \\
J013459.47+303701.9 & 17.35 &18.59 &18.37 &17.88 &17.69 &14.42 &13.73 &13.16 &12.31 \\
J013500.30+304150.9 & 18.37 &19.22 &19.30 &18.60 &19.11 &13.93 &13.24 &12.54 &11.93 \\
\\
\hline
\hline
\end{tabular}
\tablenotetext{a}{Non-detection.}
\end{center}
\end{scriptsize}
\end{table}

\pagebreak

\begin{appendix}

\section{A.~Extreme-AGB Star Variability}
\label{appendix:agb}

\begin{figure*}[b]
\begin{center}
\centerline{
\includegraphics[width=5.8cm]{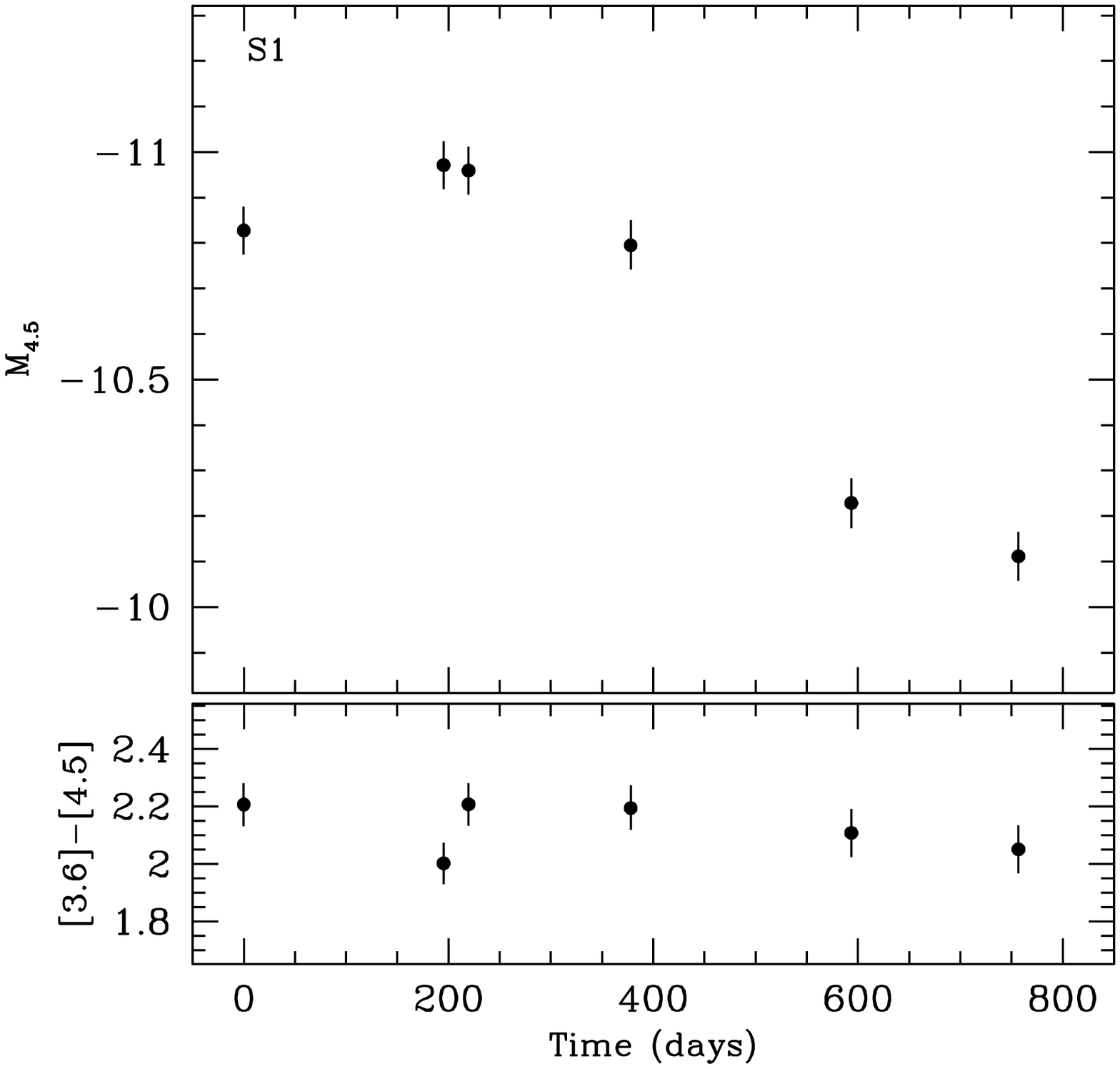} 
\includegraphics[width=5.8cm]{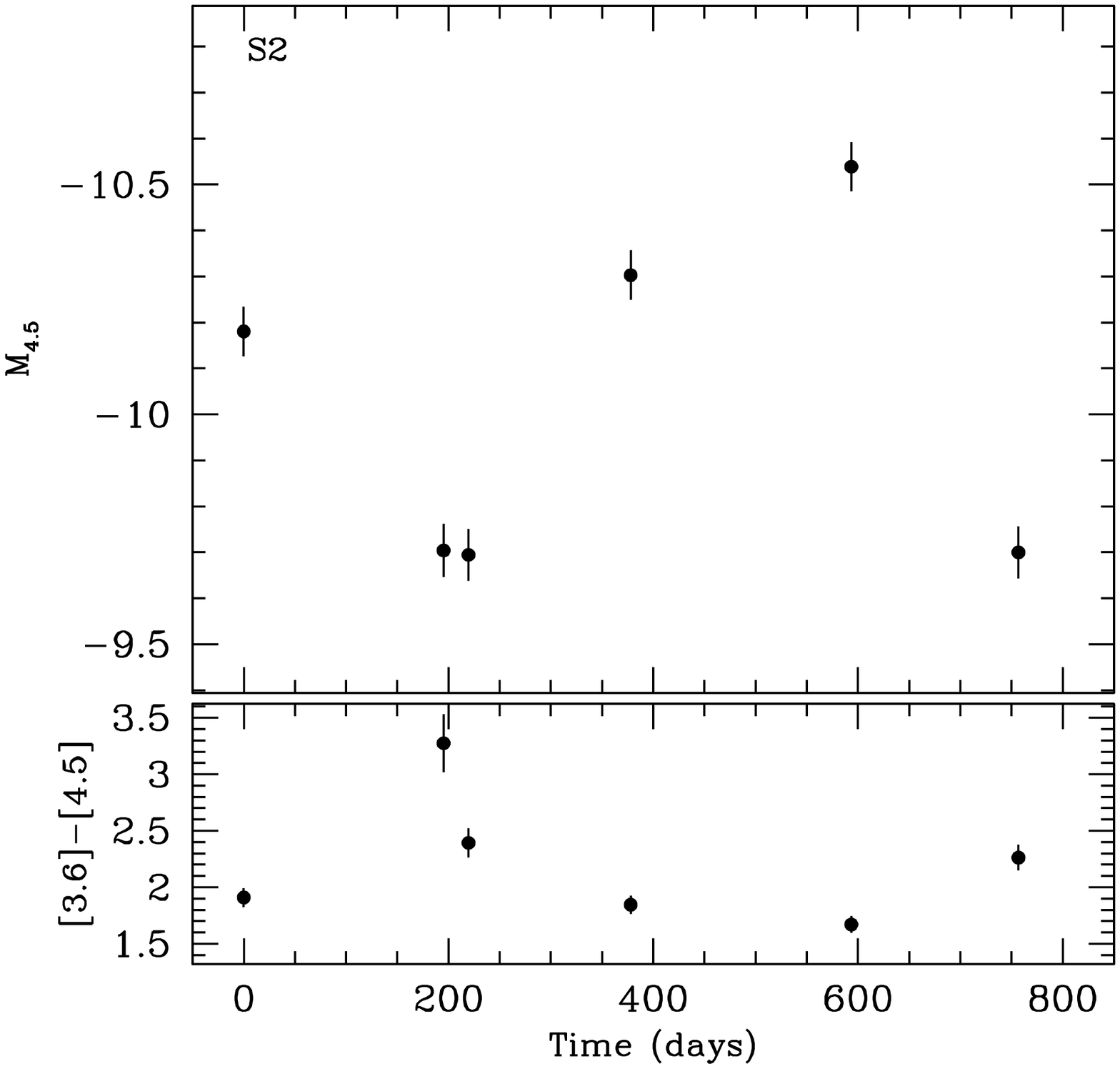}
\includegraphics[width=5.8cm]{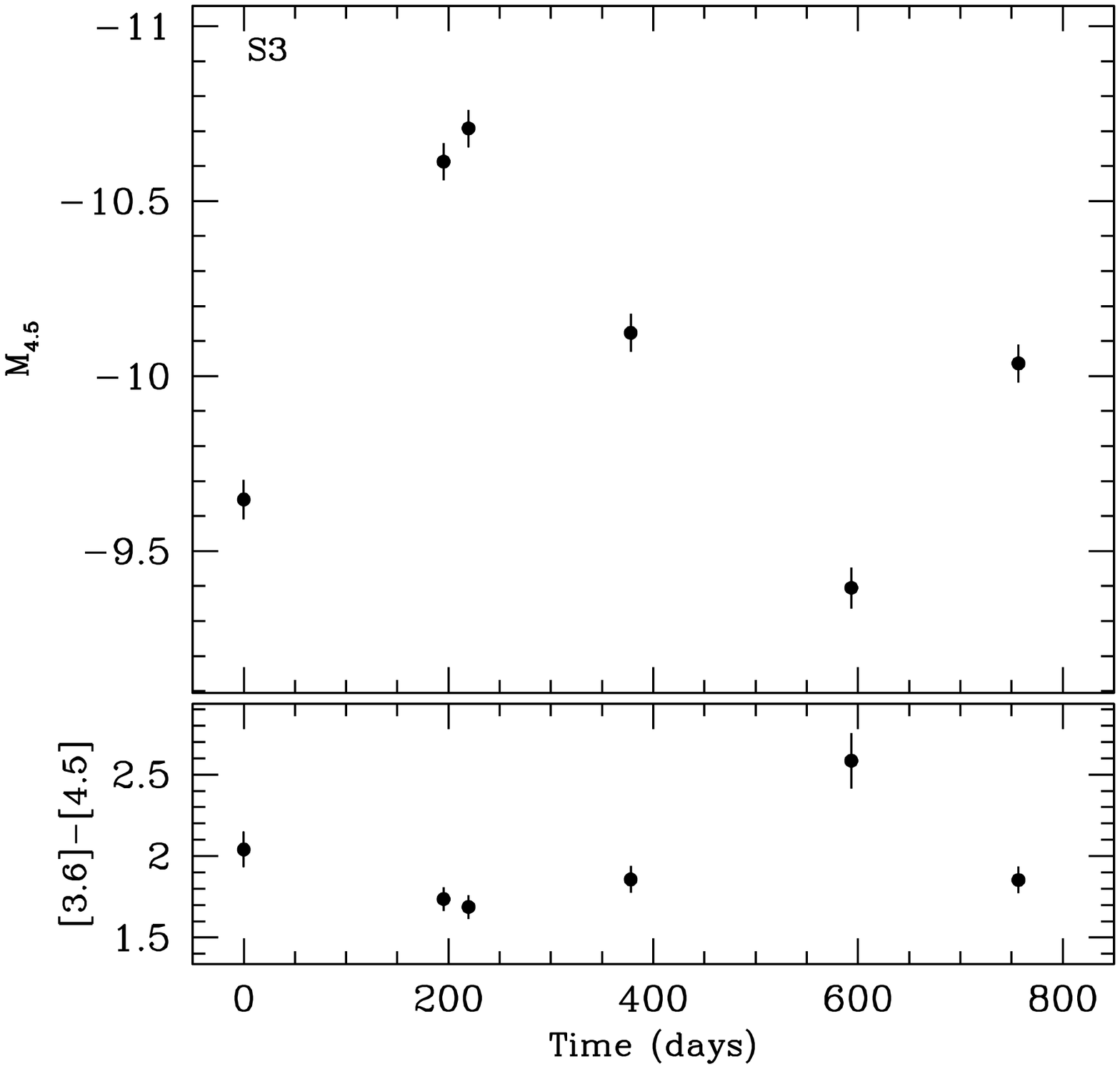}}
\centerline{
\includegraphics[width=5.8cm]{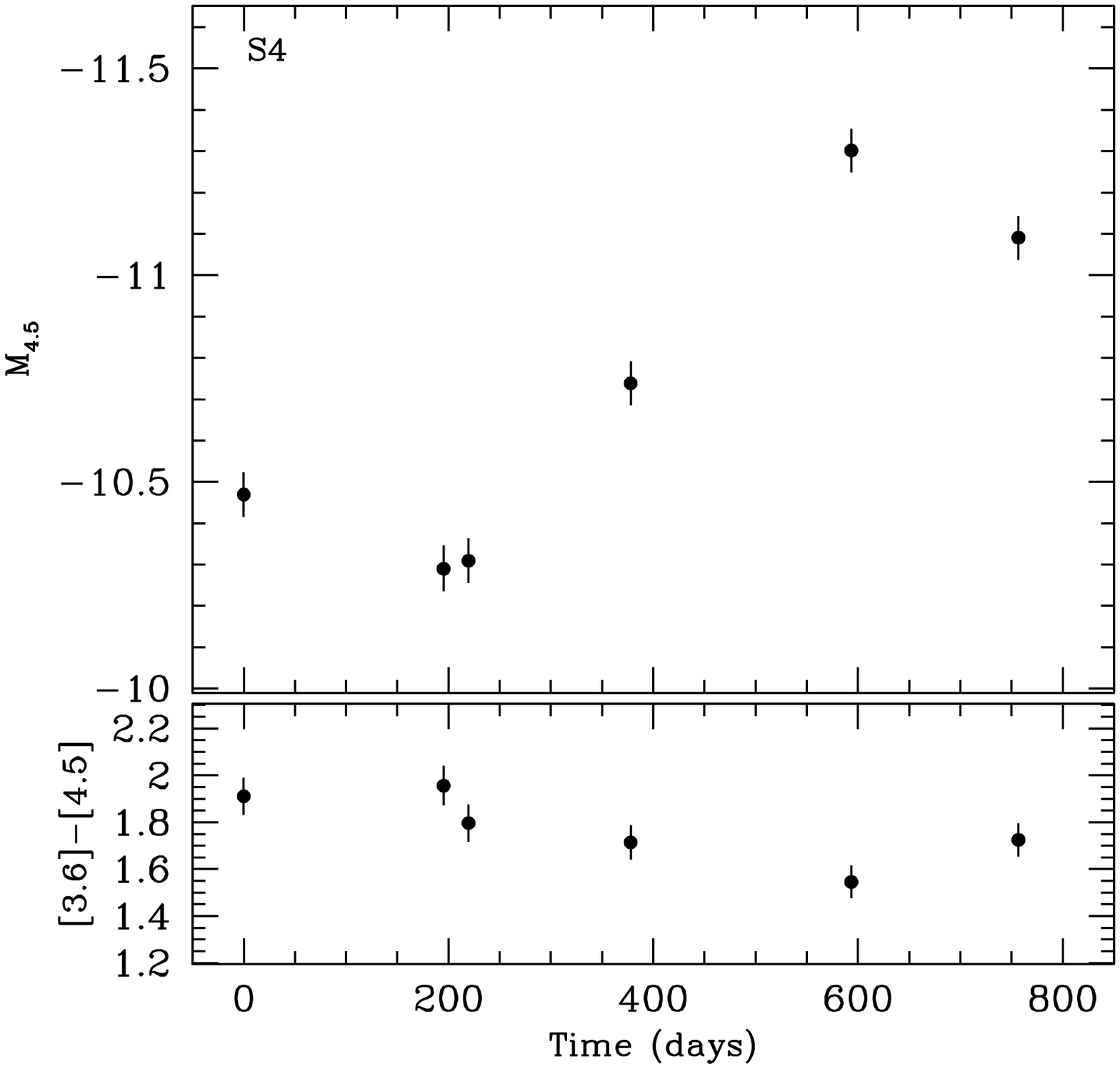} 
\includegraphics[width=5.8cm]{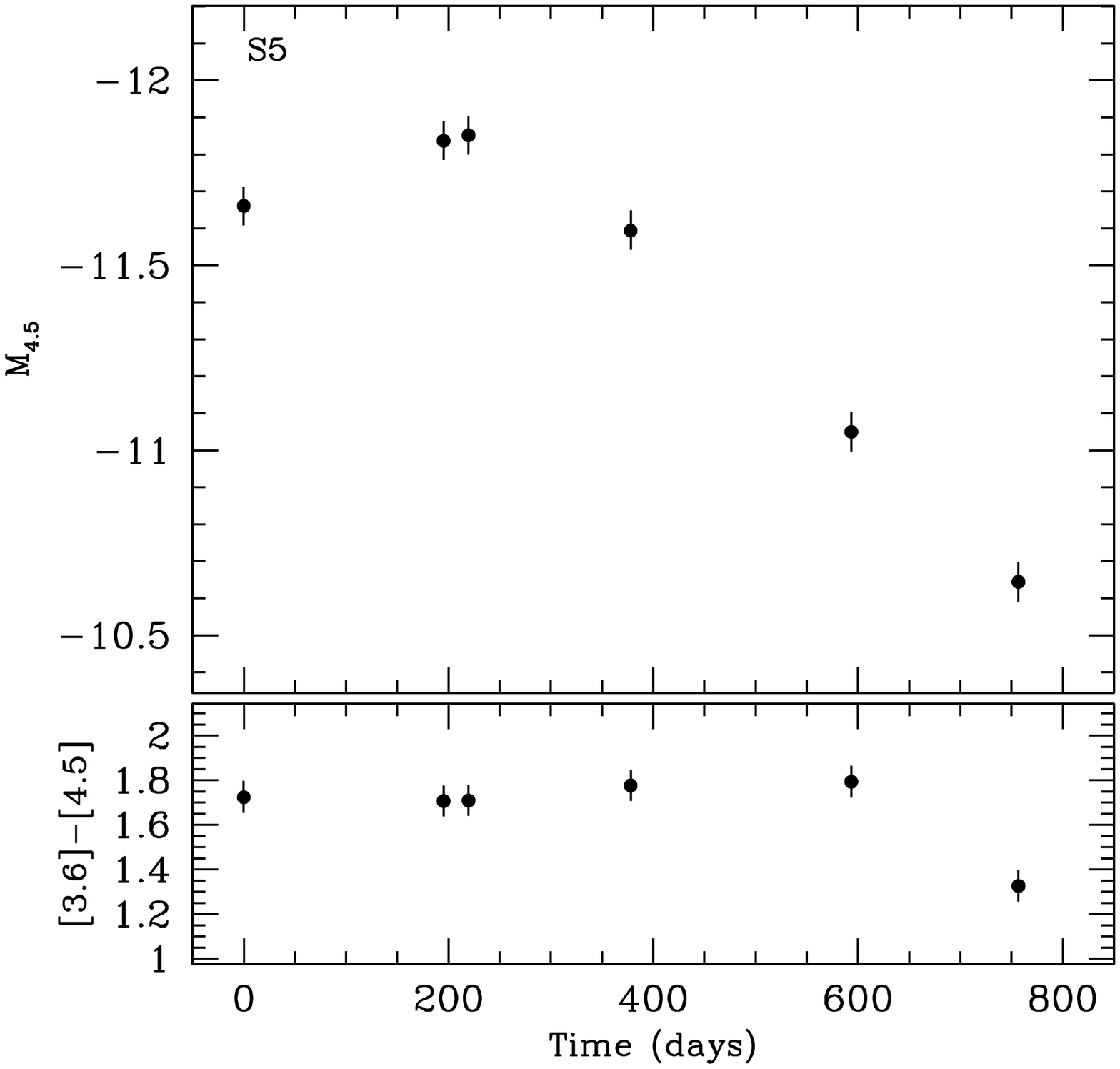}
\includegraphics[width=5.8cm]{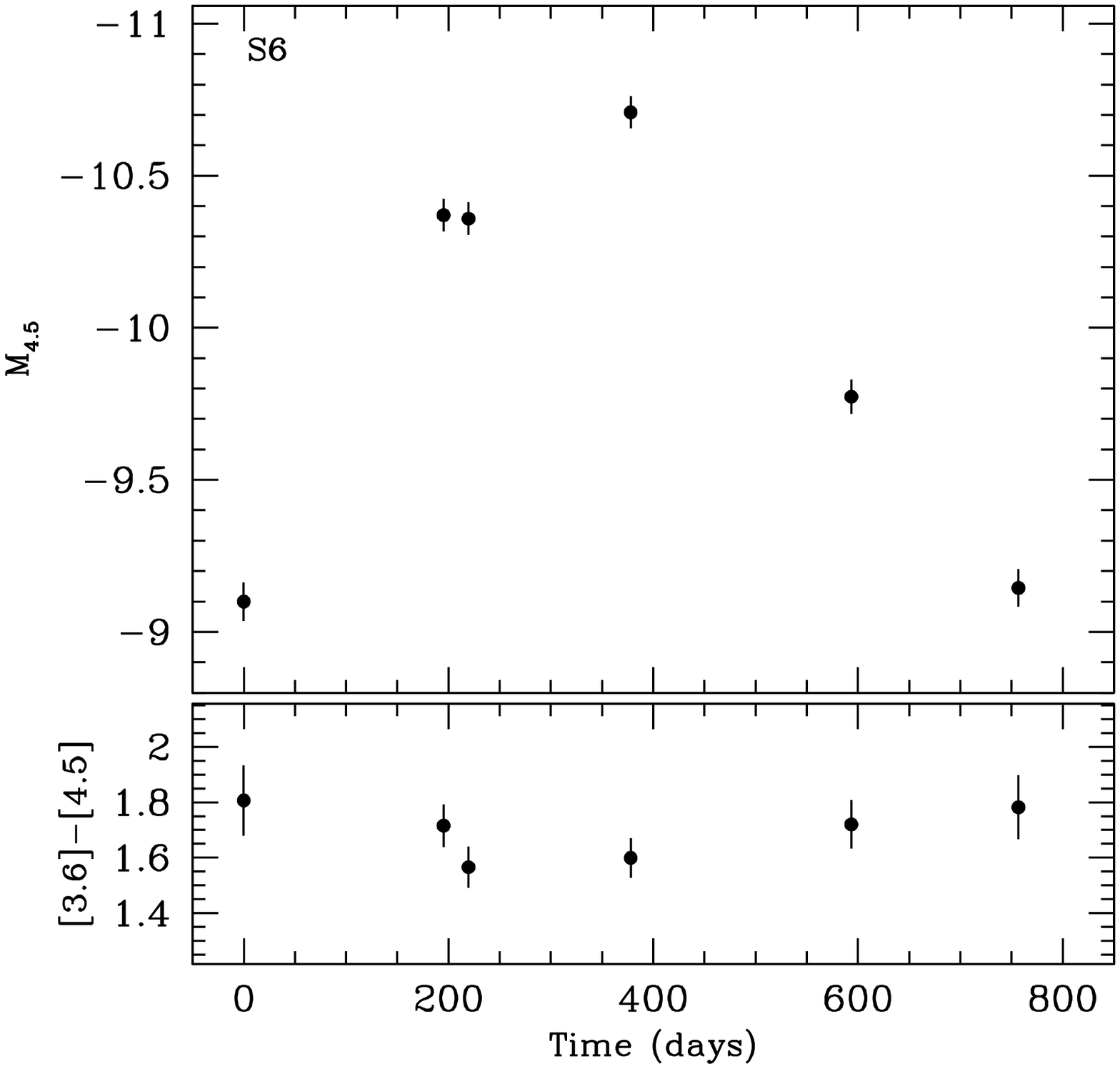}}
\centerline{
\includegraphics[width=5.8cm]{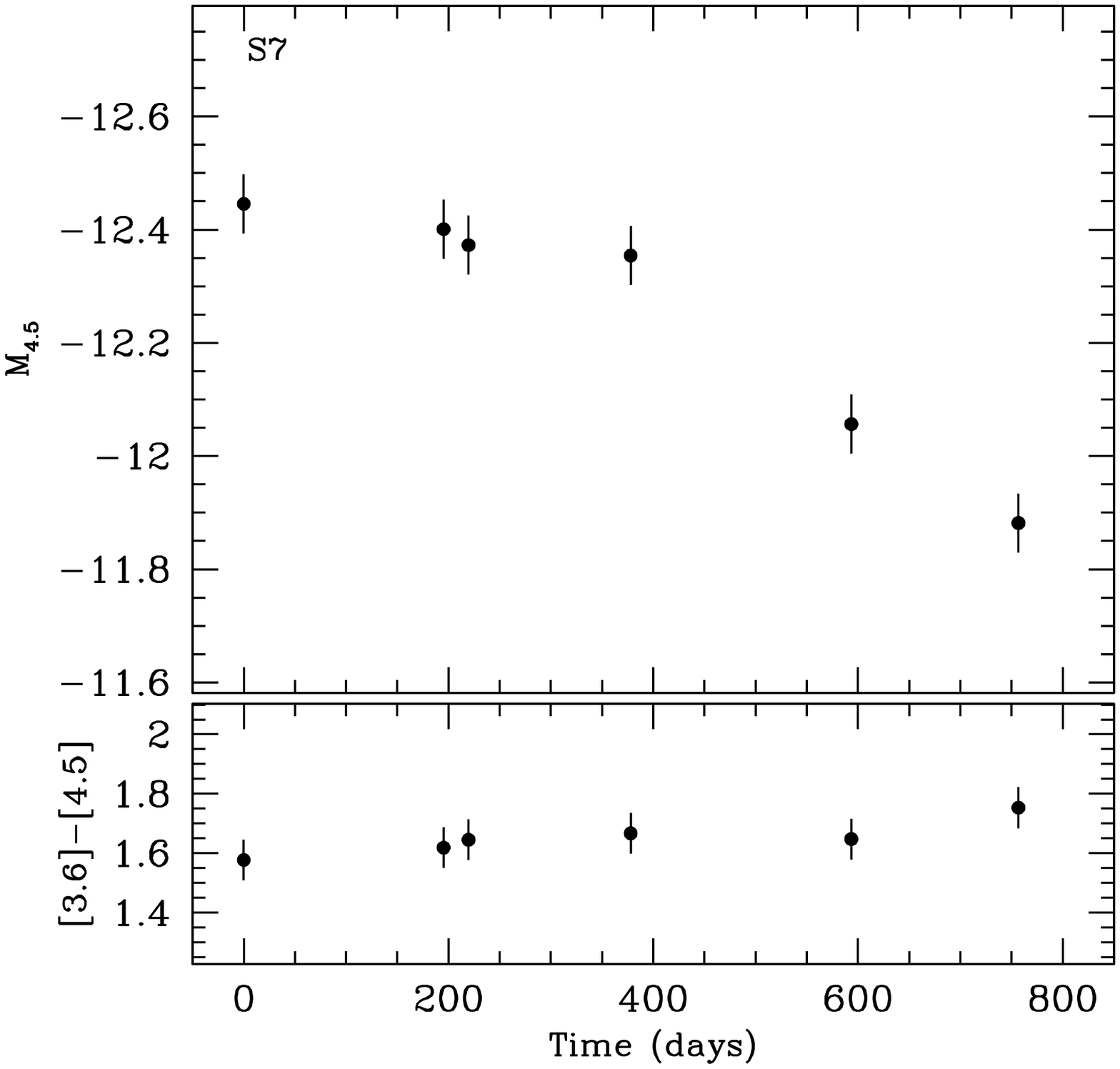} 
\includegraphics[width=5.8cm]{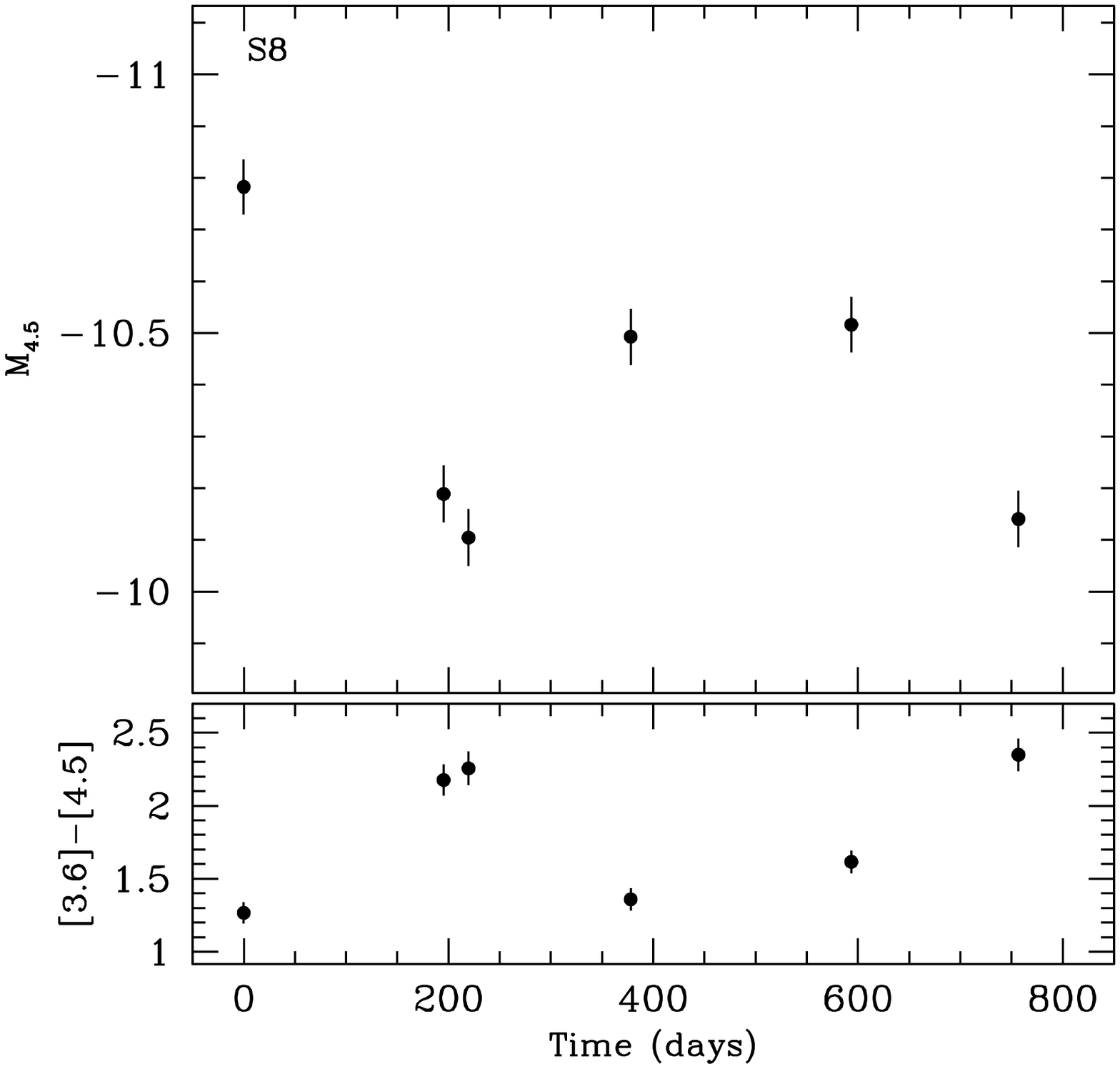}
\includegraphics[width=5.8cm]{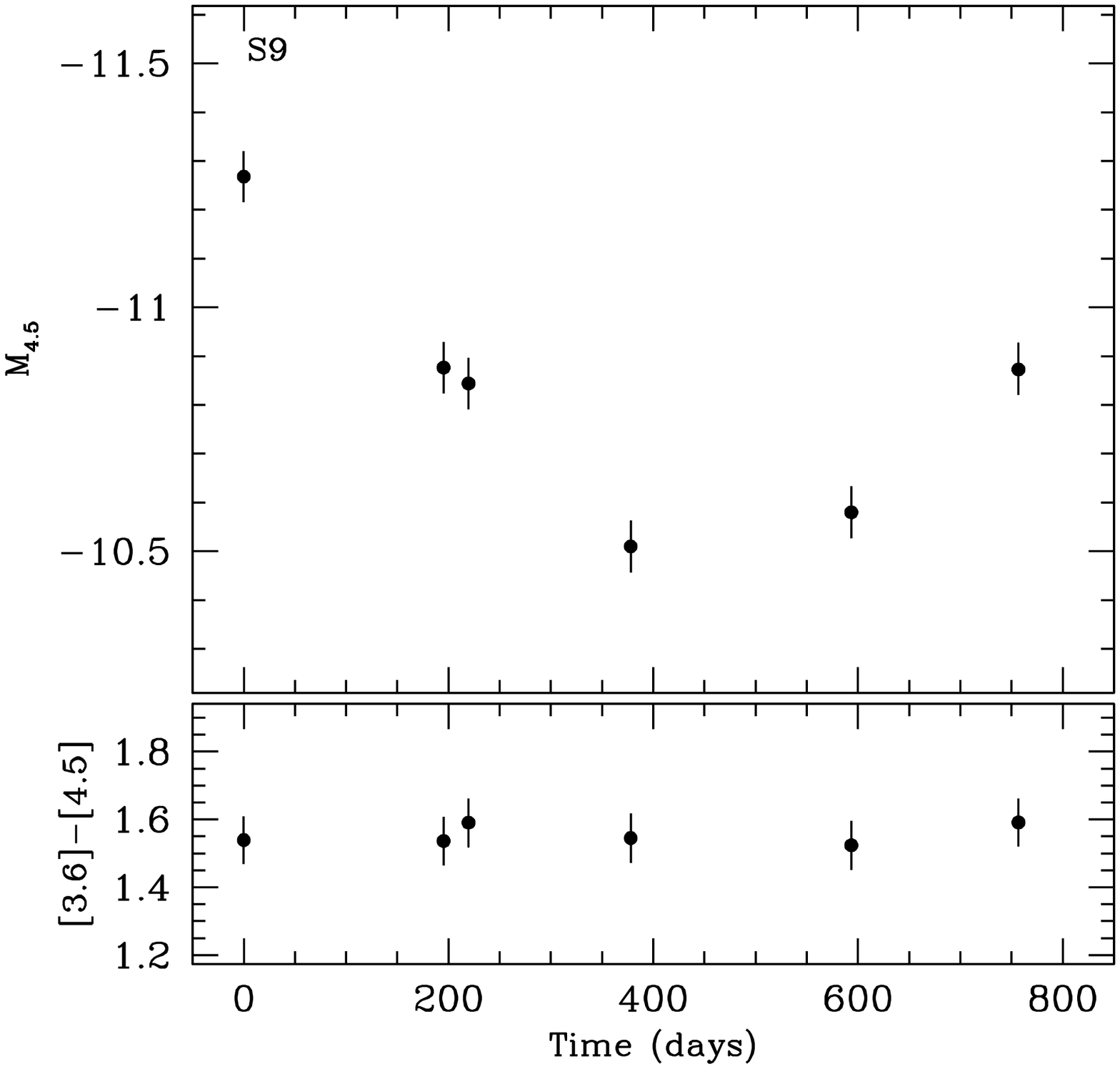}}
\caption{Lightcurves for sources S1$-$S9 of the 18 reddest
sources (sorted by color) with $M_{4.5}<-10$ and
$[3.6]-[4.5]>1.5$ in the M33 MIR color-magnitude diagram 
(open circles in Fig.~\ref{fig:cmd2}; remaining
source lightcurves are shown in Fig.~\ref{fig:agb2}).
For each source, the top and bottom panels show the absolute 4.5$\mu$m magnitude 
and the $[3.6]-[4.5]$ color variation, respectively, as a function of time.
More than a magnitude variation on a timescale of $100-1000$ days is
common.  Local minima in 4.5$\mu$m luminosity generally correspond
to local maxima in color, although there are deviations from this 
trend.  Colors can vary by more than a magnitude, with the largest excursions
reaching  $[3.6]-[4.5]\approx2.5$.  In just one source (S2), and in 
at just one epoch, we find a $[3.6]-[4.5]$ color as large as
the progenitor of NGC 300 ($\approx2.7$; Figs.~\ref{fig:cmd} \& \ref{fig:cmd2}).
}
\label{fig:agb1}
\end{center}
\end{figure*}

\begin{figure*}[t]
\begin{center}
\centerline{
\includegraphics[width=5.8cm]{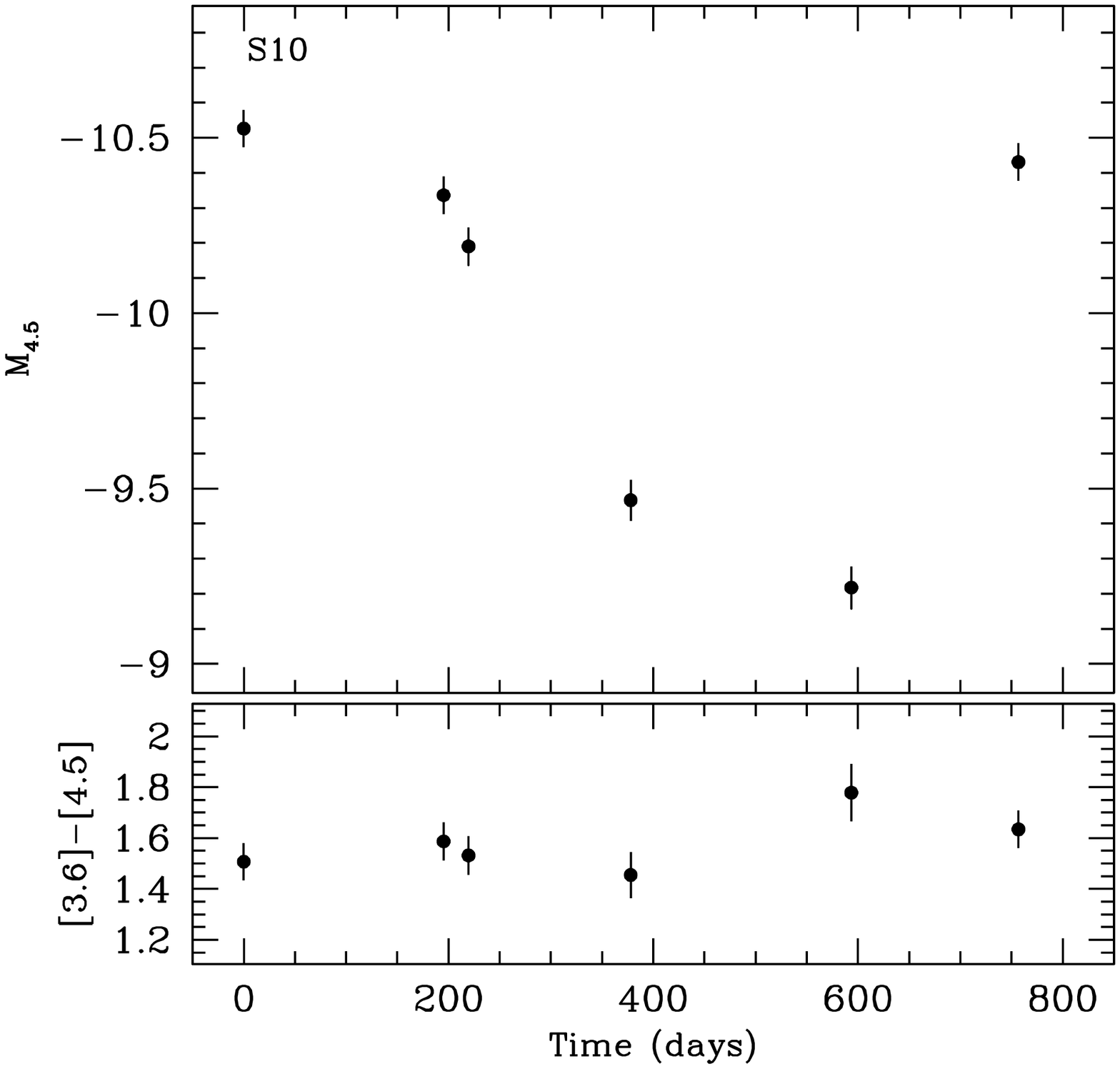} 
\includegraphics[width=5.8cm]{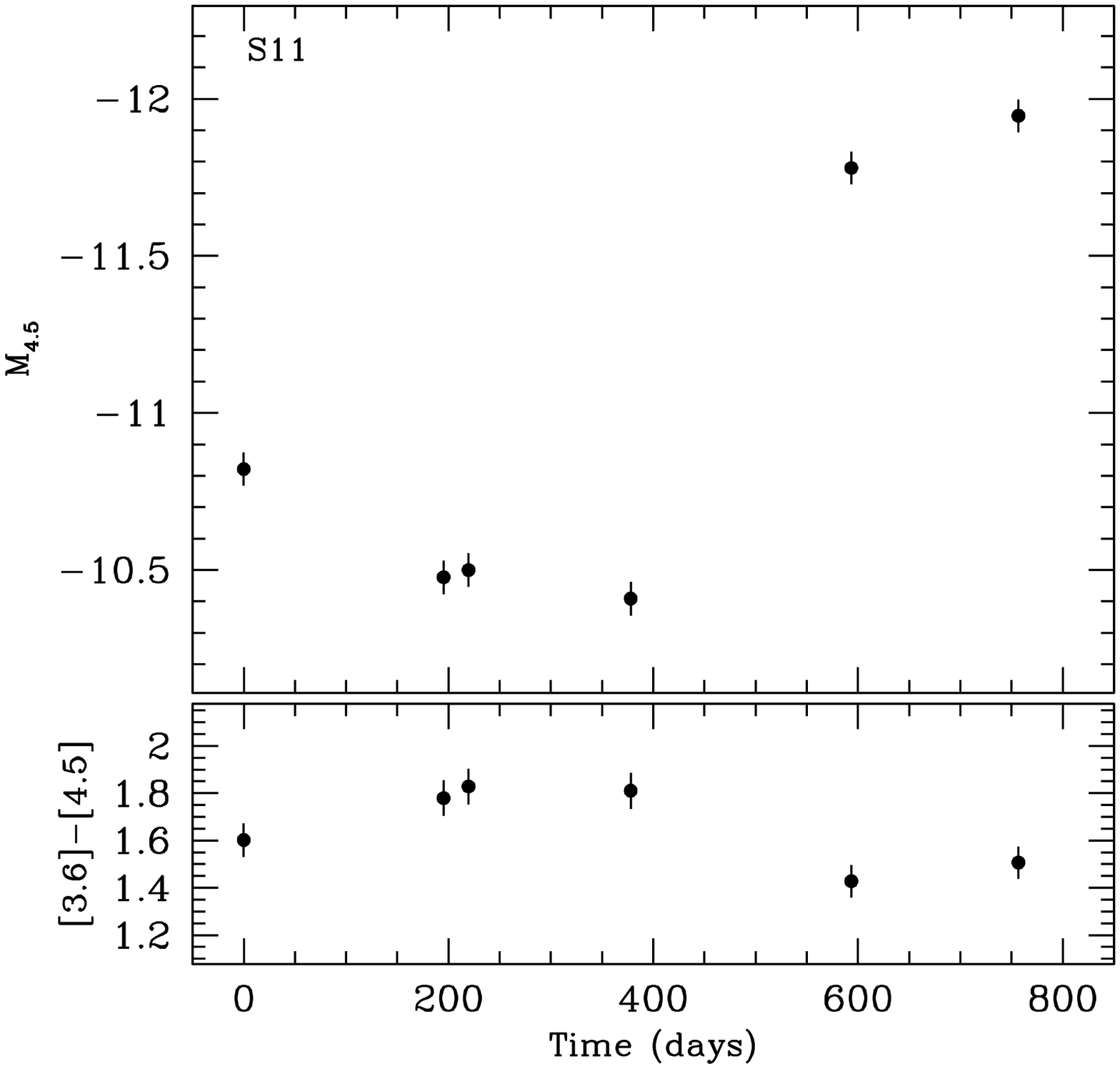}
\includegraphics[width=5.8cm]{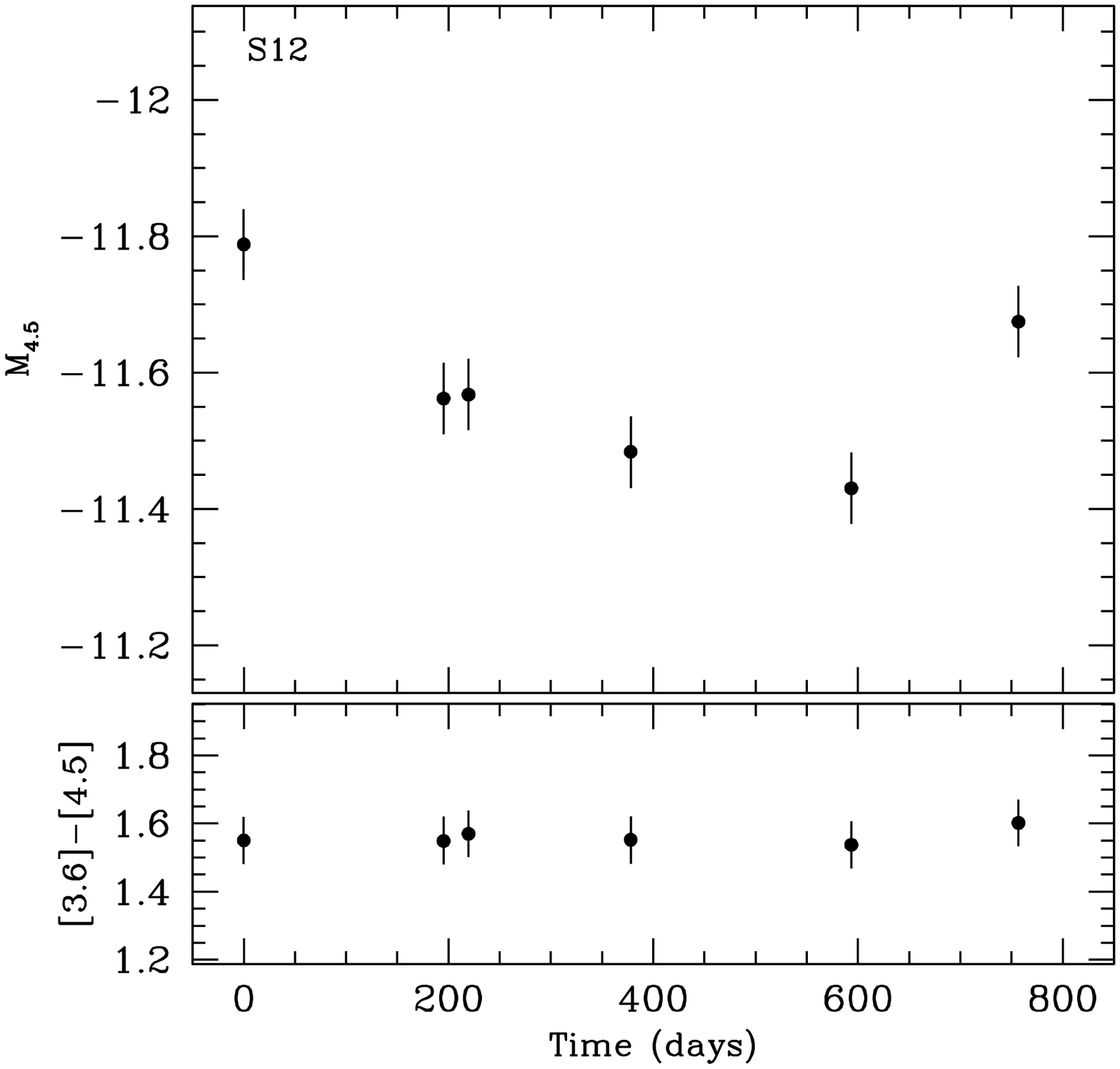}}
\centerline{
\includegraphics[width=5.8cm]{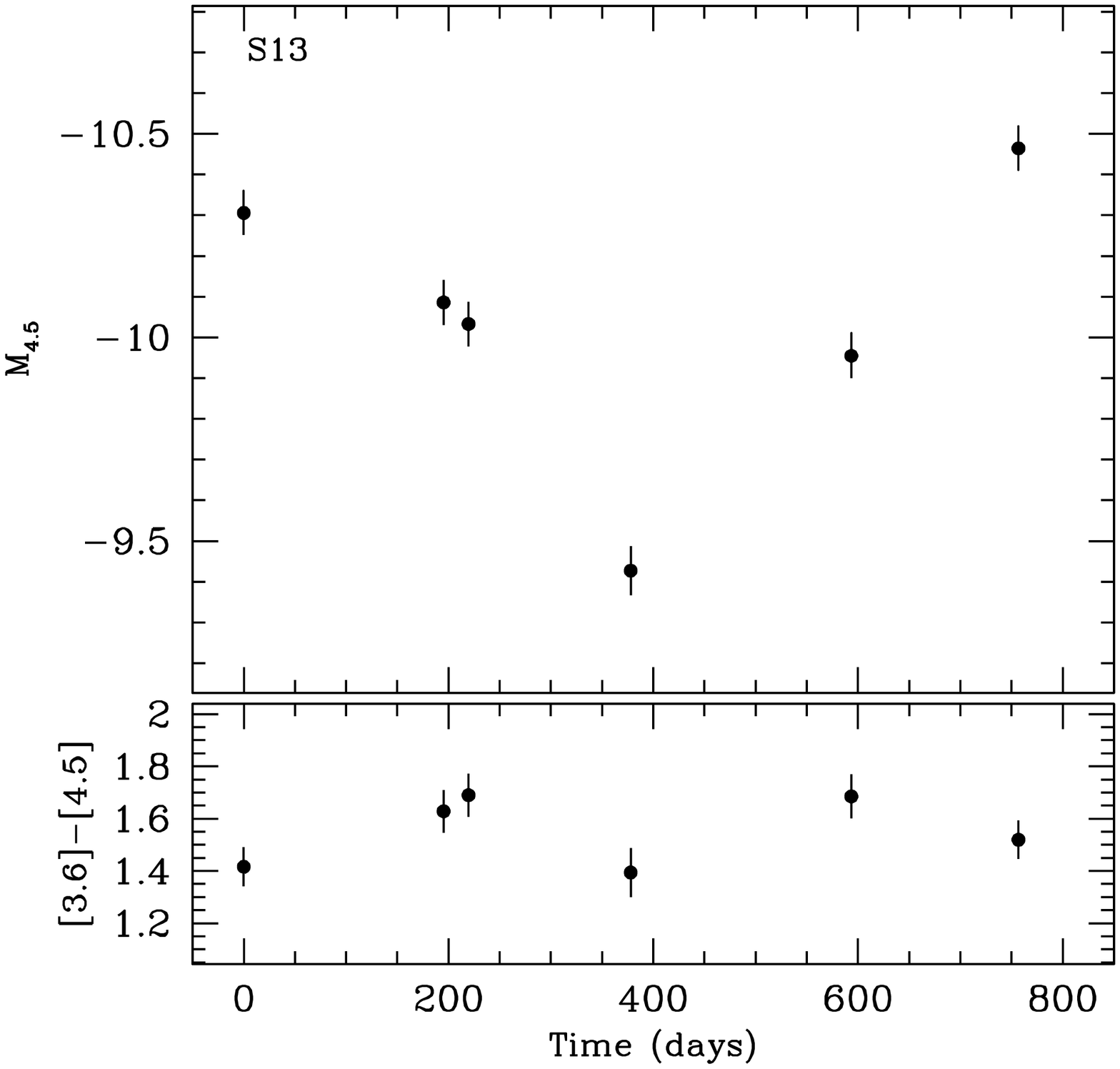} 
\includegraphics[width=5.8cm]{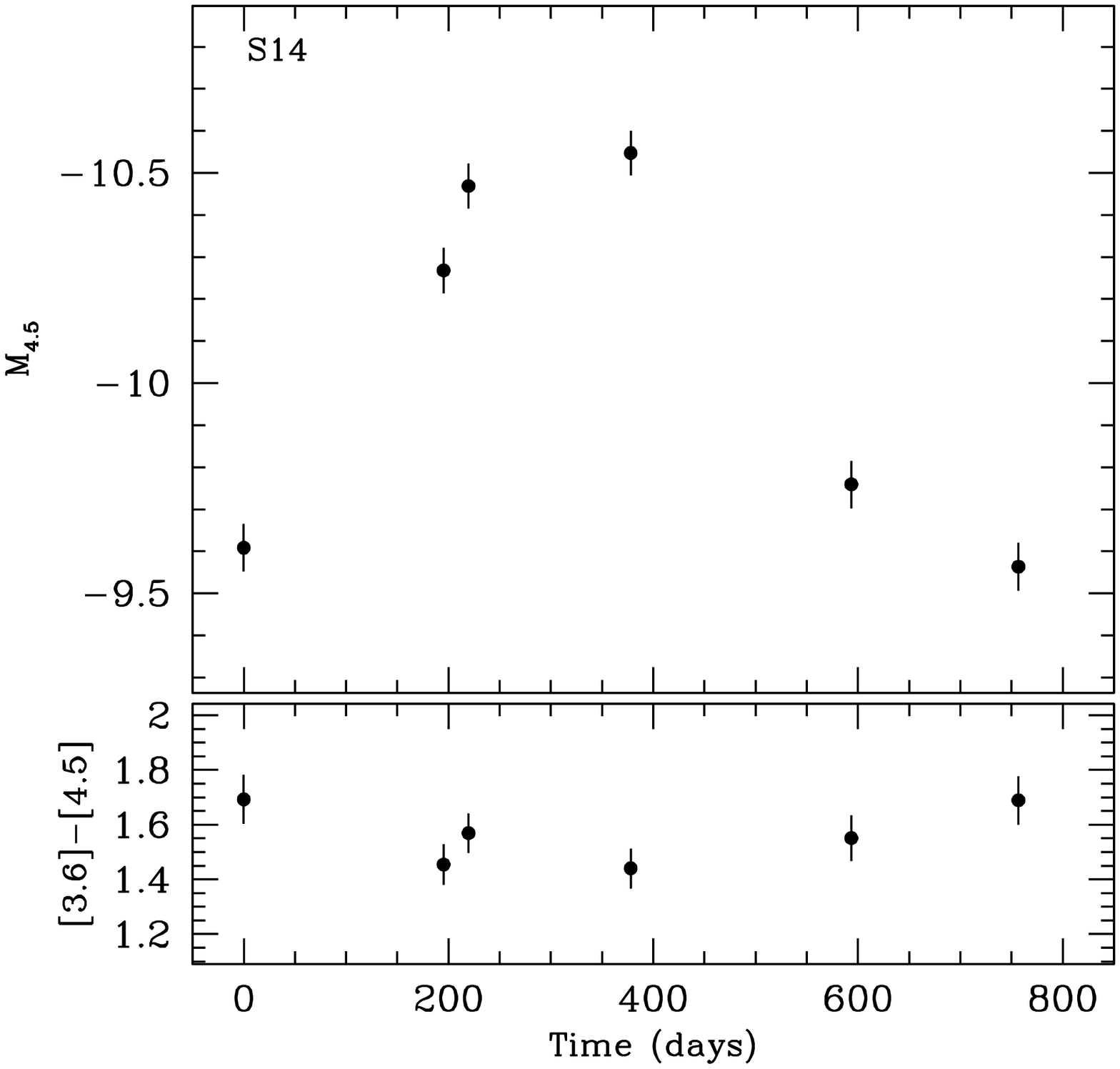}
\includegraphics[width=5.8cm]{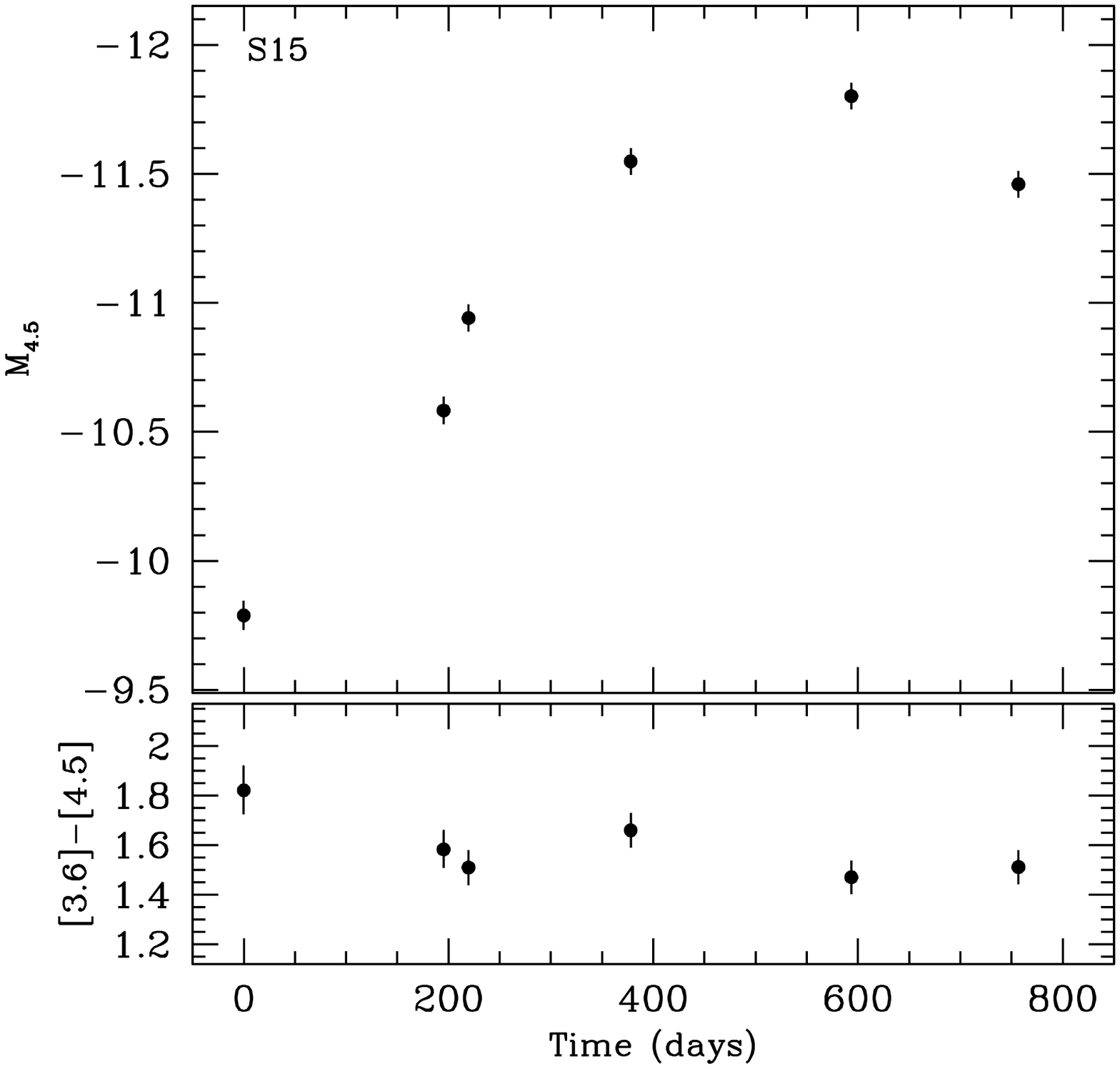}}
\centerline{
\includegraphics[width=5.8cm]{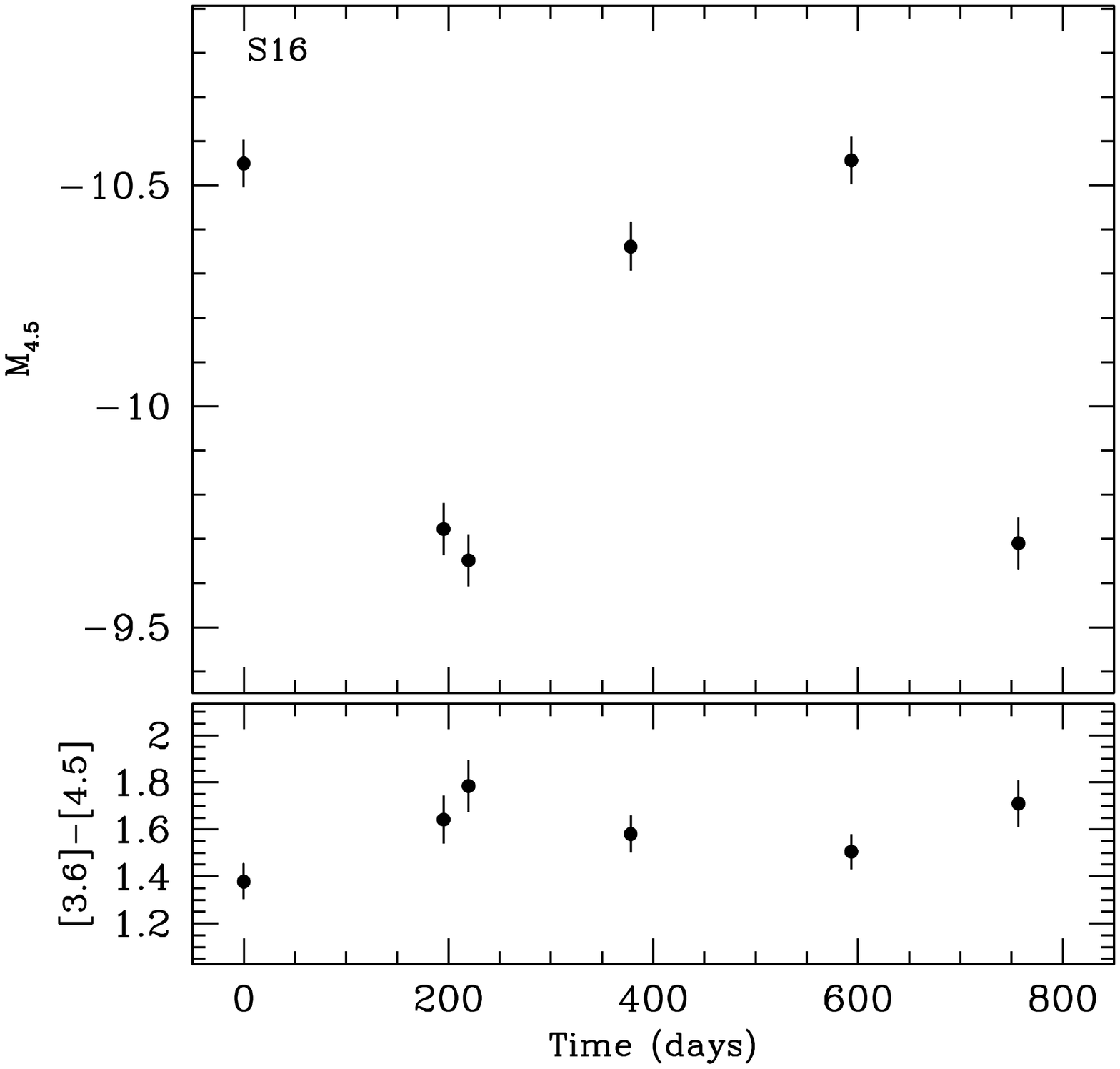} 
\includegraphics[width=5.8cm]{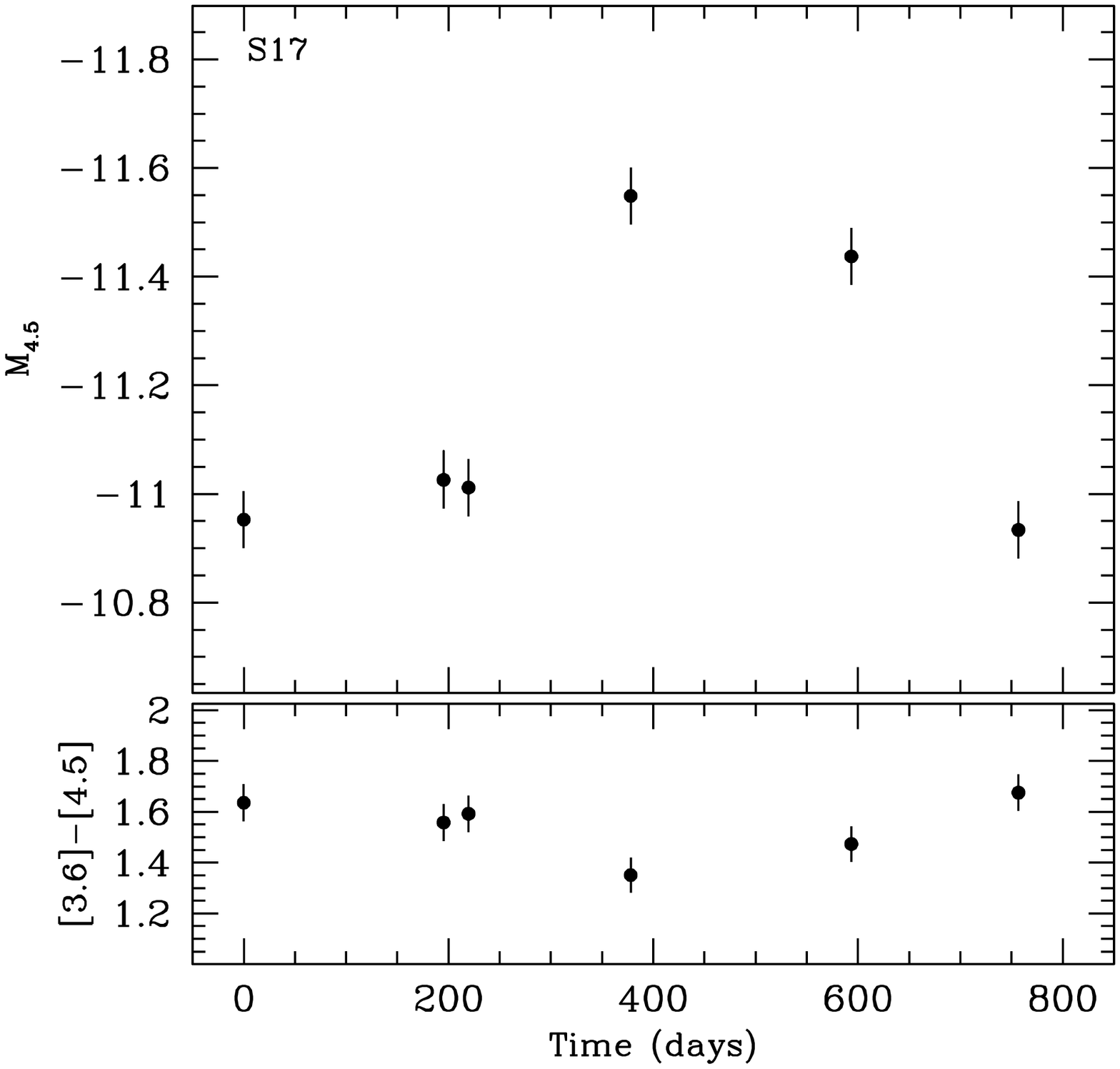}
\includegraphics[width=5.8cm]{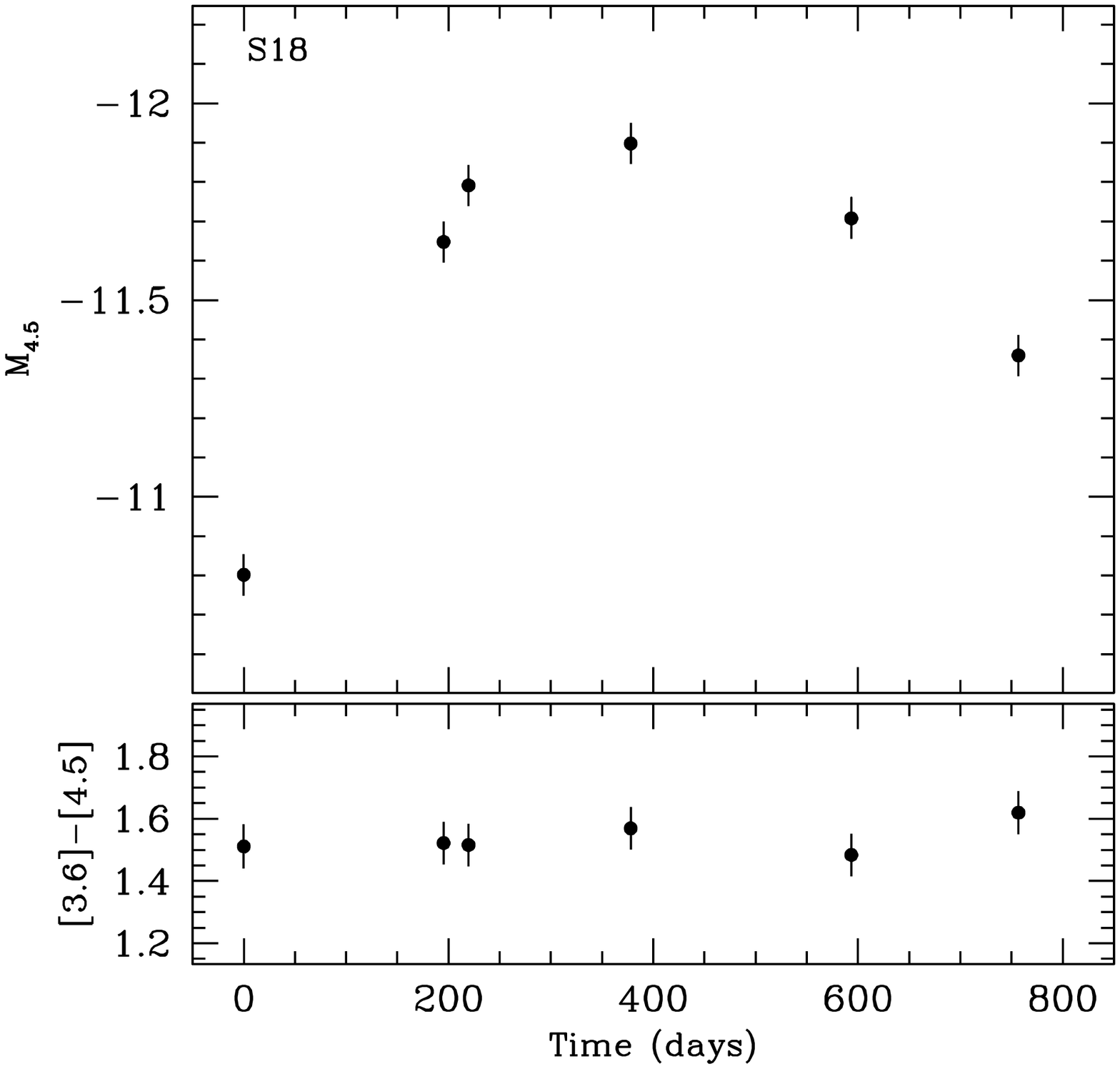}}
\caption{
Same as Figure \ref{fig:agb1}, but for sources S10-S18 of the 18 reddest
sources with $M_{4.5}<-10$ in the M33 MIR color-magnitude diagram 
(see Figs.~\ref{fig:cmd2} \& \ref{fig:sed}).}
\label{fig:agb2}
\end{center}
\end{figure*}

Section \ref{section:var} presents a discussion of the variability of 
the 18 reddest sources selected as EAGB stars (see Table \ref{table:eagb}
for their photometry).  In this Appendix (in Figs.~\ref{fig:agb1} \& \ref{fig:agb2}) 
we present the lightcurves for all 18 sources.
See the large open circles in Figures \ref{fig:cmd2}, \ref{fig:cc},
 and \ref{fig:rms}, as well as the left panel of \ref{fig:sed}
for a summary of their colors, SEDs, and RMS variability.


\section{B.~LBV Candidate Variability}
\label{appendix:lbv}

Like Appendix \ref{appendix:agb}, here we present the lightcurves for the 
16 LBV candidates from M07 that have been matched to the MIR point source
catalog, as described in \S\ref{section:m33}.  Table \ref{table:lbv}
lists photometry for these sources.
See Figures \ref{fig:cmd2},\ref{fig:cc}, \ref{fig:sed}, and \ref{fig:rms}
for a summary of their colors, SEDs, and RMS variability properties.

\begin{figure*}[b]
\begin{center}
\centerline{
\includegraphics[width=5.8cm]{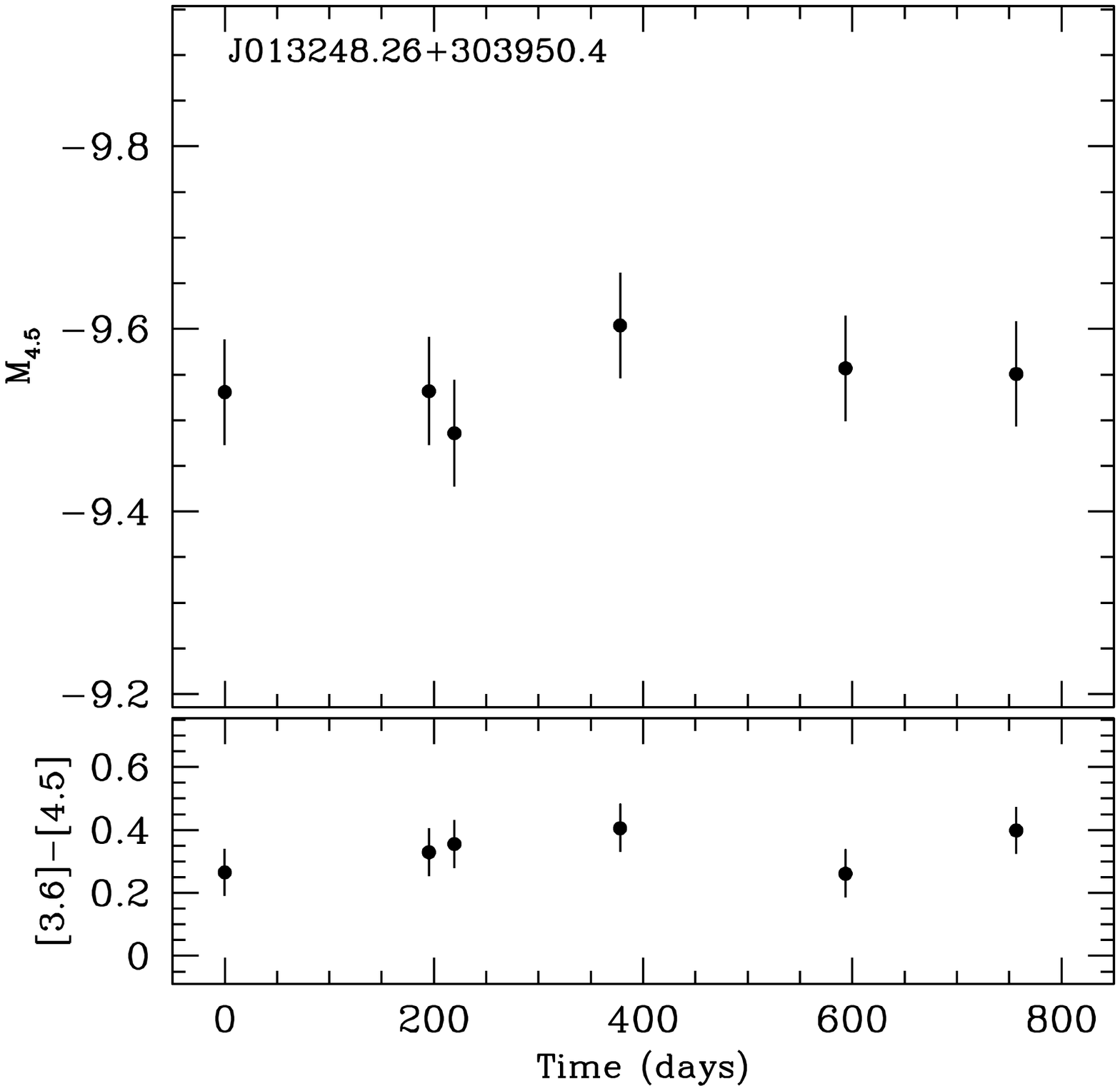} 
\includegraphics[width=5.8cm]{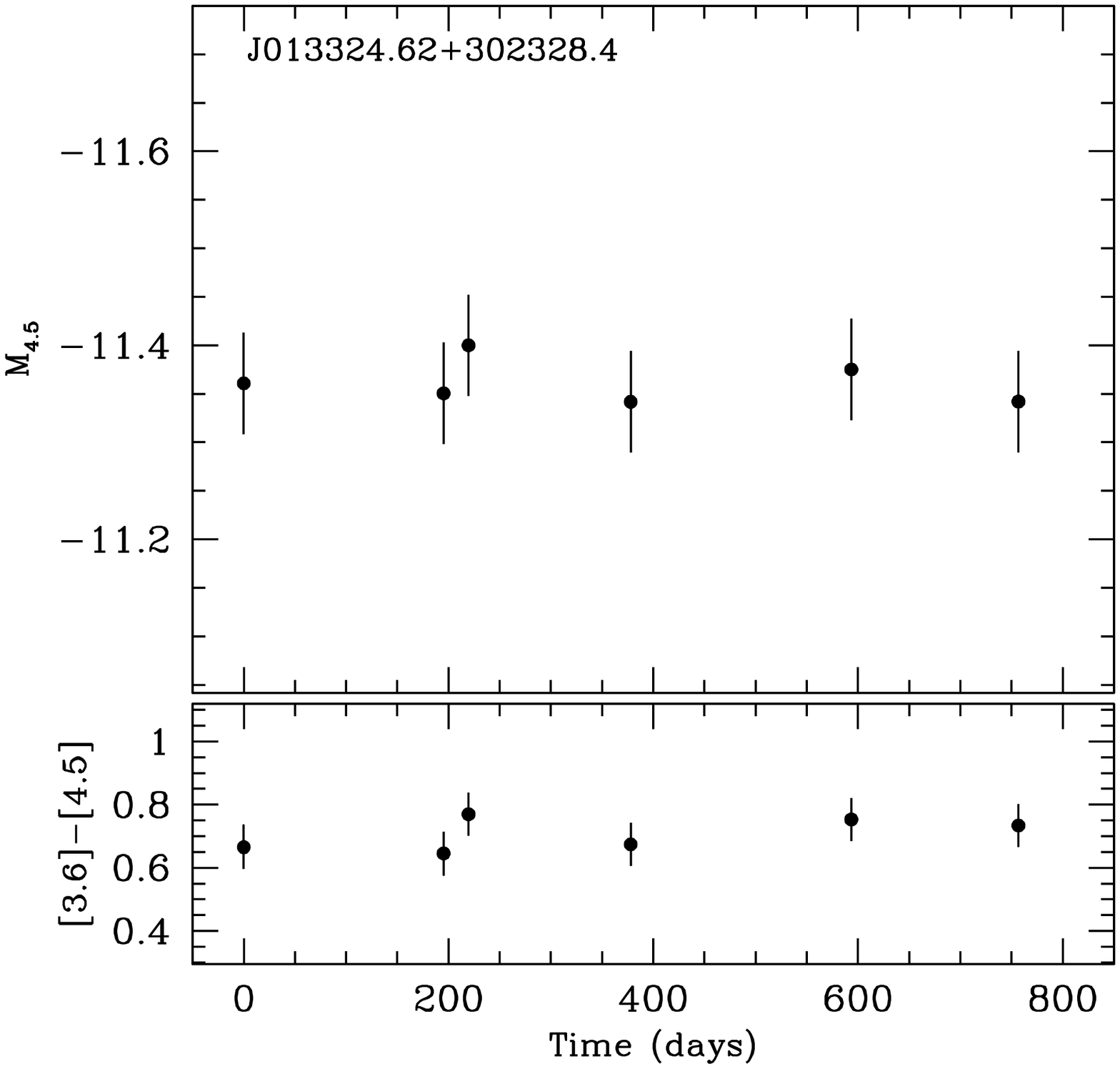}
\includegraphics[width=5.8cm]{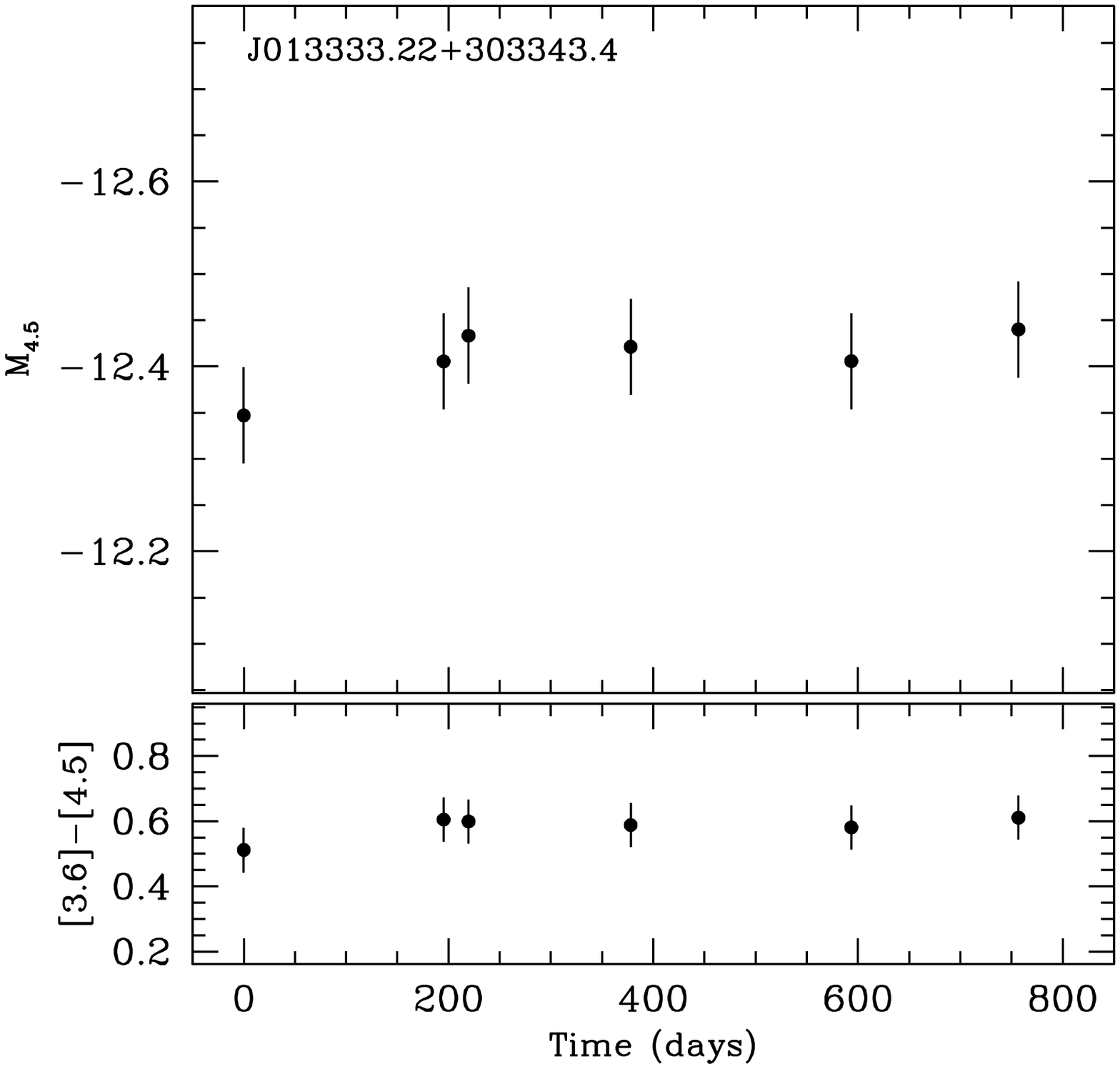}}
\centerline{
\includegraphics[width=5.8cm]{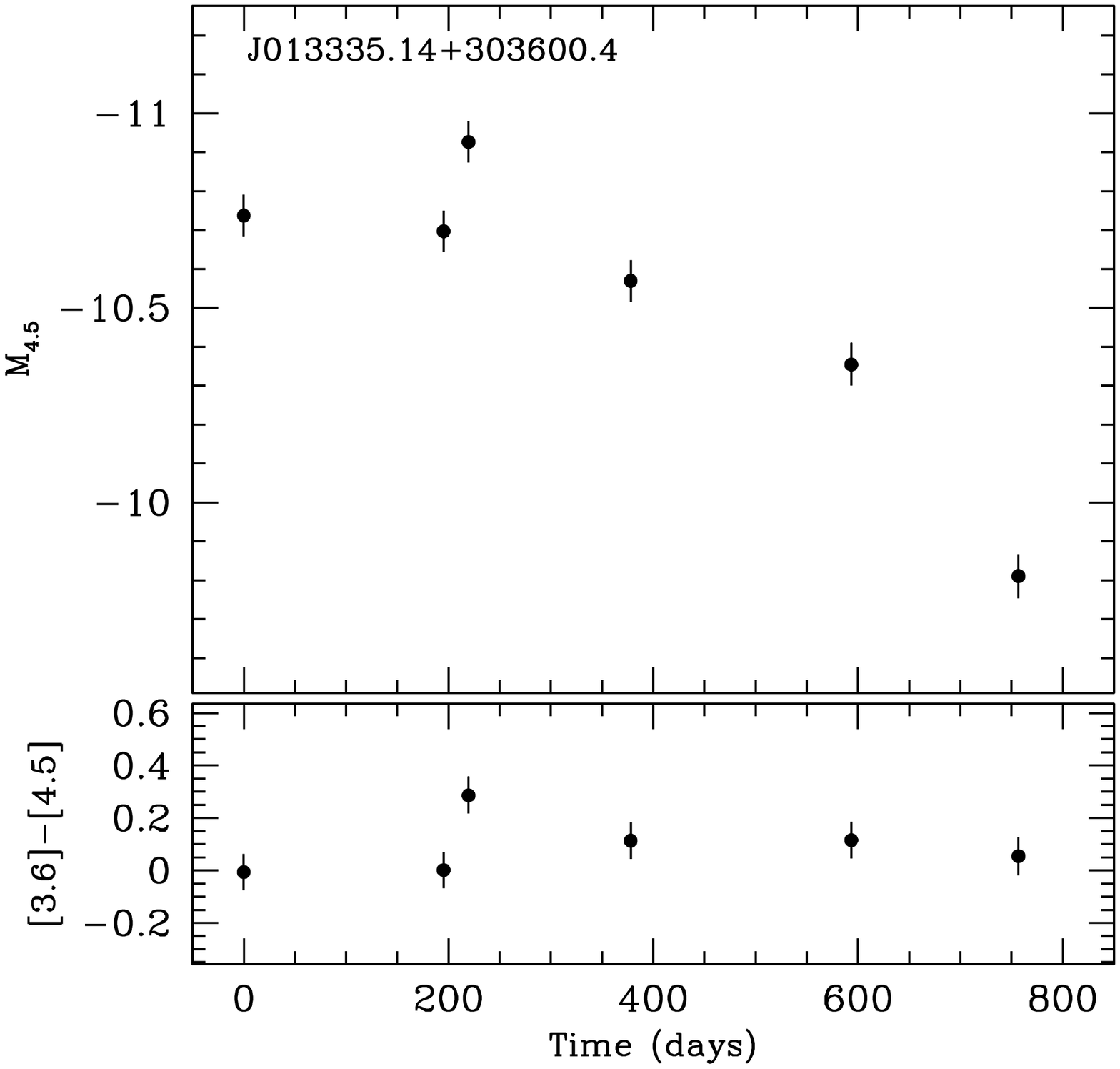} 
\includegraphics[width=5.8cm]{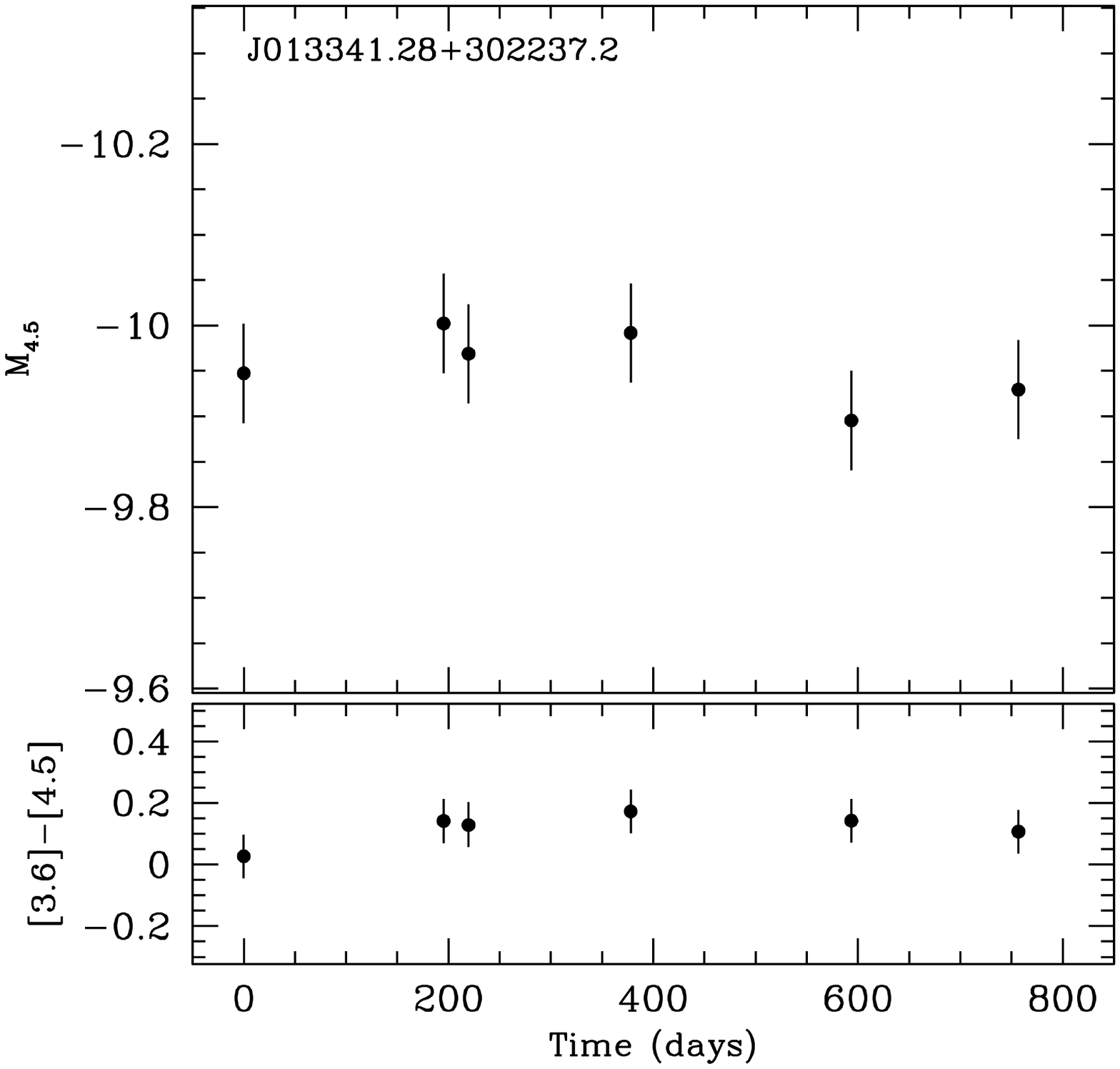}
\includegraphics[width=5.8cm]{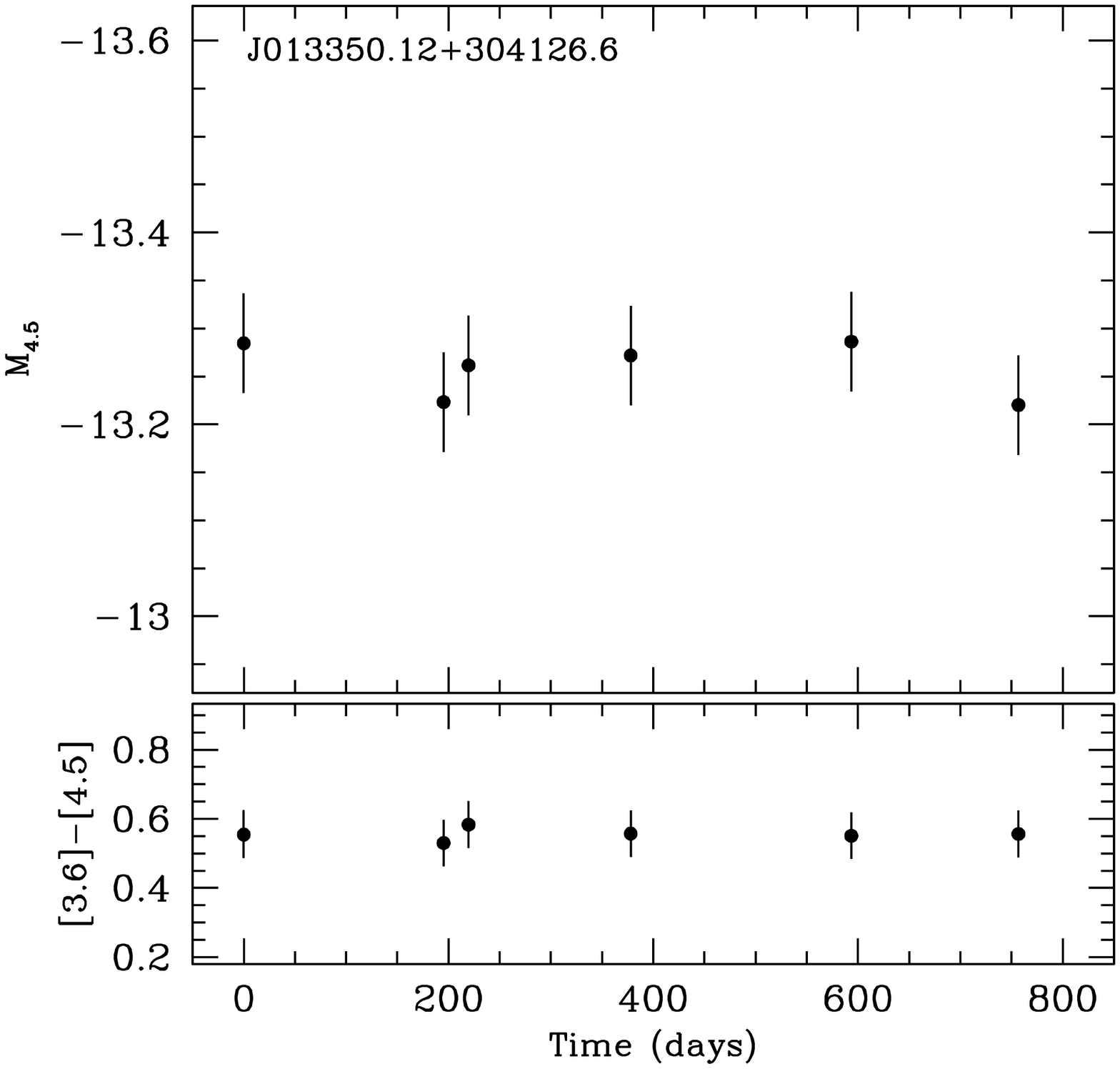}}
\centerline{
\includegraphics[width=5.8cm]{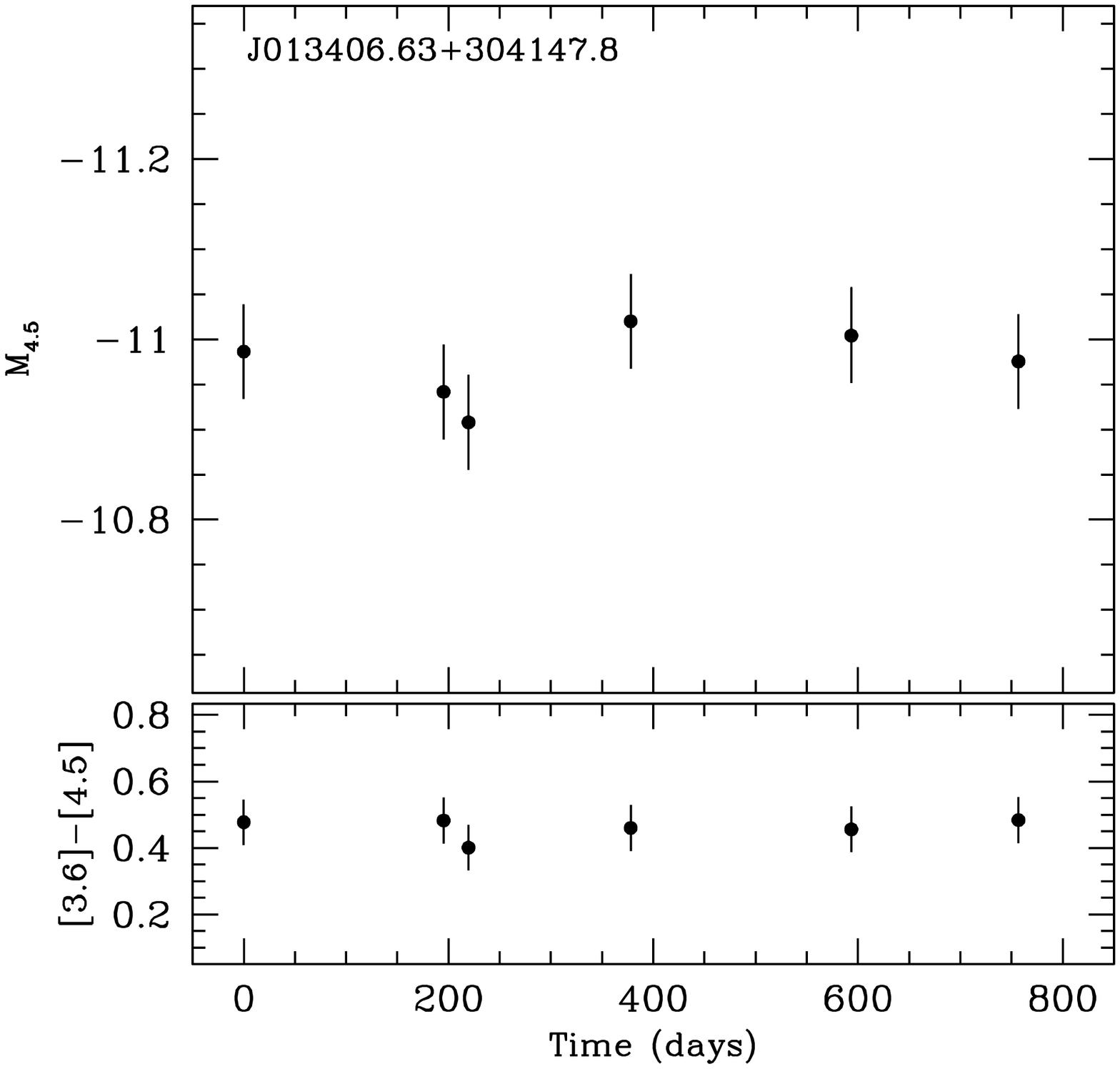} 
\includegraphics[width=5.8cm]{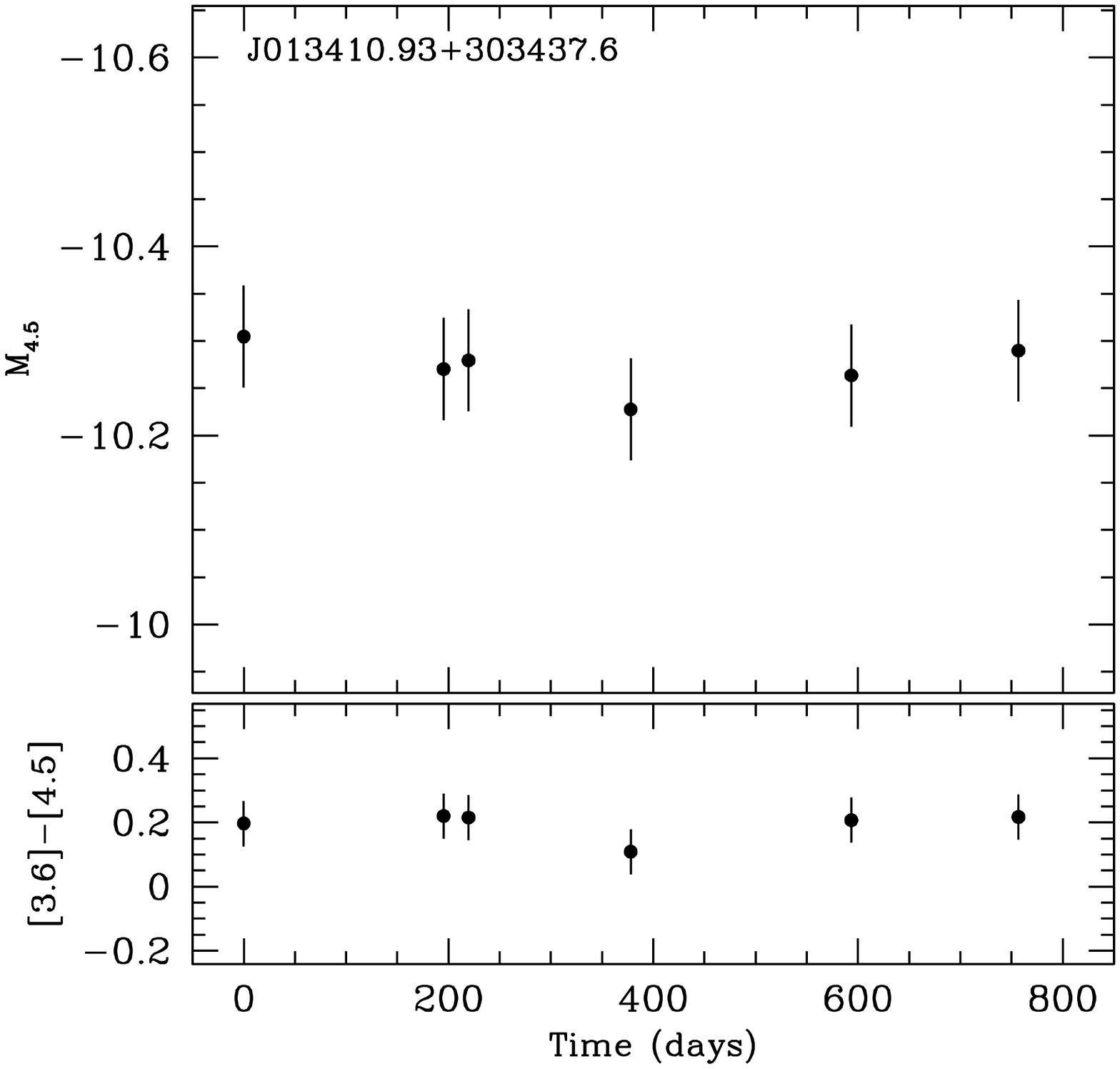}
\includegraphics[width=5.8cm]{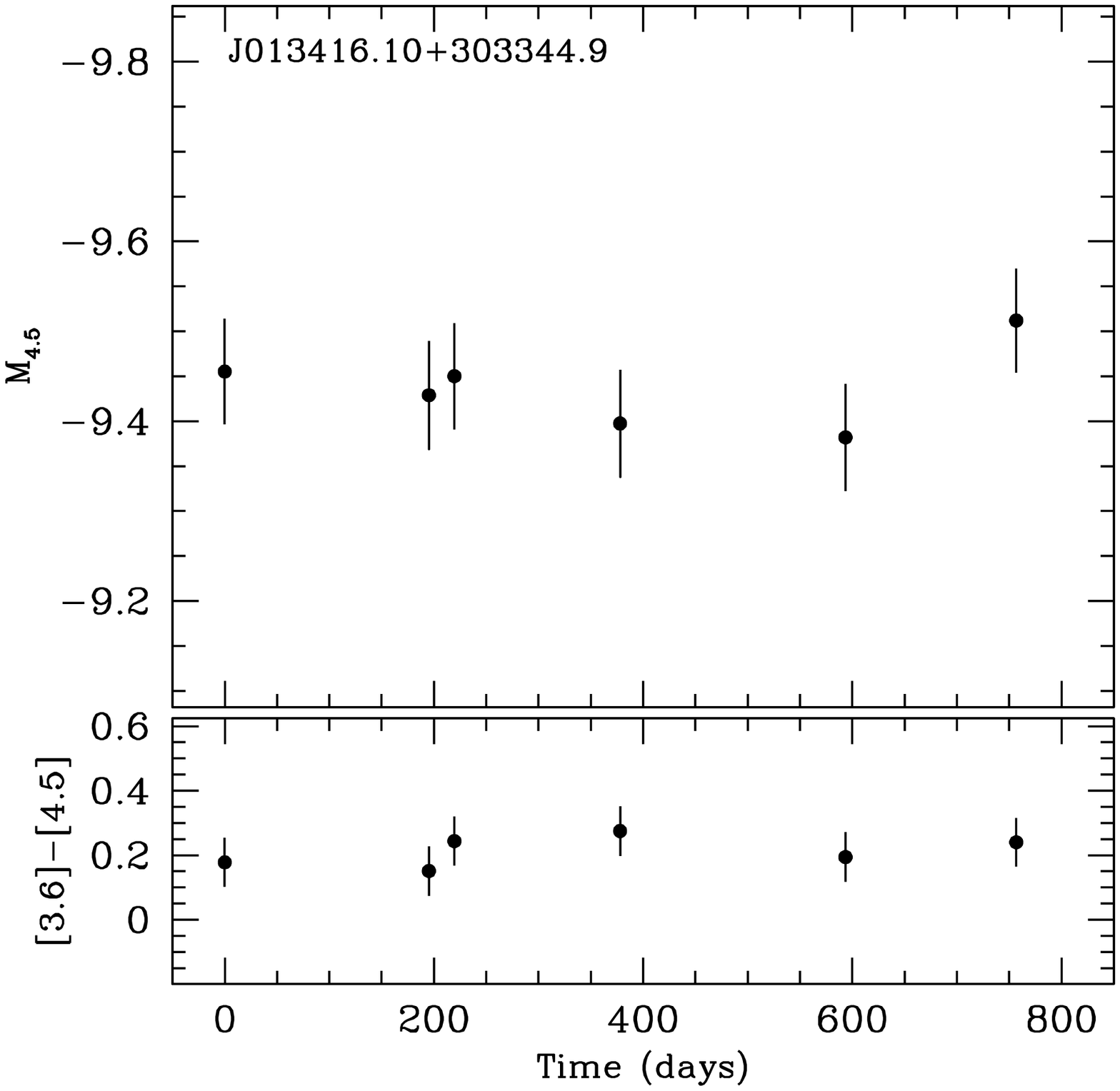}}
\caption{
Lightcurves at 4.5\,$\mu$m for 9 of the 16 LBV candidates of 
M07. Although there are several exceptions (notably J013335.14+303600.4 above
and  J013429.64+303732.1 in Fig.~\ref{fig:lbv2}), the LBV candidates do 
not vary significantly in absolute magnitude or color (see also Fig.~\ref{fig:rms}).  
As shown in Fig.~\ref{fig:cmd2} (open triangles), 
the LBV candidates with high luminosities at 4.5$\mu$m are 
characteristically more red than those with larger $M_{4.5}$. }
\label{fig:lbv1}
\end{center}
\end{figure*}
\begin{figure*}
\begin{center}
\centerline{
\includegraphics[width=5.8cm]{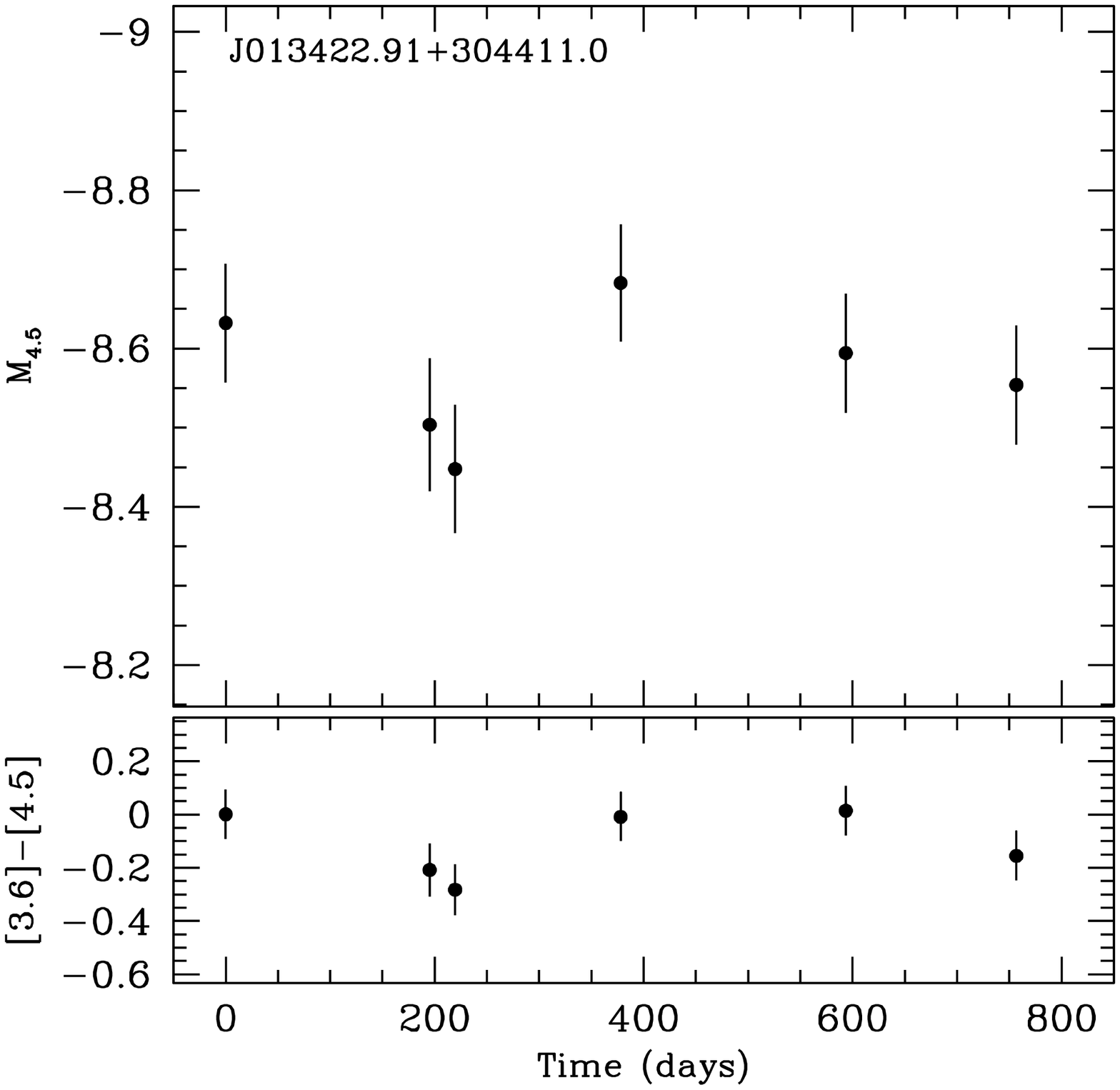} 
\includegraphics[width=5.8cm]{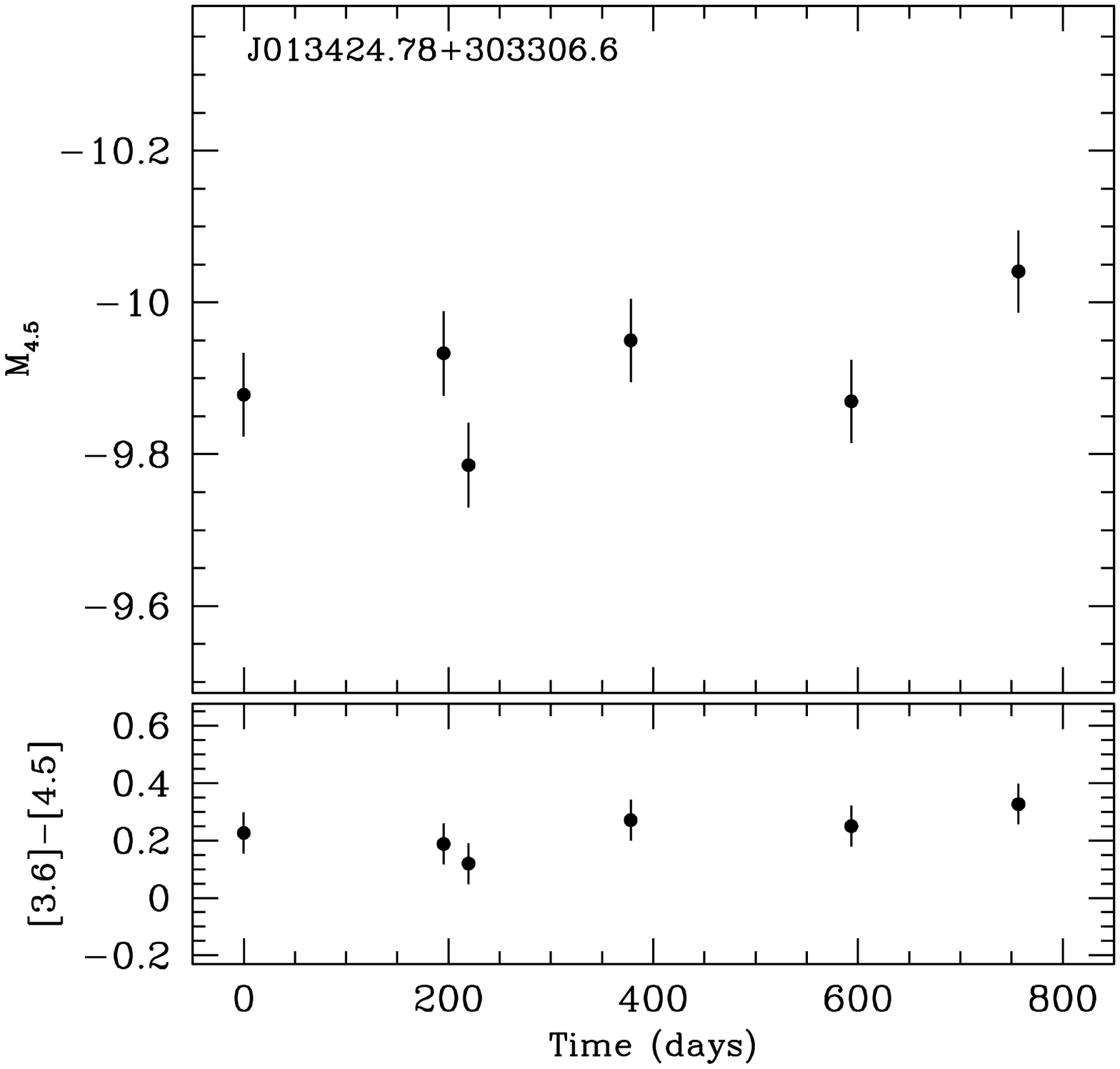}
\includegraphics[width=5.8cm]{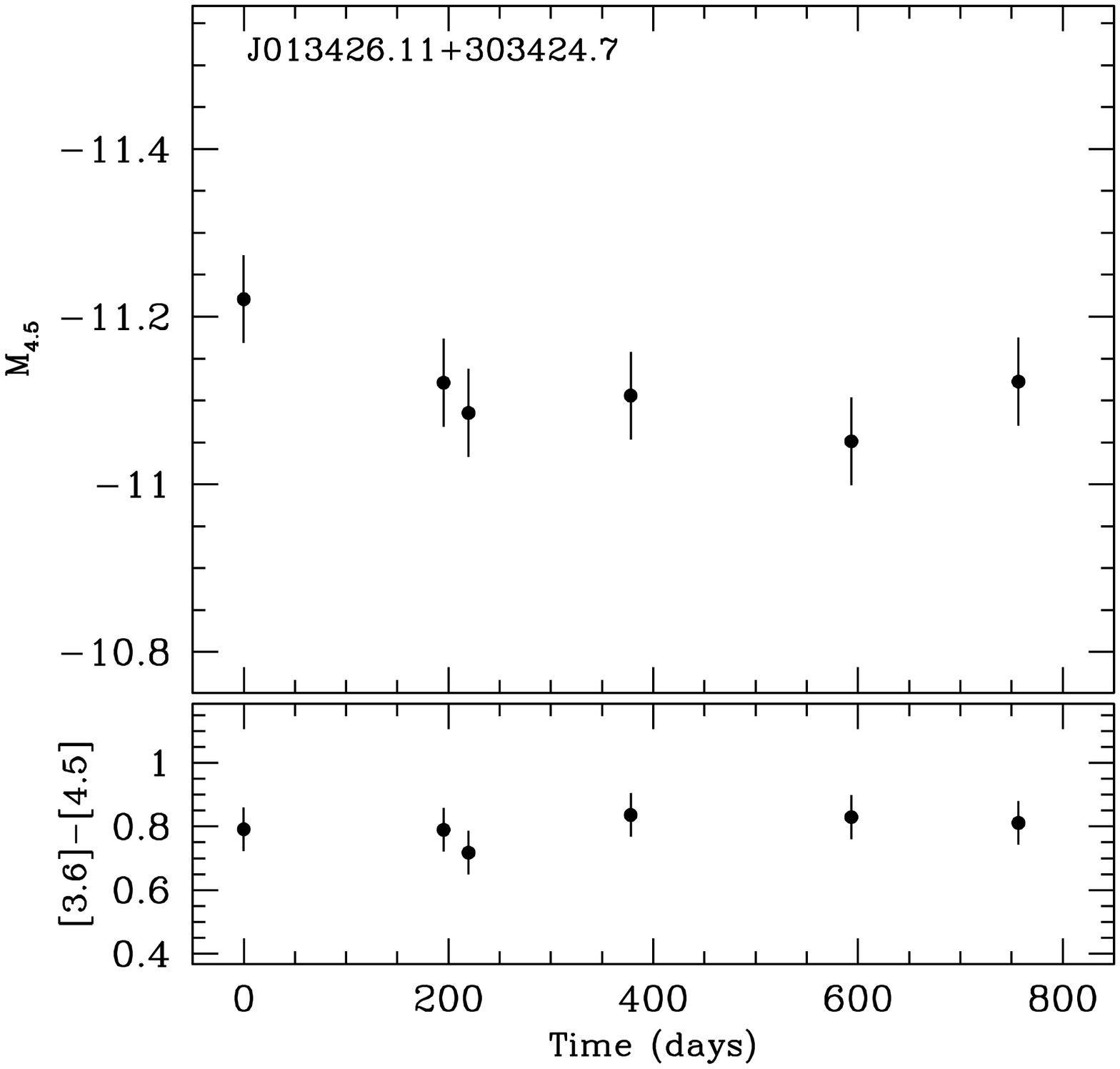}}
\centerline{
\includegraphics[width=5.8cm]{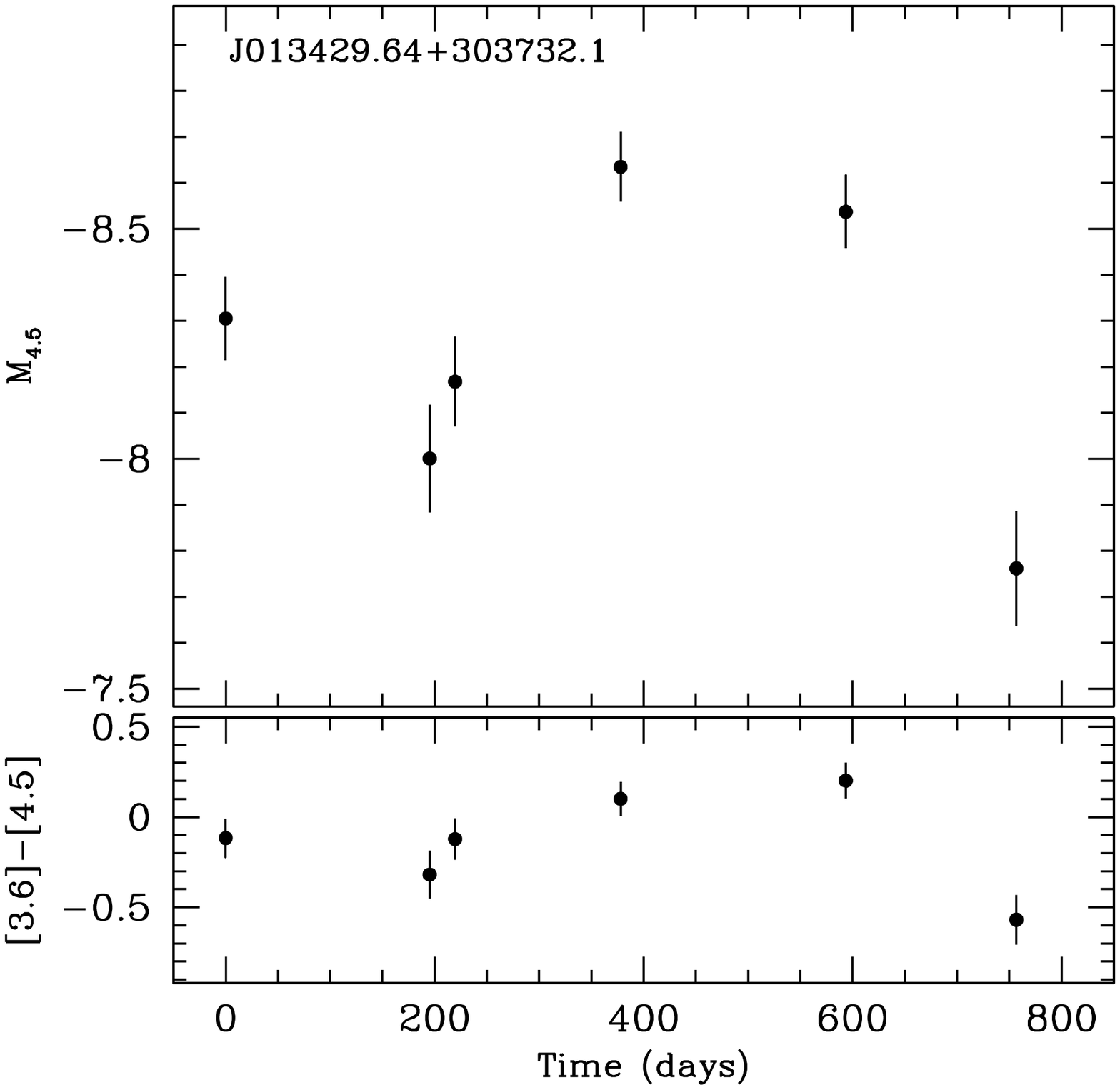} 
\includegraphics[width=5.8cm]{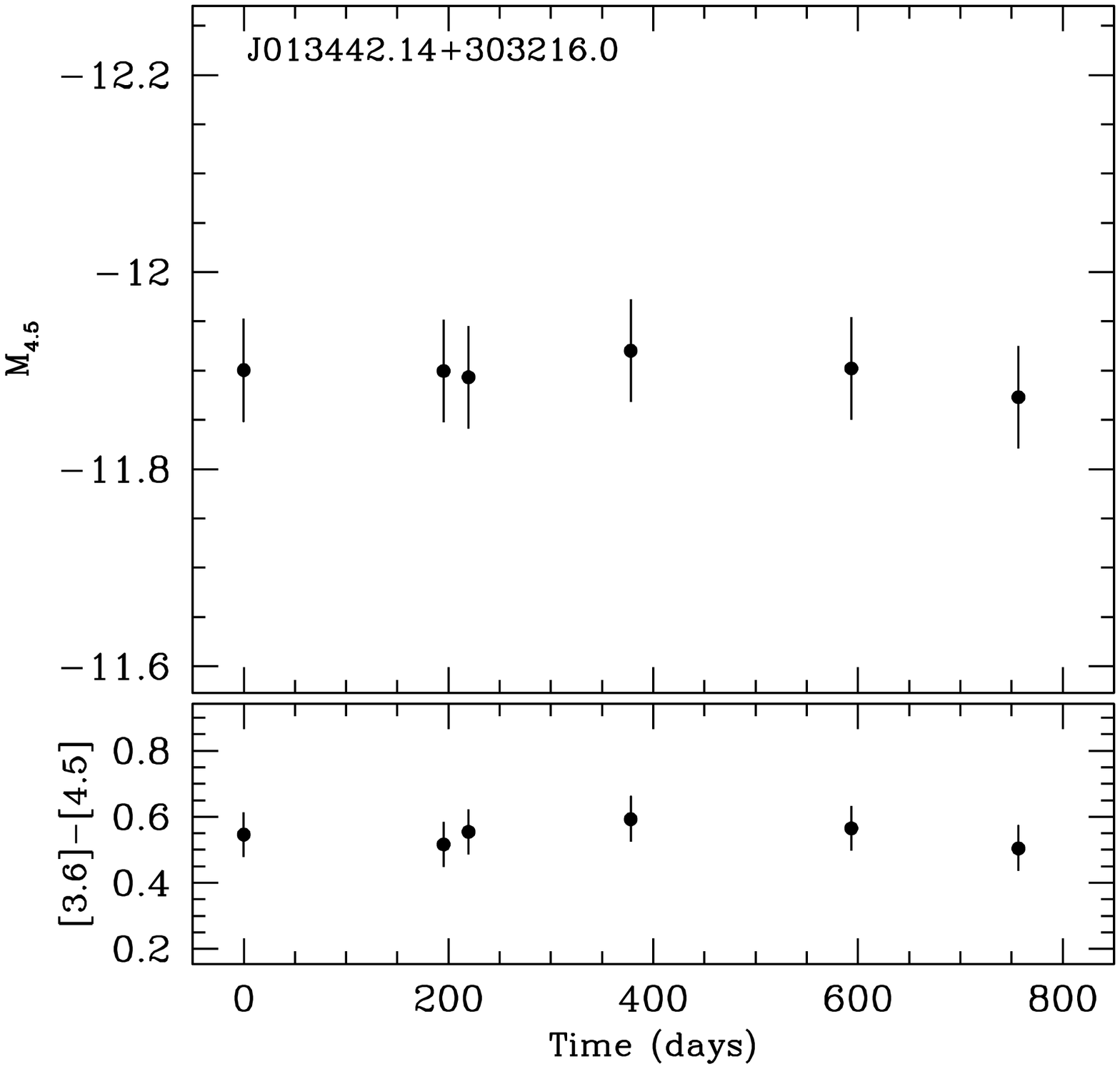}
\includegraphics[width=5.8cm]{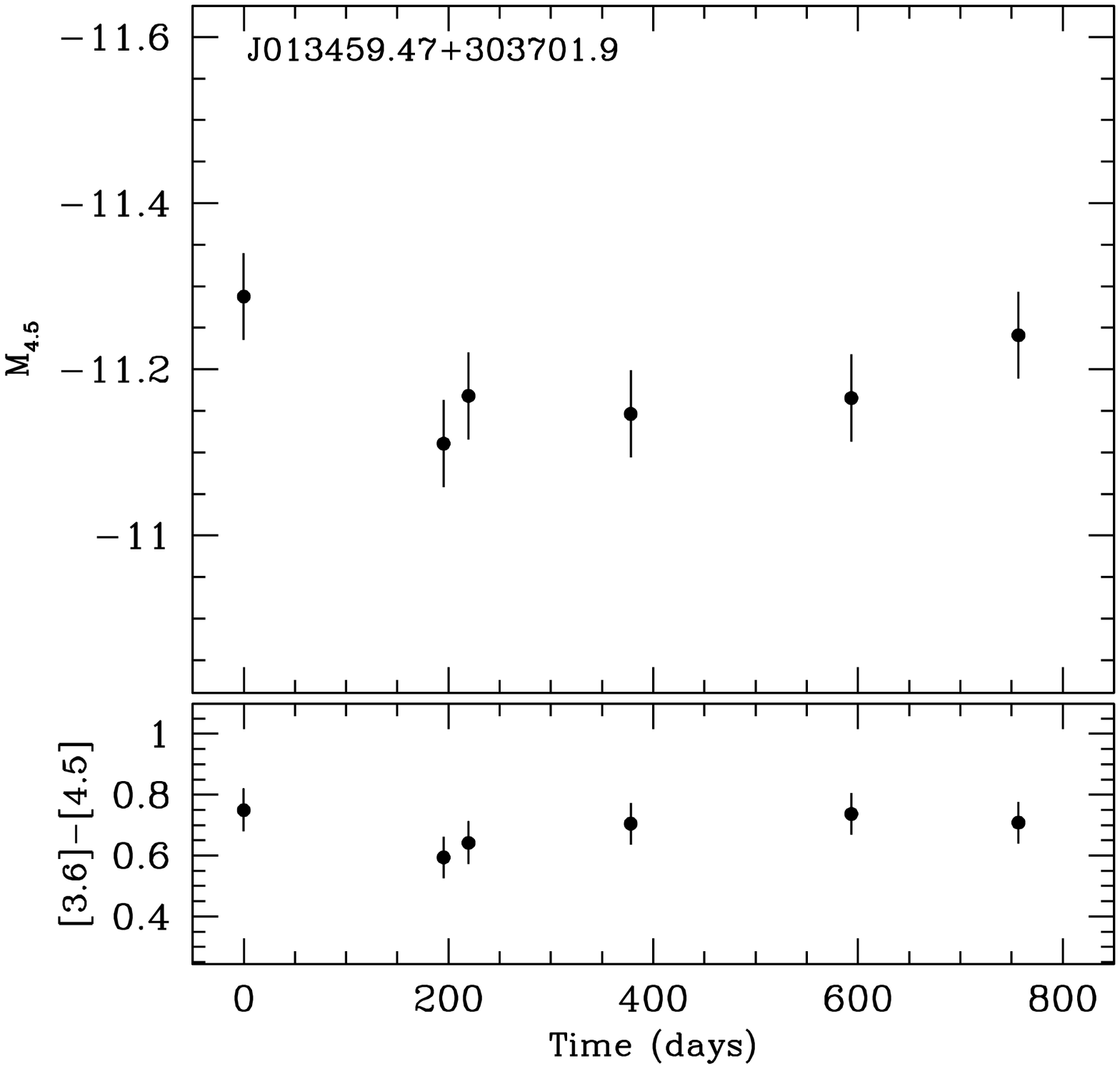}}
\centerline{
\includegraphics[width=5.8cm]{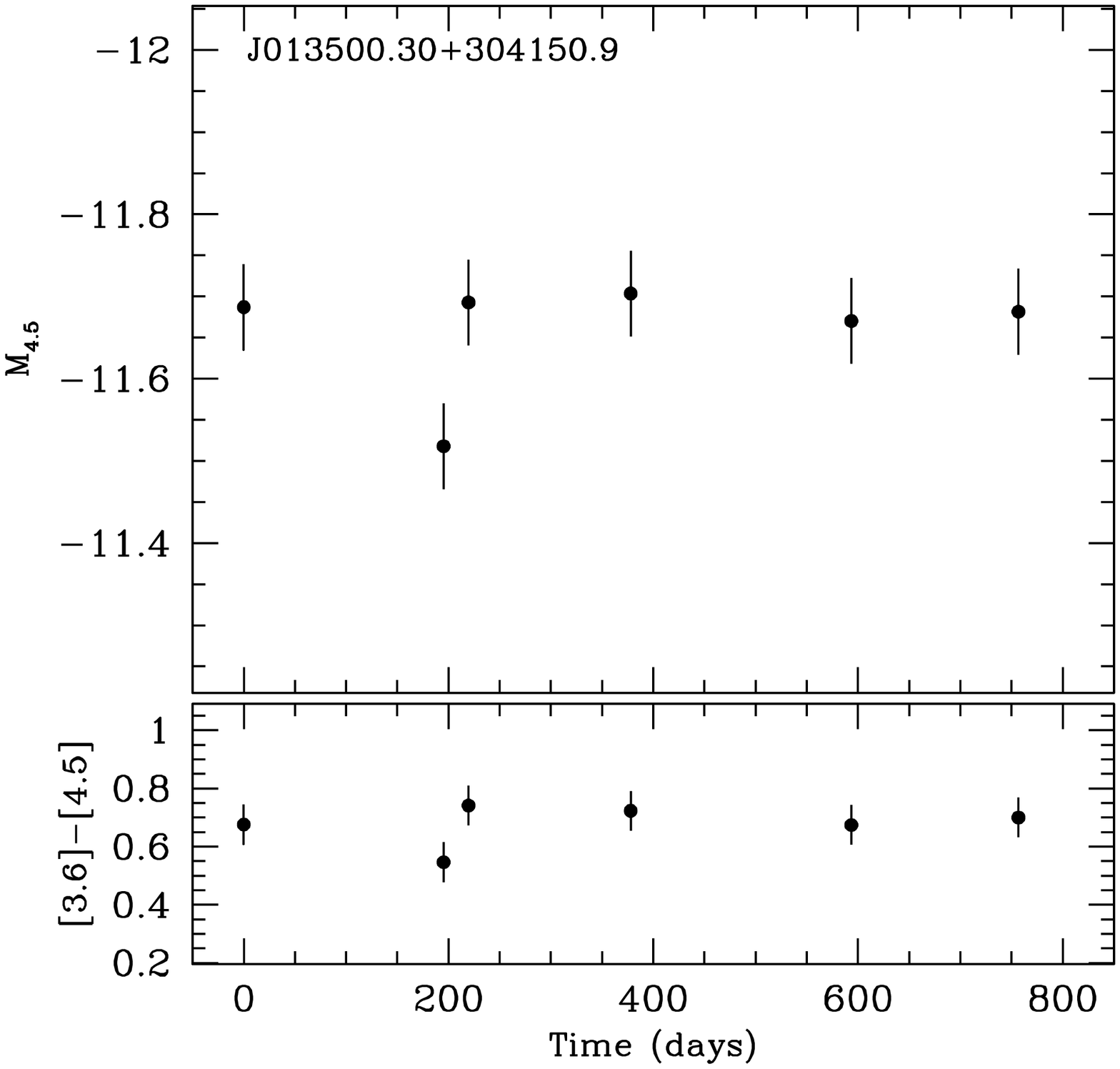}}
\caption{
Same as Figure \ref{fig:lbv1}, but for the remaining
7 LBV candidates of M07, matched to our 4.5$\mu$m catalog.}
\label{fig:lbv2}
\end{center}
\end{figure*}

\end{appendix}

\end{document}